\documentclass[fleqn,usenatbib]{mnras}
   
\usepackage{amsmath}
\usepackage{indentfirst}
\usepackage{graphicx}
\usepackage{multirow}
\usepackage{bigstrut}
\usepackage{diagbox}
\usepackage{caption}
\usepackage{afterpage}
\usepackage[figuresleft]{rotating}
\usepackage{threeparttable}
\usepackage{ulem}
\usepackage{lmodern}

\newcommand{\orcid}[1]{\href{https://orcid.org/#1}{\includegraphics[width=10pt]{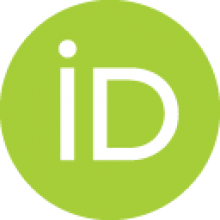}}}

\title[Detailed Mapping of the Galactic Disk Structure]{Detailed Mapping of the Galactic Disk Structure in the Solar Neighborhood through LAMOST K Dwarfs}

\author[Tang et al.]{
Xi-Can Tang$^{1}$\orcid{0009-0004-5137-0092},Hao Tian$^{2,3}$\thanks{E-mail:tianhao@nao.cas.cn}\orcid{0000-0003-3347-7596},Jing Li$^{1}$\thanks{E-mail:lijing@bao.ac.cn}\orcid{0000-0002-4953-1545},Bing-qiu Chen$^{4}$\orcid{0000-0003-2472-4903},Yi-Rong Chen$^{1}$,Chao Liu$^{2,3,6}$\orcid{0000-0002-1802-6917},
\newauthor
Dan Qiu$^{2,6}$\orcid{0000-0002-8280-4808}
\\
$^{1}$School of Physics and Astronomy, China West Normal University, 1 ShiDa Road, Nanchong 637002, China\\
$^{2}$Key Laboratory of Space Astronomy and Technology, National Astronomical Observatories, CAS, Beijing 100101,China\\
$^{3}$Institute for Frontiers in Astronomy and Astrophysics, Beijing Normal University, Beijing 102206,  China\\
$^{4}$South-Western Institute for Astronomy Research, Yunnan University, Kunming 650500, China\\
$^{5}$School of Astronomy and Space Science, University of the Chinese Academy of Sciences, Beijing 101408, China\\
$^{6}$University of Chinese Academy of Sciences, 100049, China\\
}
\date{Accepted 2024 April 23. Received 2024 April 23; in original form 2023 December 11}
   
\pubyear{2024}
\begin{document}

\label{firstpage}
\pagerange{\pageref{firstpage}--\pageref{lastpage}}
\maketitle

\begin{abstract}
The Galactic disk is one of the main components of the Milky Way, which contributes most of the luminosity. Its structure is essential for understanding the formation and evolution
of the Milky Way.  Using 174,443 K-type dwarf stars observed by both  LAMOST and Gaia DR3, we study the disk density profile in the local volume within 1,200 pc. 
In the azimuthal dimension, we find strong asymmetric signal of the thin disk. The surface density and the scale height of the southern disk significantly change versus the azimuthal angle at the same galactocentric distance $R$. 
Meanwhile, in the vertical dimension, the scale height of the northern disk has quite different trend than that of the southern one. The scale height of the southern disk shows a decreasing trend with  $\phi\sim-2.5^\circ$, and change to an increasing one with $\phi\sim5.0^\circ$. 
Meanwhile, the scale height of the northern disk has a consistently smaller increase. Finally, we divide the entire sample into three subsamples based on metallicity and all three subsamples show significant non-axisymmetric and north-south asymmetric signals in the Galactic disk. Furthermore, we find that the scale height of the metal-poor ([Fe/H] $<$ -0.4 dex) subsample in the northern disk is greater than that of the metal-rich ([Fe/H] $>$ -0.1 dex) subsample. However, in the southern disk, the scale height exhibits varying relationships across different metallicity slices.
\end{abstract}


\section{Introduction} \label{sec:intro}
Studying the structure of the Galactic disk in the solar neighbourhood provides valuable insights into the formation, 
stellar populations, distribution of dark matter, and evolutionary history of our Galaxy. From a macroscopic 
perspective, it is widely accepted that the density profile of stars in the Milky Way disk can be described
by either exponential \citep{gilmore1983new, juric2008milky, bovy2012spatial, bland2016galaxy, liu2017mapping} 
or ${\rm sech}^2$ \citep{van1988three, van2011galaxy, wang2018mapping, wang2020mapping} functions both radially
and vertically. The Milky Way disk can be divided into two main components: the thin disk and the thick disk
\citep{gilmore1983new}. Recent studies consistently show that the scale height ($h_z$) of 
young stellar populations is smaller compared to that of older populations, with the scale height of the 
thin disk ranging from $0.2$ to $0.45$ kpc, which is less than the range of $0.7$ to $1.2$ kpc for the thick
disk \citep{juric2008milky, carlin2012algorithm, liu2012chemo, bovy2012spatial, Xiang2018msto, Yu2021rc}. 
However, research on the scale length ($h_r$) of the Milky Way depends more on the stellar samples, leading
to more complex conclusions. \citet{wang2018mapping}, using LAMOST red giant branch, obtained a scale length
of $h_{r,\text{thin}} = 2.13 \pm 0.23$ kpc for the thin disk and $h_{r,\text{thick}} = 2.72 \pm 0.57$ kpc 
for the thick disk. Similarly, \citet{juric2008milky} also reported a thinner scale length for the thin 
disk ($h_{r,\text{thin}}=2.6$ kpc) compared to the thick disk ($h_{r,\text{thick}}=3.6$ kpc). However, 
some studies yielded contradictory conclusions. Based on the assumption that the element abundance crudely 
represented stellar age, \cite{bovy2012spatial}, using G dwarfs from SDSS/SEGUE, found that the scale 
length of young stellar populations was approximately 4.5 kpc, greater than the 2 kpc for older populations. 
Subsequently, \citet{wan2017red} also confirmed that the scale length of young stellar populations 
in the Galactic disk was approximately 4.7 kpc, exceeding the 3.4 kpc for older populations.

In the solar vicinity, \cite{ferguson2017milky} studied the scale heights of the thin and thick disks by analyzing K and M dwarfs with a stellar density profile described by the ${\rm sech}^2(z/2H)$ function. They found a scale height of $h_\mathrm{z,thin} = $ 0.242 $\pm$ 0.004\,kpc for the thin disk and $h_\mathrm{z,thick} = $ 0.734 $\pm$ 0.026\,kpc for the thick disk. These results were consistent with the previous findings of \citet{kent1991galactic}, who estimated a scale height of 0.247\,kpc at $R$ = 8\,kpc. Using a sample of late-type stars and employing a Bayesian approach, \citet{dobbie2020bayesian} found that the scale height of the northern thin disk is approximately $25\thinspace\%$ larger than that of the southern thin disk. Furthermore, \citet{yu2021flare} investigated the north-south asymmetrical structure near the Sun. They concluded that the scale heights of the southern and northern disks are 140\,pc and 220\,pc, respectively, at $R$ = 8.5 kpc. Moreover, they employed OB-type stars to demonstrate that the vertical position of the Galactic mid-plane ($Z_0$) is within 100\,pc, but it is nearly zero ($Z_0$ = 0) at the solar location.

However, for observers situated within the Milky Way disk, it presents a significant challenge to directly construct a comprehensive stellar profile of the Galaxy. Previous studies have commonly adopted a method where the distribution of stars in the $R$-$Z$ plane is characterized within the integrated azimuth. Through the analysis of kinematics, dynamics, and chemical abundance, researchers have made noteworthy observations of refined asymmetric structures in the Milky Way, including the warp \citep{2006A&A...451..515M,2018MNRAS.481L..21P,chen2019intuitive,wang2020mapping}, flare \citep{lopez2002old,2006A&A...451..515M,lopez2014flare,bovy2016stellar}, ripples \citep{newberg2002ghost,morganson2016mapping,price2015reinterpretation,xu2015rings}, snail shell \citep{antoja2018dynamically,xu2020exploring,2020ApJ...890...85L}, and others. Consequently, it is crucial to focus on describing the azimuthal properties of the Galactic disk and leverage potential asymmetric substructures to enhance our understanding of the dynamic origins and evolution of the Milky Way.

The Large Sky Area Multi-Object Fiber Spectroscopic Telescope \citep[LAMOST;][]{cui2012large,zhao2012lamost,luo2015first} Low Resolution Spectroscopic Survey (LRS, R $\sim$ 1,800) has successfully obtained more than 11 million spectra spanning the wavelength range from 3690 to 9100 \AA\ since 2011. This extensive dataset from LAMOST provides us with a remarkable opportunity to investigate the detailed structure of the Milky Way disk. In a recent study (Tian et al. in preparation; Paper I), K giants from LAMOST Data Release 7 (DR7) were utilized to create stellar distribution maps in both the azimuthal and radial directions of the Milky Way. They revealed an asymmetric distribution of stars in the azimuthal direction, specifically within the range of $R$ $<$ 12\,kpc, which gradually diminishes as the Galactocentric distance increases, particularly towards the outer disk. Additionally, a pronounced exponential relationship was observed when examining the azimuthal angle relative to the Galactocentric distance, whether derived from linear density or scale height gradients. 

In this study, our primary objective is to investigate the intricate nature of the Galactic disk, specifically in the solar neighbourhood, taking into account the asymmetric characteristics present in both the northern and southern Galactic disks. To achieve this, we will employ high-sampling-rate and high-distance-precision samples of K dwarfs obtained from LAMOST data release 9 (DR9). The paper is organized as follows. In Section \ref{sec:data}, we outline our methodology for selecting the K dwarf samples from LAMOST DR9, while in Section \ref{sec:se}, we discuss how we reconstructed the stellar density along a given line-of-sight. Additionally, in Section \ref{sec:profile}, we provide a concise overview of how we model number densities. Moving forward, we present the results and discussions of the asymmetric substructures of the disk in the solar neighborhood in Sections \ref{sec:result} and \ref{sec:discussion}, respectively. Finally, we conclude the paper with our findings in the last section.

\section{Data} \label{sec:data}

K dwarfs are late-type stars with low luminosity. They present a unique opportunity to explore
the three-dimensional structures of the Galactic disk on small scales in the solar neighborhood.
The spectra of K dwarfs showcase prominent absorption lines of metal oxides \citep{lena2013observational}, 
enabling their differentiation from other stellar types based on spectral analysis. There are 1,233,258 K-type stars in LAMOST DR9 
selected according to the spectra. The parameters of these 
stars are derived using the LAMOST Stellar Parameter pipeline \citep[LASP;][]{wu2014automatic}, 
with precision values of 110 K for effective temperature ($T_\mathrm{eff}$), 0.19 dex for surface 
gravity (log $g$), and 0.11 dex for metallicity ([Fe/H]). Notably, [Fe/H] holds particular 
significance in characterizing the chemical compositions of the Galactic disk.

We cross-match our selected K dwarf catalogue with the $Gaia$ EDR3 \citep{brown2021gaia} and Two Micron All-Sky Survey \citep[2MASS;][]{skrutskie2006two} using a search radius of $3\arcsec$ to obtain reliable photometric ($G, B_P, R_P, J, H, K$ magnitudes) and astrometric data. The $Gaia$ G-band magnitudes of our sample stars range from 11 to 18\,mag, corresponding to mean parallax errors of 0.018 to 0.120 mas \citep{lindegren2021gaia}. Considering the small parallax errors and zero-point bias, along with the limited depth of K dwarfs detected by LAMOST, we can directly adopt the astrometric distances of $Gaia$ ($r_{\mathrm{geo}}$) as the distances of our catalogued K dwarfs.

  \begin{figure}
    \centering
    \includegraphics[height=0.35\textwidth]{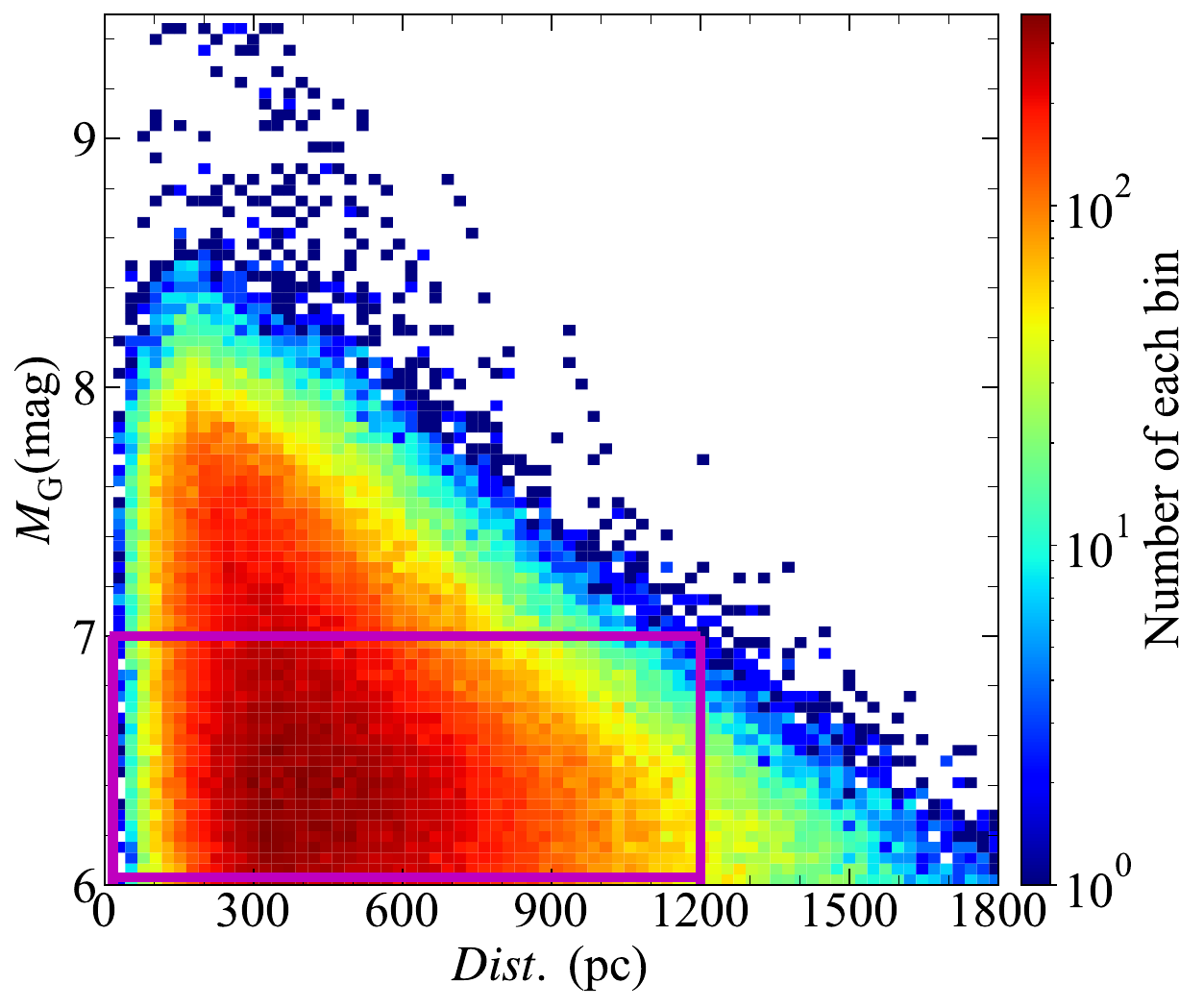}
    \caption{Distribution of all K dwarfs selected from LAMOST DR9 in the $M_\mathrm{G}$ versus distance space. The colours represent the observed star count for each pixel, while the selection of the K dwarf sample is indicated by the magenta box.}
    \label{fig:dist_mg}
\end{figure}

In this study, we calculate the intrinsic colors $(G_\mathrm{BP}-G_\mathrm{RP})_\mathrm{0}$ of K dwarfs by utilizing the reddening $E(B-V)$ obtained from the 3D dust maps of \citet{green20193d} and the reddening coefficients provided by \citet{casagrande2018use}. The calculation is performed using the formula $(G_\mathrm{BP}-G_\mathrm{RP})_\mathrm{0}=(G_\mathrm{BP}-G_\mathrm{RP})-1.339E(B-V)$. Furthermore, we determine the $Gaia$ G-band absolute magnitudes ($M_\mathrm{G}$) using the equation $M_\mathrm{G} = G-5\mathrm{lg}\,d+5-A_G$, where we adopt the G-band extinction coefficient from \citet{chen2019three} to derive the G-band extinction, given by $A_\mathrm{G}=1.91E(B-V)$. 

To ensure the exclusion of misidentified sources among the K dwarf candidates and to address stellar selection effects, we apply the following criteria:
\begin{enumerate}
\item LAMOST $g$-band signal-to-noise ratio SNRg $>$ 20.0.
\item $Gaia$ Renormalised Unit Weight Error RUWE $<$ 1.4.
\item $Gaia$ parallax ($\varpi$) and parallax errors ($\sigma_\varpi$) satisfy $\varpi$/$\sigma_\varpi$ $>$ 5.0 and $\varpi$ $>$ 0.0\,mas.
\item $Gaia$ photometric uncertainties $\sigma_{G_\mathrm{BP}}$, $\sigma_{G_\mathrm{RP}}$, $\sigma_{G}$ $<$ 0.1\,mag.
\item $G$-band magnitudes $G$ $<$ 18.3\,mag.
\end{enumerate}

The decline in the number of observable K dwarf stars becomes evident as magnitude and distance increase, indicating varying levels of data completeness for stars with different absolute magnitudes. In order to mitigate the impact of these selection effects and ensure the inclusion of the maximum range of observable distances for all stars, we incorporate additional essential criteria based on the colour-magnitude diagram (CMD) and distance-magnitude diagram obtained from the $Gaia$ mission (see Fig.\ref{fig:dist_mg}). For K dwarf stars, the following additional criteria are employed:
\begin{enumerate}
\item $0.4 < (G_\mathrm{BP}-G_\mathrm{RP})_0 < 2.2$\,mag;
\item $6.0 < M_\mathrm{G} < 7.0$\,mag;
\item $Dist. < 1,200$\,pc.
\end{enumerate}

\begin{figure}
    \centering
    \includegraphics[width=0.40\textwidth]{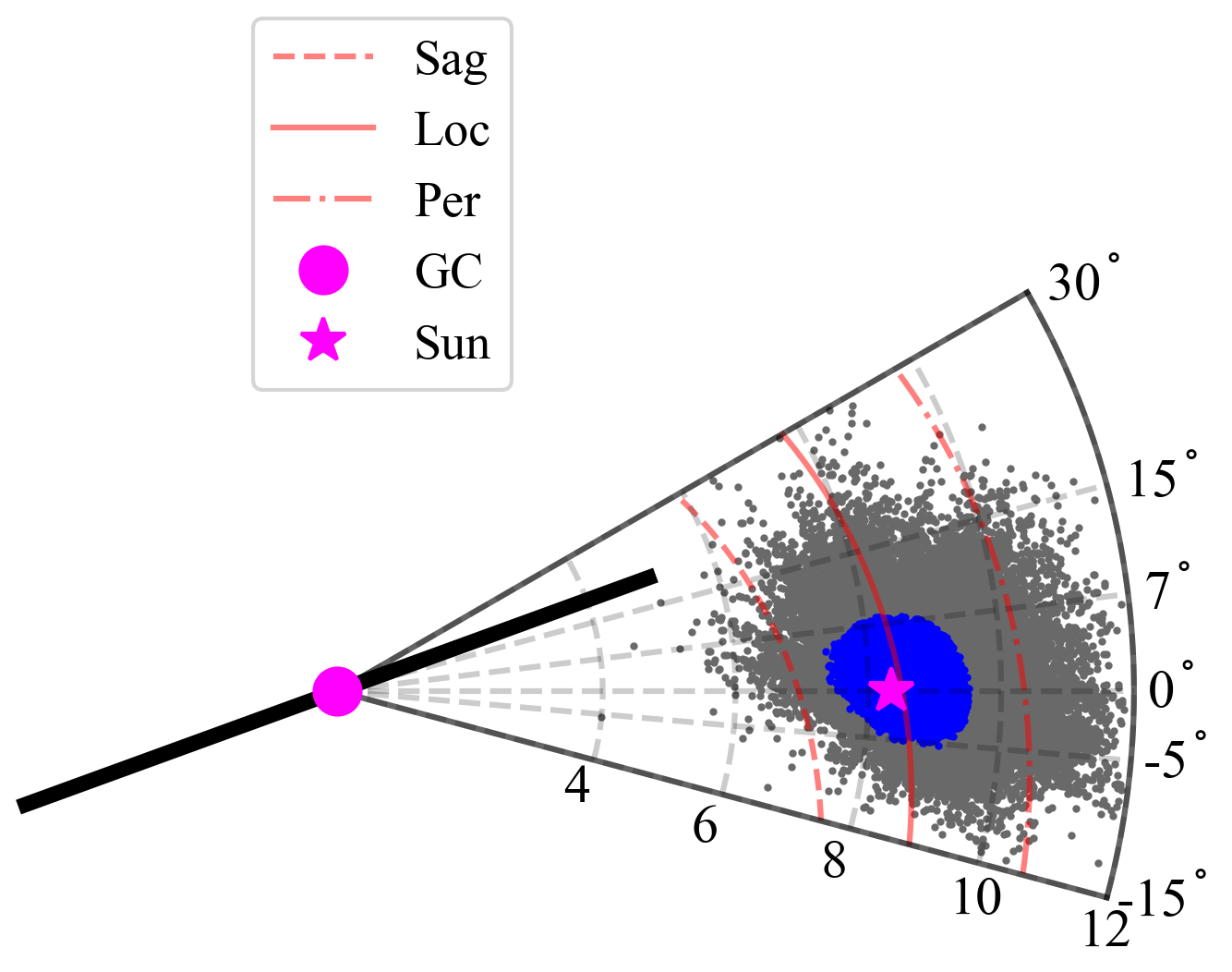}
    \caption{The spatial distribution of K dwarfs in the $R-\phi$ plane. 
    Blue dots represent the K dwarfs selected in this study. And grey dots represent 
    all K dwarfs from LAMOST DR9. The pentagram denotes the Sun's position, while solid magenta dot
    denotes the Galactic center, and black line indicates the Galactic Bar. Spiral arms are represented by the 
    red lines.}
    \label{fig:dk_RPhi_fan}
\end{figure}

By applying the selection criteria mentioned above, we acquire a total of 174,334 K dwarfs for our final sample. We adopt the solar Galactocentric radius and vertical position as
$(R_{\sun}, z_{\sun})$ = (8.34, 0.027)\,kpc \citep{reid2014trigonometric,chen2001stellar},
and calculate the cylindrical coordinates of the K dwarf stars centered at the Galactic center, with the Sun 
set $\phi=0^\circ$, and the rotation direction as the positive direction of the azimuthal angle $\phi$.
The spatial distribution of the selected K dwarfs is illustrated in Fig.~\ref{fig:dk_RPhi_fan}, 
and the final K dwarf sample is primarily concentrated in the range
of $\phi$ = -5$^\circ$ to 7$^\circ$ and $R$ = 7.5 to 9.5 kpc.

\begin{figure*}
    \centering
    \includegraphics[width=0.95\textwidth]{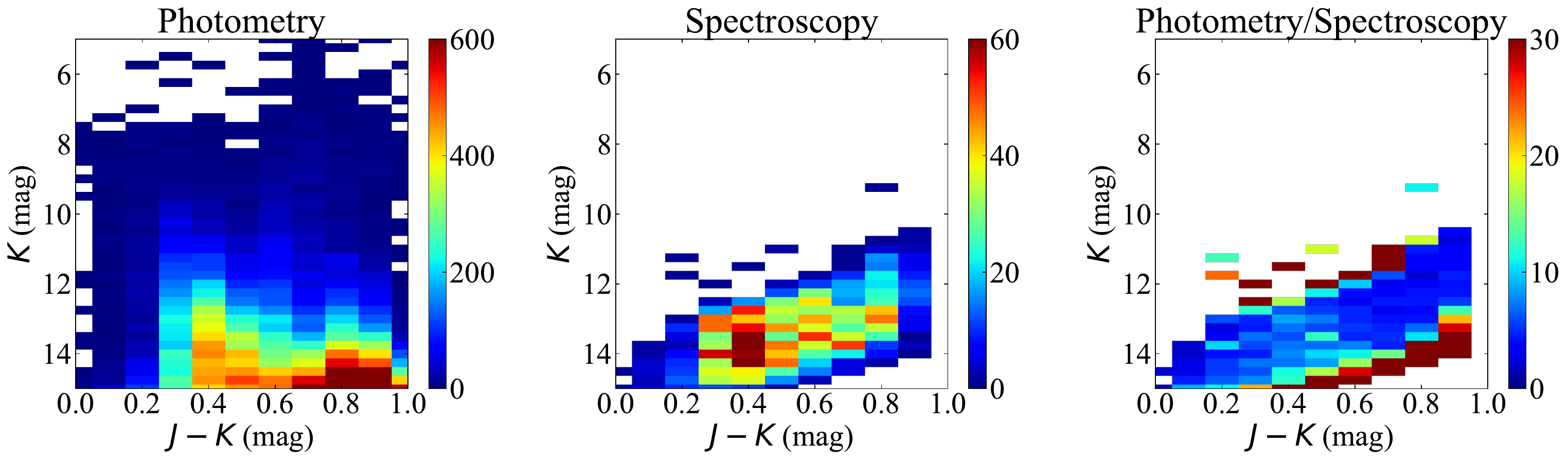}
    \includegraphics[width=0.95\textwidth]{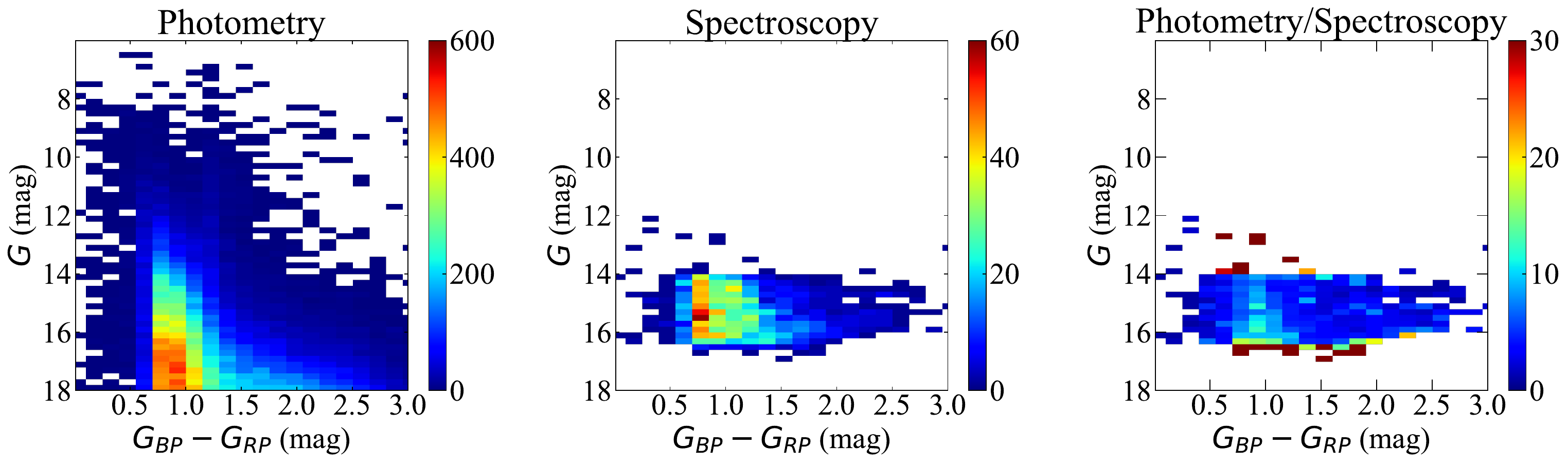}
    \caption{An exemplary demonstration of the selection function analysis presented for the LAMOST plate 'GAC073S05B1', utilizing the 2MASS and $Gaia$ DR3 selection coefficients for comparison. The upper panels depict the results obtained from the 2MASS dataset. The left panel illustrates the photometric CMD diagram. The middle panel presents the spectroscopically selected data along the same line-of-sight. The right panel displays the selection function $S^{-1}(J-K, K, l, b)$. The lower panels provide the analogous information, but for the $Gaia$ dataset, with the selection function $S^{-1}(G_\mathrm{BP}-G_\mathrm{RP}, G, l, b)$.}
    \label{fig:sf}
\end{figure*}

\section{Stellar Number Density} \label{sec:se}
 
In this study, we employ a Bayesian method similar to previous studies \citet{liu2017mapping} and \citet{wang2018mapping,wang2020mapping} to derive stellar number density and correct the selection bias of LAMOST. We utilize photometric data, such as 2MASS \citep{skrutskie2006two} and $Gaia$, to obtain the selection function of our sample. In this section, we provide a concise overview of how the Bayesian algorithm is employed to obtain the stellar density. Furthermore, we compare the results obtained by correcting selection effects using the \textit{Gaia} and 2MASS datasets.
 
\subsection{The Bayesian Method} \label{subsec:subse}

The details of the adopted Bayesian methods can be found in \citet{liu2017mapping}. In this section, we present a concise summary of this approach. Within a specified small volume $(D, D+\Delta D)$ and a color-magnitude box $(c, m)$ (2MASS: $(J-K, K)$, Gaia: $(G_\mathrm{BP}-G_\mathrm{RP}, G)$), with line-of-sight coordinates given in Galactic coordinates $(l, b)$, the probability of detecting a star in both the LAMOST ($p_\mathrm{sp}(D \vert c, m, l, b)$) and photometric surveys ($p_\mathrm{ph}(D \vert c, m, l, b)$) is equivalent, as given by,
\begin{equation}
\begin{aligned}
p_\mathrm{sp}(D \vert c, m, l, b) = p_\mathrm{ph}(D \vert c, m, l, b).
\end{aligned}
\end{equation}
In accordance with Bayes' theorem, the stellar density for the photometric $v_\mathrm{ph}$ and LAMOST $v_\mathrm{sp}$ can be related as follows:
\begin{align}
v_\mathrm{sp}\left(D \vert c, m, l, b\right) = v_\mathrm{ph}\left(D \vert c, m, l, b\right) \cdot S\left( c, m, l, b\right),
\end{align}
where $S$ represents the selection function, given by,
\begin{align}
S\left( c, m, l, b\right) &= \frac{\int_{0}^{\infty} v_\mathrm{sp}\left(D \vert c, m, l, b\right) \Omega D^2 dD}{\int_{0}^{\infty} v_\mathrm{ph}\left(D \vert c, m, l, b\right) \Omega D^2 dD} \nonumber\\
&=\frac{n_\mathrm{sp}\left( c, m, l, b\right)}{n_\mathrm{ph}\left( c, m, l, b\right)},
\end{align}
where $\Omega$ is the solid angle of the line-of-sight, $n_\mathrm{sp}(c, m, l, b)$ denotes the number of spectroscopic stars within a given $(c, m, l, b)$ bin, while $n_\mathrm{ph}(c, m, l, b)$ provides the same information but for photometric data.

Considering the relatively limited number of stars observed by LAMOST in a specific line-of-sight direction
compared to the photometric survey, and the need to account for errors arising from distance, we utilize Kernel 
Density Estimation (KDE) to describe $v_\mathrm{sp}$ as follows:
\begin{align}
v_\mathrm{sp}(D \vert c, m, l, b)=\frac{1}{\Omega D^2} \Sigma_{i}^{n_\mathrm{sp}(c, m, l, b)} p_{i}(D),
\end{align}
where $p_{i}(D)$ represents the probability density function of the $i$-th star in the line-of-sight.

\begin{figure}
    \centering
    \includegraphics[width=0.40\textwidth]{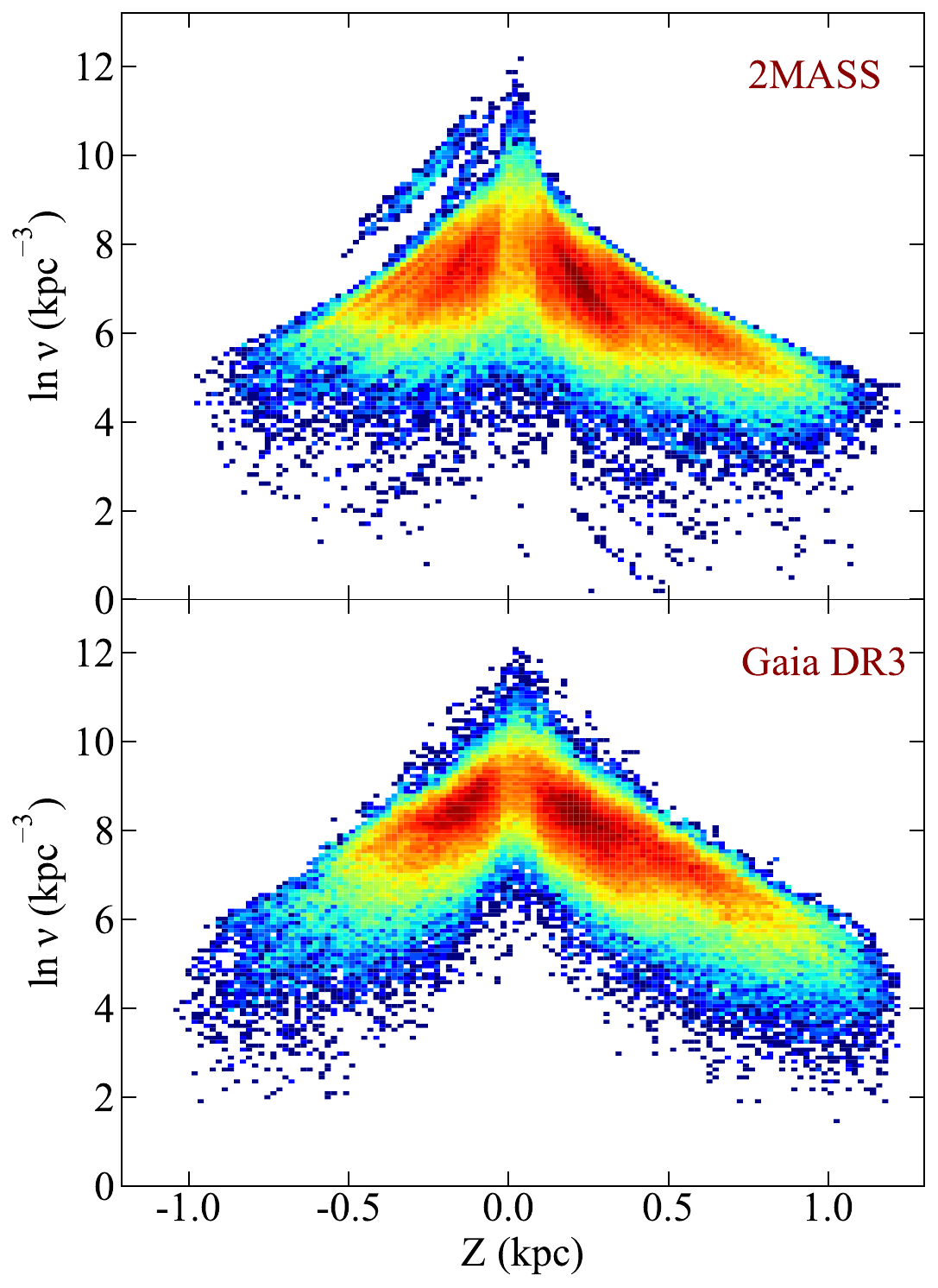}
    \caption{The stellar densities for our selected K dwarfs corrected using the 2MASS (upper panel) and the $Gaia$ (bottom panel) selection function.}
    \label{fig:znu_2g}
\end{figure} 

\subsection{Selection Function Correction} \label{se_g2}

The selection function is calculated for each plate in LAMOST. We first adopt the position of the central object of each LAMOST plate to gather data from the corresponding photometric surveys in the same region, employing a radius of $2.5\degr$. Considering the limiting magnitudes and photometric qualities of 2MASS and $Gaia$, we apply restrictions to the photometric samples. Specifically, we only include stars with photometric errors smaller than 0.1\,mag in the $J$, $K$, $G$, $B_P$, and $R_P$ bands. Additionally, we adopt that the $K$ magnitude is less than 14.3\,mag and the $G$ magnitude is less than 18.3\,mag. The computation of the selection function for an example plate in the LAMOST survey is illustrated in Fig.~\ref{fig:sf}. Regarding 2MASS CMD, we employ bin sizes of $\Delta (J-K) = 0.1$\,mag and $\Delta K = 0.25$\,mag for each plate. For $Gaia$ CMD, we adopt the bin sizes of $\Delta (G_\mathrm{BP}-G_\mathrm{RP}, G) = 0.15$\,mag and $\Delta G = 0.2$\,mag.

Fig.~\ref{fig:znu_2g} illustrates the vertical distribution of number densities after applying selection function corrections using the 2MASS and $Gaia$ datasets. When using the 2MASS data as a reference for bias correction, a distinct deviation is noticeable in the upper-left quadrant of the $\mathrm{ln}v$ - $Z$ distribution. We speculate that this discrepancy may arise from a region of elevated density in the 2MASS CMD located around ($J-K$, $K$) $\approx$ (0.8, 14.0)\,mag (visible in the upper panel of Fig.~\ref{fig:sf}), primarily populated by K and M dwarfs. However, the reliable photometric data of the 2MASS K band is limited to magnitudes smaller than 14.3 mag, leading to subtle discrepancies between these two regions and potentially resulting in biased corrections to the K dwarf stellar number density.

Considering that the magnitude limit for source detection by $Gaia$ is $G$ = 20.7 mag \citep{hodgkin2021gaia}, which surpasses that of LAMOST, and our sample's maximum depth is only 1,200\,pc, we conclude that $Gaia$ DR3 data is comprehensive under these conditions. Furthermore, based on the results shown in Fig.~\ref{fig:znu_2g}, the distribution using $Gaia$ selection function appears to be more convergent and better conforms to an exponential pattern in terms of morphology. Thus, we deduce that utilizing $Gaia$ DR3 selection function can more effectively correct the stellar density of K dwarfs in LAMOST DR9. Finally, by leveraging the selection function for each plate and combining it with the aforementioned KDE method, we obtain the corrected stellar density.

\section{Modelling number density} \label{sec:profile}
\subsection{Subsamples} \label{subsec:subsamp}

\begin{figure}
    \centering
    \includegraphics[width=0.40\textwidth]{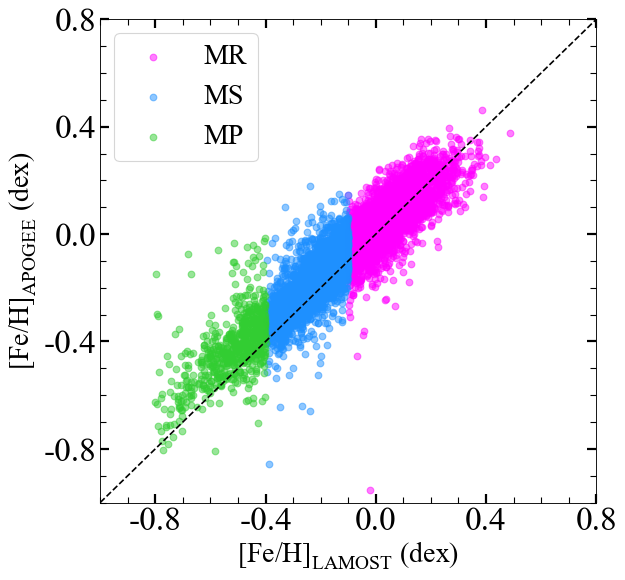}
    \caption{The comparison of metallicity between LAMOST and APOGEE of 7422 common K dwarfs. The colours represent K dwarf subsamples, where MR represents the metal-rich abundance sample, MS denotes the standard metallicity abundance sample, and MP represents the metal-poor abundance sample.}
    \label{fig:feh_mrsp}
\end{figure}

The primary objective of this study is to investigate the three-dimensional structure of the Galactic disk 
in the solar neighborhood and explore the potential relationship between the asymmetric structure of the 
Milky Way disk and metallicity. To achieve these, we initially meticulously collect K dwarf samples with 
reliable metallicity measurements ($\sigma_{\mathrm{[Fe/H]}}$ / $\mathrm{[Fe/H]}$ $<$ 0.2). Subsequently,
we categorize them into three subgroups based on their metallicity values: $\mathrm{[Fe/H]}$ $>$ $-$0.1 
dex (Metal-Rich, MR), $-$0.4 $<$ $\mathrm{[Fe/H]}$ $<$ $-$0.1 dex (Metal-Standard, MS), and
$\mathrm{[Fe/H]}$ $<$ $-$0.4 dex (Metal-Poor, MP). 
In Figure \ref{fig:feh_mrsp}, we show the comparison of metallicity between LAMOST and APOGEE for 
the 7422 common K dwarf stars. Table \ref{tab:sta} presents
the statistical results of metallicity for all 
metallicity subsamples of K dwarfs, along with the number of stars. 
The errors between the MR and MS samples are negligible, whereas for
the MP sample, LASP-derived [Fe/H] values exhibit a slight underestimation. Nevertheless, we maintain
confidence in utilizing LASP-calculated metallicity to examine the relationship between Galactic disk 
structure and metallicity, particularly in analyzing disparities between the MR and MP samples.

In the subsequent analysis, we further divide each of the selected K dwarf subsamples (AK, MR, MS, 
and MP) into multiple bins based on their Galactocentric distance $(R)$ and azimuthal angle $(\phi)$.
Taking into account the distribution of the sample ratio and considering Poisson noise, we select 
four central values for $R$: 7.60 kpc, 8.40 kpc, 8.80 kpc, and 9.25 kpc. For $\phi$, we choose the 
central values: $-2.5^\circ$, $0.0^\circ$, $2.5^\circ$, and $5.0^\circ$. The detailed binning 
scheme can be found in Tables~\ref{tab:dk_anu}-~\ref{tab:dk_hs}.

\begin{table*}
    \centering
    \caption{Statistical data of metallicity subsamples of K dwarfs}
    \begin{threeparttable}
    \begin{tabular}{ccccccc}
    \hline
    \hline
        ~ & [Fe/H](dex) & $\mu_\mathrm{[Fe/H]}$(dex) & $\sigma_\mathrm{[Fe/H]}$(dex) & $\Delta_\mathrm{[Fe/H]}$(dex) & Number & Abbr. \\ \hline
        K Dwarfs & (-0.8,0.6) & -0.158 & 0.222 & -0.037 & 174,443 & AK \\ 
        Metal-Rich & (-0.1,0.6) & -0.011 & 0.104 & -0.015 & 39,329 & MR \\ 
        Metal-Standard & (-0.4,-0.1) & -0.226 & 0.084 & -0.052 & 73,214 & MS \\ 
        Metal-Poor & (-0.8,-0.4) & -0.548 & 0.138 & -0.097 & 24,565 & MP \\ \hline\hline
    \end{tabular}
\begin{tablenotes}
\footnotesize
\item[\textbf{$\bullet$ $\Delta_\mathrm{[Fe/H]}=\mathrm{[Fe/H]}_\mathrm{LAMOST}-\mathrm{[Fe/H]}_\mathrm{APOGEE}$}]
\end{tablenotes}
\end{threeparttable}
\label{tab:sta}  
\end{table*}

\subsection{Disk Model}\label{subsec:profile}
\begin{figure}
    \centering
    \includegraphics[width=0.40\textwidth]{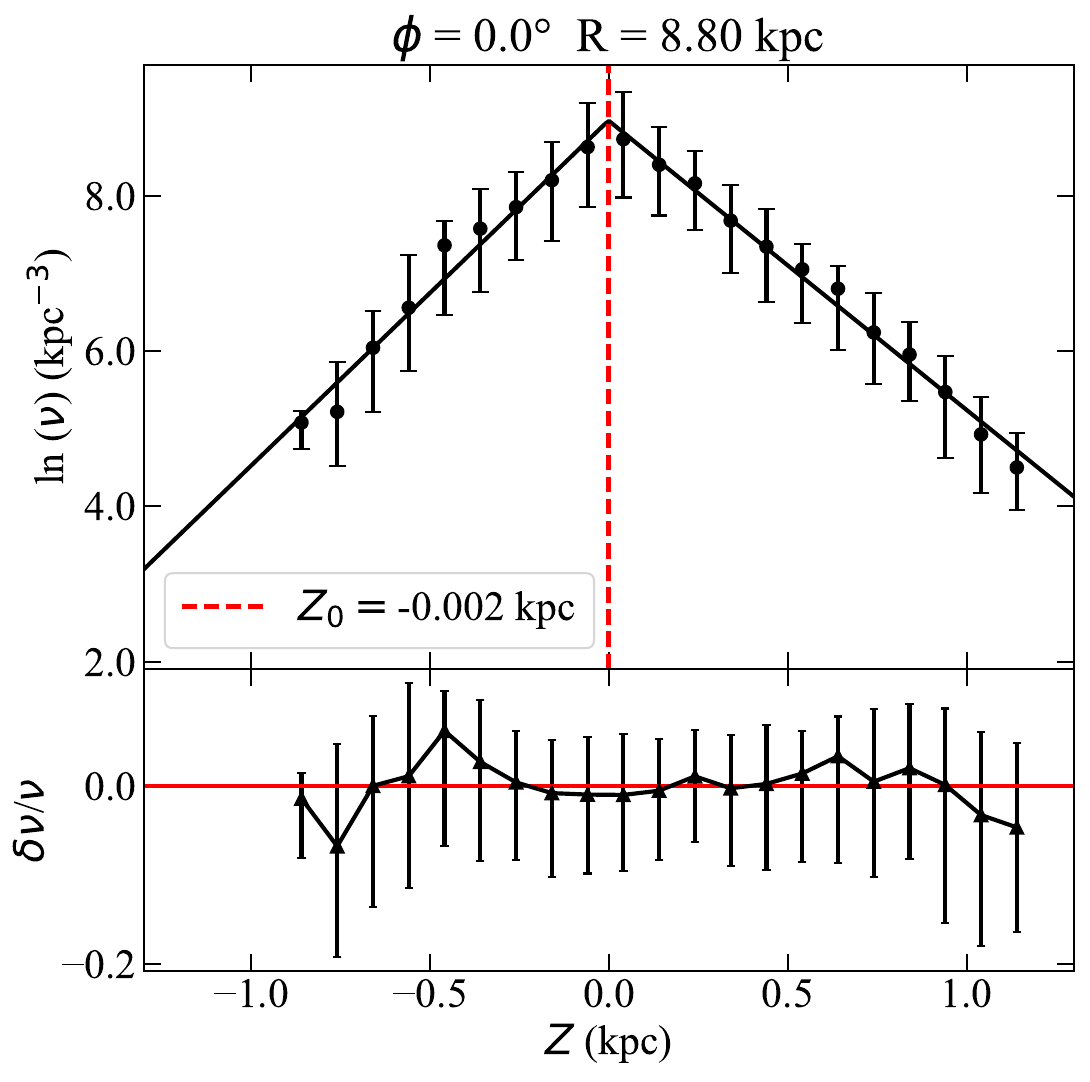}
    \includegraphics[width=0.40\textwidth]{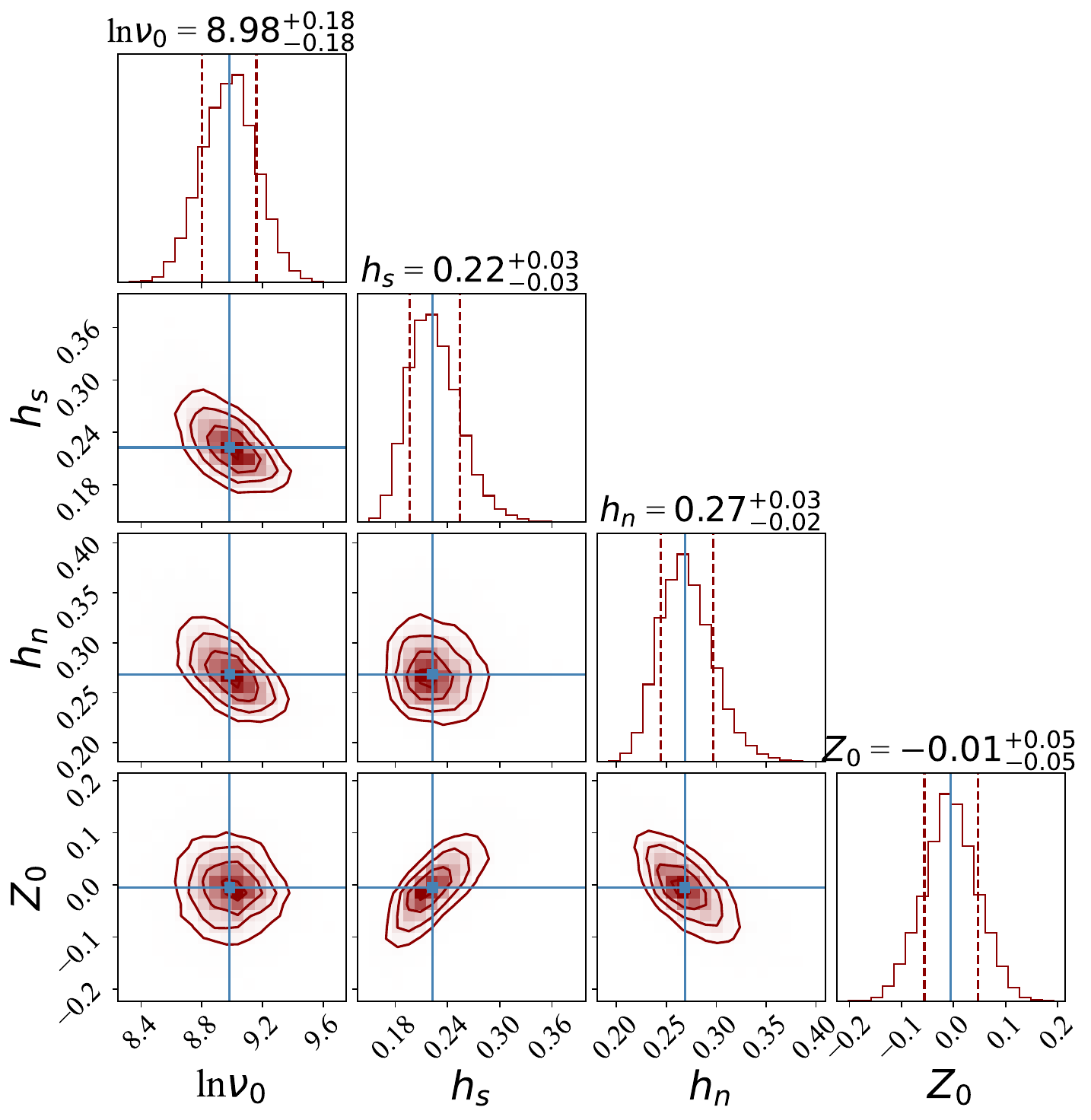}
    \caption{Fitting of the stellar number density for an example AK subsample located at the ($R$,$\phi$) = (8.8 kpc, 0.0$^\circ$). The upper panel shows the best-fitting model (black line) and observed data (black dots with error bars). The red vertical dotted line indicates the fitted position of the mid-plane of the Galactic disk. The bottom part of the top panel displays the relative residuals, $\delta \nu/\nu$ = ($\nu_\mathrm{obs}$ - $\nu_\mathrm{model}) / \nu_\mathrm{model}$. The lower panel presents the likelihood distributions of the parameters $(\mathrm{ln}\nu_0~(\mathrm{kpc}^{-3}), h_\mathrm{n}~\mathrm{(kpc)}, h_\mathrm{s}~\mathrm{(kpc)}, $ and $ Z_0~\mathrm{(kpc)})$ obtained from the MCMC simulation.}
    \label{fig:mcmc}
\end{figure}

In our study, we adopt an exponential model to describe the vertical distribution of the Galactic disk stellar number density. Our samples exhibit a narrow metallicity distribution ranging from $-$0.8 to 0.6\,dex, and their maximum distances are confined within 1,200\,pc. To investigate the structure of the Milky Way disk, we adopt a single disk component, specifically the Galactic thin disk, and introduce the free parameter $Z_0$ to characterize the position of the Galactic mid-plane. The model can be given by,
\begin{align}
\nu(Z|(R, \phi))=\nu_0(R, \phi)\exp\left(-\frac{|Z-Z_0|}{h_\mathrm{z}(R, \phi)}\right),
\end{align}
where $\nu_0(R, \phi)$ represents the stellar density at $Z$ = 0, and $h_\mathrm{z}(R, \phi)$ corresponds to the scale heights of the northern ($h_\mathrm{n}$) and southern ($h_\mathrm{s}$) Galactic disk. We adopt uniform priors for $h_\mathrm{z}(R, \phi)$ and $Z_0$ over the ranges of 0.0 to 1.0\,kpc and $-$1.0 to 1.0\,kpc, respectively.

To obtain the best-fitted parameters ($\ln \nu_0, h_\mathrm{n}, h_\mathrm{s}$, $Z_0$), we first perform binned statistics on the vertical distances and densities for each $(R,\phi)$ bin. The median values are selected as the vertical stellar density profiles along the $Z$ bins. Subsequently, we employ the Markov Chain Monte Carlo (MCMC) method to determine the maximum value of the posterior probability for the four free parameters. The likelihood function of the density distribution is defined as follows:
\begin{gather}
\begin{aligned}
    \mathcal{L}(\{\nu_\mathrm{obs}(Z_i|(R, \phi))\}|\nu_0,h_\mathrm{n},h_\mathrm{s}, Z_0)=\\
    \prod_i \exp(-\frac{1}{2}(\nu_\mathrm{obs}(Z_i|(R, \phi))-\\
    \nu_\mathrm{model}(Z_i|R, \phi,\nu_0,h_\mathrm{n},h_\mathrm{s}, Z_0))^2).
\end{aligned}
\end{gather}
where $Z_i$ and $\nu_\mathrm{obs}$ represent the $i$-th $Z$ bin and the average stellar density in that specific bin, respectively. Moreover, $\nu_\mathrm{model}$ denotes the density given by the disk model.

As an illustration, Fig.~\ref{fig:mcmc} presents the fitting results obtained for a specific AK subsample located at the ($R$, $\phi$) = (8.8 kpc, 0.0$^\circ$) bin. To thoroughly examine the uncertainties associated with distance estimation and selection bias, we employ the Bootstrap method to derive the uncertainties of the fitted parameters.

\begin{figure}
    \centering
    \includegraphics[width=0.45\textwidth]{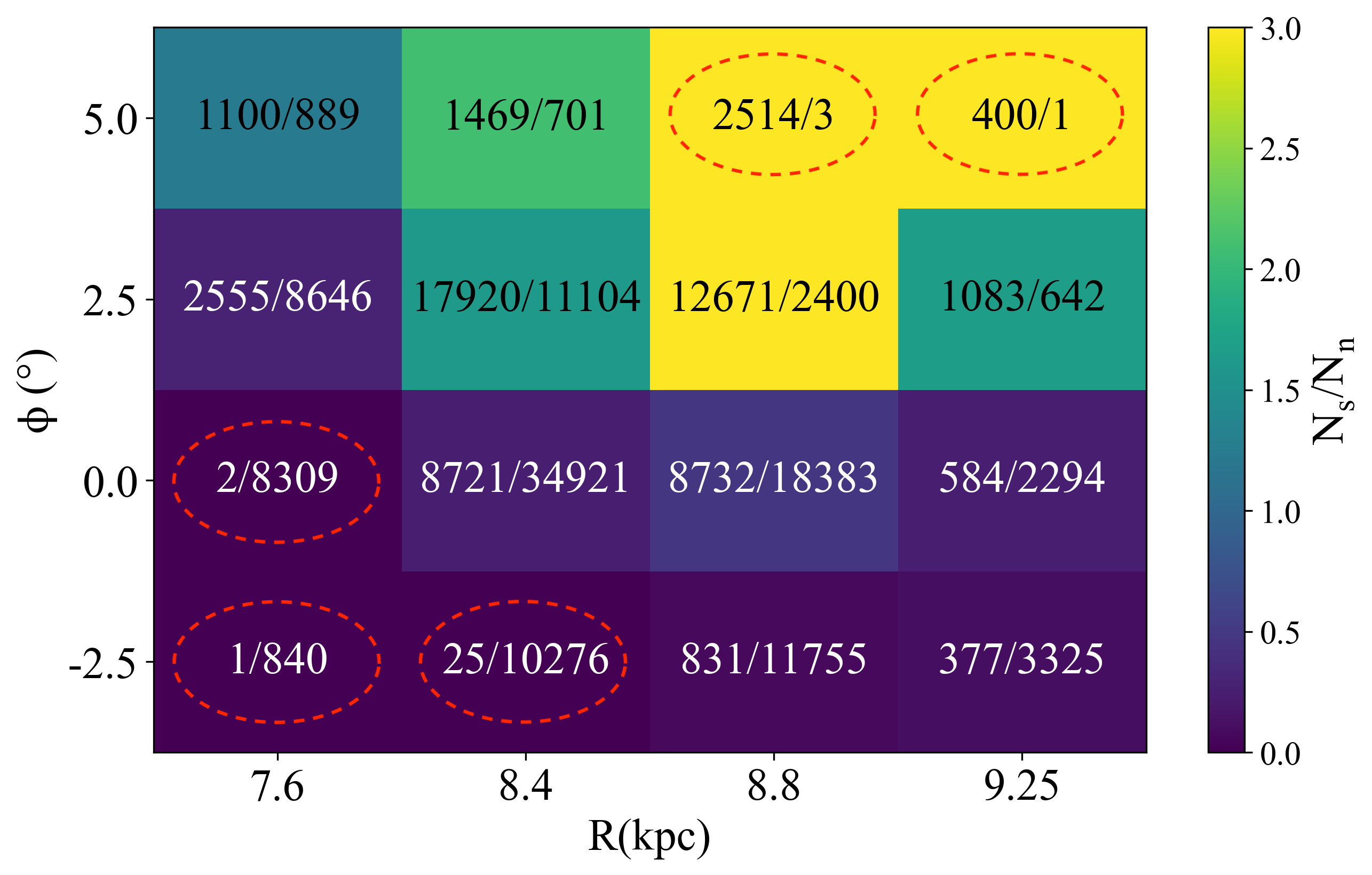}
    \caption{The stellar distribution of LAMOST AK dwarf samples across the southern and northern celestial hemispheres is illustrated in the bins, with the numbers representing the observed counts of LAMOST K dwarfs in the southern $(Z<0)$ and northern $(Z>0)$ hemispheres.}
\label{fig:dk_RPhi}
\end{figure}

\section{The Galactic Disk Structure} \label{sec:result}
Figure \ref{fig:dk_RPhi} illustrates the number of observed stars in each (R, $\phi$) bin of the AK sample. It should be noted that there are insufficient stars in the southern hemisphere in the bins $(R, \phi) = (7.6 \, \mathrm{kpc}, -2.5^\circ)$, $(7.6 \, \mathrm{kpc}, 0.0^\circ)$, and $(8.4 \, \mathrm{kpc}, -2.5^\circ)$, and in the northern hemisphere in the bins $(R, \phi) = (8.8 \, \mathrm{kpc}, 5.0^\circ)$ and $(9.25 \, \mathrm{kpc}, 5.0^\circ)$. The best fitted values are also provided for these 16 bins, but the small uncertainties are caused by the bootstrap method. For all these bins, only the scale heights on one side with enough stars are well constrained. These bins are highlighted by the red dashed ellipses in the following discussions.

\begin{figure*}
    \centering
    \includegraphics[width=0.95\textwidth]{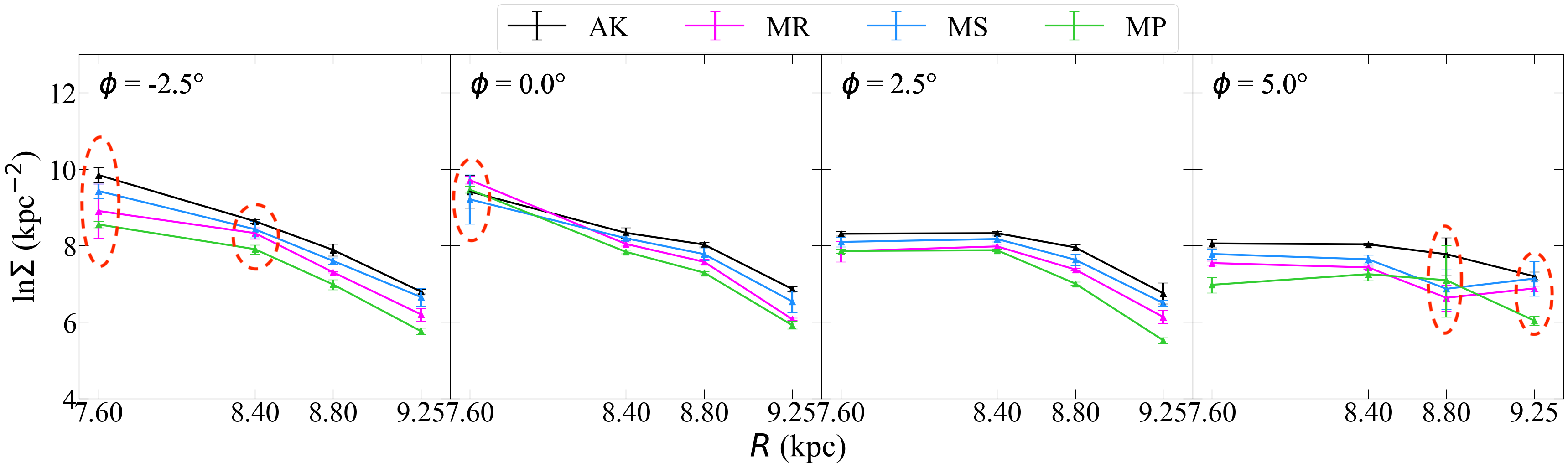}
    \caption{Surface density variation for different metallicity samples with $R$ at a fixed $\phi$. } 
\label{fig:dk_anu_R}
\end{figure*}

\begin{figure}
    \centering
    \includegraphics[width=0.33\textwidth]{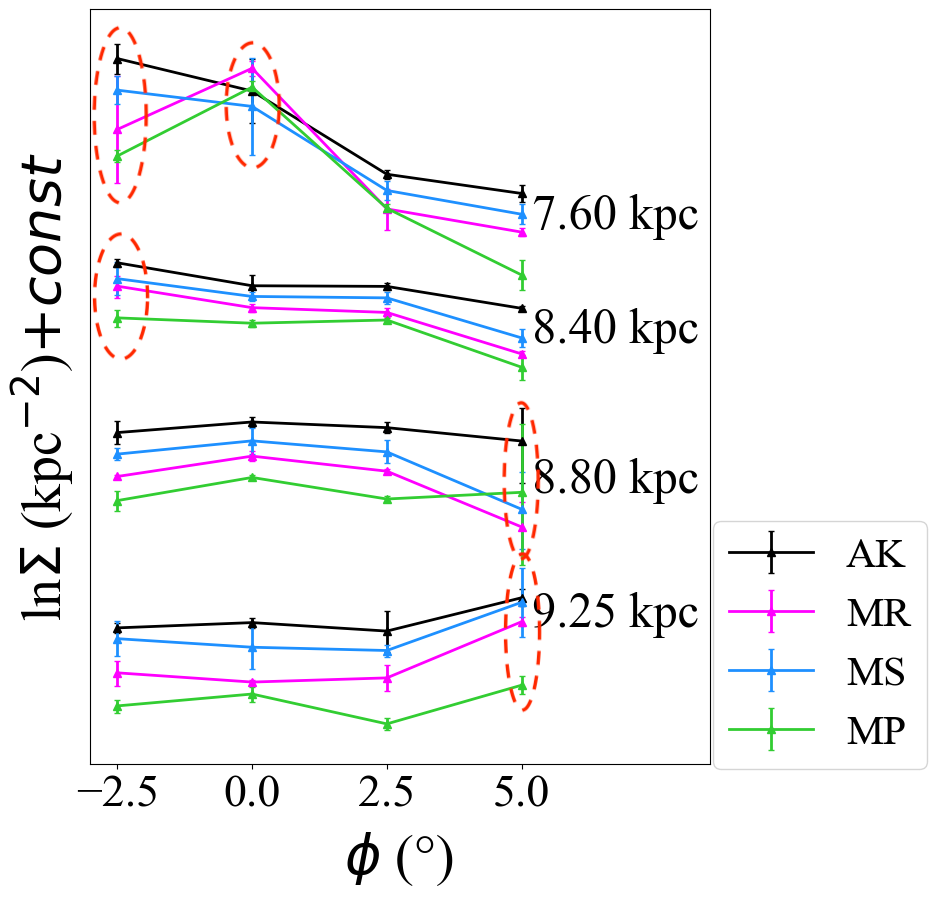}
    \caption{Surface density variation for different metallicity samples with $\phi$ at a fixed $R$.}
\label{fig:dk_anu_phi}
\end{figure}

\begin{figure*}
    \centering
    \includegraphics[width=0.9\textwidth]{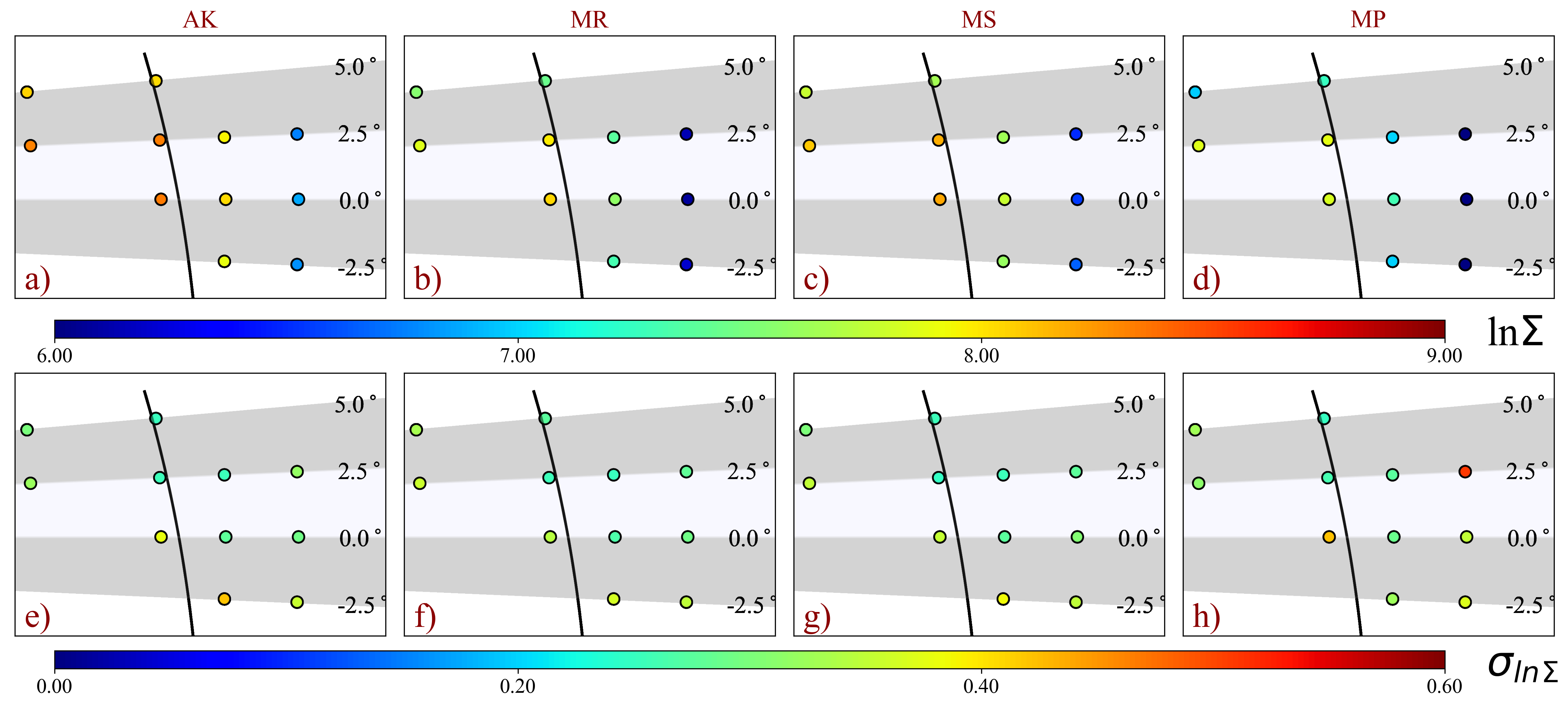}
    \caption{Surface density maps of the Galactic disk for different metallicity samples. The black dashed lines mark the positions of the Local Arm as given by \citep{chen2019galactic}. The upper panels show the surface density distribution, while the lower panels depict the corresponding errors.} 
\label{fig:dk_anu_fan}
\end{figure*}

\subsection{Surface Density} \label{subsec:dp}
Figure~\ref{fig:dk_anu_R} depicts the relationship between surface density and $R$ for K dwarf samples with different metallicities. Excluding the points with 
very low stellar counts in either side of the Galactic disks (marked with red dashed ellipses), the overall trend reveals a gradual decrease in surface density with the increase in $R$, except for a specific bin at $(R, \phi)$ = (9.25 kpc, 5.0$^\circ$) where the MP sample exhibits  a slight increasing trend in surface density. 

Figure \ref{fig:dk_anu_phi} illustrates the variation of surface density with $\phi$. For bins at $R=$ 7.6 and 8.4 kpc, the surface density exhibits a weakly decreasing trend with the azimuthal angle. However, at larger Galactocentric distance $R$, the surface density almost keeps a constant value. 

Fig.~\ref{fig:dk_anu_fan} shows the spatial distributions of the surface density of different metallicity K dwarf subsamples in the \(R\) - \(\phi\) plane. The position of the Local Arm is adopted from \citet{chen2019galactic}. The figure provides a visual representation of variations in surface density with respect to both $R$ and $\phi$ changes. Interestingly, there is a gradual decrease in density within the Local Arm as $\phi$ increases, while outside of it, the surface density remains unchanged.

\begin{table}
  \centering
  \caption{Fitted surface density (ln$\Sigma$, kpc$^{\scriptscriptstyle -2}$) for the AK sample.}
    \begin{tabular}{@{}c|llll}
    \hline
    \hline
    \diagbox[width=4.5em,trim=l]{$R$ (kpc)}{$\phi $ ($^\circ$)}   & (-4.0,-1.0) & (-1.0,1.0) & (1.0,4.0) & (4.0,6.0) \bigstrut\\
    \hline
    (7.0,8.2) & 9.84$^{+0.20}_{-0.19}$ & 9.41$^{+0.43}_{-0.42}$ & 8.31$^{+0.06}_{-0.06}$ & 8.06$^{+0.10}_{-0.11}$ \bigstrut[t]\\
    (8.2,8.6) & 8.64$^{+0.05}_{-0.05}$ & 8.33$^{+0.14}_{-0.14}$ & 8.32$^{+0.05}_{-0.04}$ & 8.03$^{+0.03}_{-0.02}$ \\
    (8.6,9.0) & 7.89$^{+0.15}_{-0.16}$ & 8.03$^{+0.06}_{-0.06}$ & 7.95$^{+0.07}_{-0.07}$ & 7.78$^{+0.43}_{-0.57}$ \\
    (9.0,9.5) & 6.80$^{+0.06}_{-0.06}$ & 6.87$^{+0.07}_{-0.07}$ & 6.76$^{+0.26}_{-0.28}$ & 7.20$^{+0.11}_{-0.12}$ \bigstrut[b]\\
    \hline
    \hline
    \end{tabular}%
  \label{tab:dk_anu}%
\end{table}%

\begin{table}
  \centering
  \caption{The position of Galactic mid-plane ($Z_0$, kpc) for the AK sample.}
    \begin{tabular}{@{}c|rrrr}
    \hline
    \hline
    \diagbox[width=4.5em,trim=l]{$R$ (kpc)}{$\phi$ ($^\circ$)}  & (-4.0,-1.0) & (-1.0,1.0) & (1.0,4.0) & \multicolumn{1}{l}{(4.0,6.0)} \bigstrut\\
    \hline
    (7.0,8.2) & -0.10$^{+0.00}_{-0.01}$ & -0.13$^{+0.06}_{-0.07}$ & 0.00$^{+0.00}_{-0.01}$ & 0.02$^{+0.03}_{-0.02}$ \\
    (8.2,8.6) & 0.02$^{+0.01}_{-0.00}$ & 0.02$^{+0.02}_{-0.01}$ & 0.03$^{+0.00}_{-0.00}$ & 0.07$^{+0.01}_{-0.01}$ \\
    (8.6,9.0) & 0.03$^{+0.01}_{-0.00}$ & -0.02$^{+0.00}_{-0.00}$ & 0.02$^{+0.00}_{-0.00}$ & 0.02$^{+0.06}_{-0.06}$ \\
    (9.0,9.5) & -0.01$^{+0.00}_{-0.00}$ & 0.04$^{+0.01}_{-0.01}$ & -0.02$^{+0.00}_{-0.00}$ & 0.25$^{+0.00}_{-0.00}$ \bigstrut[b]\\
    \hline
    \hline
    \end{tabular}%
  \label{tab:dk_z0}%
\end{table}%

\begin{table}
  \centering
  \caption{Scale height on the south side ($h_s$, kpc) of disk for the AK sample.}
    \begin{tabular}{@{}c|rrrr}
    \hline
    \hline
    \diagbox[width=4.5em,trim=l]{$R$ (kpc)}{$\phi$ ($^\circ$)} & (-4.0,-1.0) & (-1.0,1.0) & (1.0,4.0) & \multicolumn{1}{l}{(4.0,6.0)} \bigstrut\\
    \hline
    (7.0,8.2) & 0.50$^{+0.00}_{-0.00}$ & 0.52$^{+0.01}_{-0.01}$ & 0.20$^{+0.00}_{-0.00}$ & 0.19$^{+0.01}_{-0.00}$ \bigstrut[t]\\
    (8.2,8.6) & 0.54$^{+0.00}_{-0.01}$ & 0.26$^{+0.04}_{-0.03}$ & 0.24$^{+0.00}_{-0.00}$ & 0.19$^{+0.01}_{-0.00}$ \\
    (8.6,9.0) & 0.26$^{+0.03}_{-0.03}$ & 0.20$^{+0.01}_{-0.00}$ & 0.22$^{+0.00}_{-0.00}$ & 0.19$^{+0.02}_{-0.03}$ \\
    (9.0,9.5) & 0.19$^{+0.02}_{-0.02}$ & 0.22$^{+0.01}_{-0.02}$ & 0.26$^{+0.02}_{-0.03}$ & 0.40$^{+0.03}_{-0.03}$ \bigstrut[b]\\
    \hline
    \hline
    \end{tabular}%
  \label{tab:dk_hs}%
\end{table}%

\begin{table}
  \centering
  \caption{Scale height on the north side ($h_n$, kpc) of disk for the AK sample.}
    \begin{tabular}{@{}c|rrrr}
    \hline
    \hline
    \diagbox[width=4.5em,trim=l]{$R$ (kpc)}{$\phi$ ($^\circ$)} & (-4.0,-1.0) & (-1.0,1.0) & (1.0,4.0) & \multicolumn{1}{l}{(4.0,6.0)} \bigstrut\\
    \hline
    (7.0,8.2) & 0.19$^{+0.00}_{-0.01}$ & 0.23$^{+0.01}_{-0.00}$ & 0.24$^{+0.00}_{-0.00}$ & 0.19$^{+0.01}_{-0.00}$ \bigstrut[t]\\
    (8.2,8.6) & 0.24$^{+0.00}_{-0.01}$ & 0.25$^{+0.00}_{-0.00}$ & 0.24$^{+0.00}_{-0.00}$ & 0.19$^{+0.00}_{-0.01}$ \\
    (8.6,9.0) & 0.24$^{+0.00}_{-0.01}$ & 0.26$^{+0.00}_{-0.00}$ & 0.22$^{+0.00}_{-0.00}$ & 0.37$^{+0.15}_{-0.15}$ \\
    (9.0,9.5) & 0.33$^{+0.01}_{-0.01}$ & 0.25$^{+0.00}_{-0.01}$ & 0.29$^{+0.02}_{-0.02}$ & 0.52$^{+0.00}_{-0.01}$ \bigstrut[b]\\
    \hline
    \hline    
    \end{tabular}%
  \label{tab:dk_hn}%
\end{table}%

\begin{figure*}
    \centering
    \includegraphics[width=0.95\textwidth]{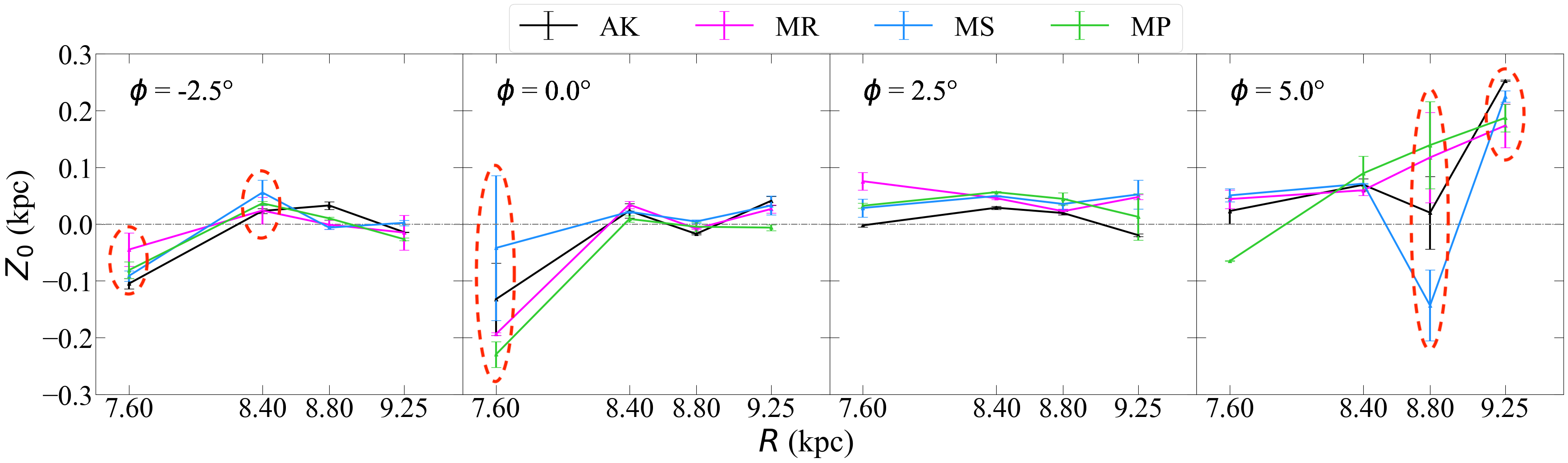}
    \caption{Similar to Fig.~\ref{fig:dk_anu_R}, but for the variations of the position of the Galactic mid-plane. Dashed lines represent $Z$ = 0 kpc.}
\label{fig:dk_z0_R}
\end{figure*}

\begin{figure}
    \centering
    \includegraphics[width=0.33\textwidth]{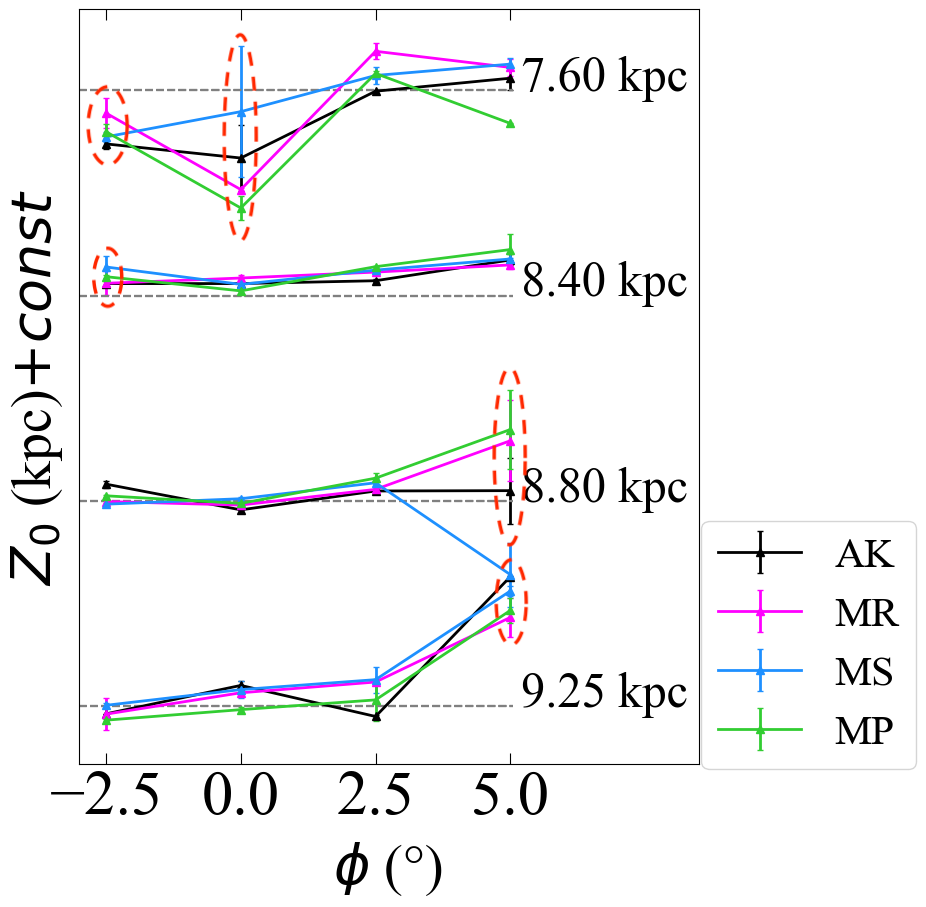}
    \caption{Similar to Fig.~\ref{fig:dk_anu_phi}, but for the variations of the position of the Galactic mid-plane and dashed lines represent $Z$ = 0 kpc.}
\label{fig:dk_z0_phi}
\end{figure}

\begin{figure*}
    \centering
    \includegraphics[width=0.9\textwidth]{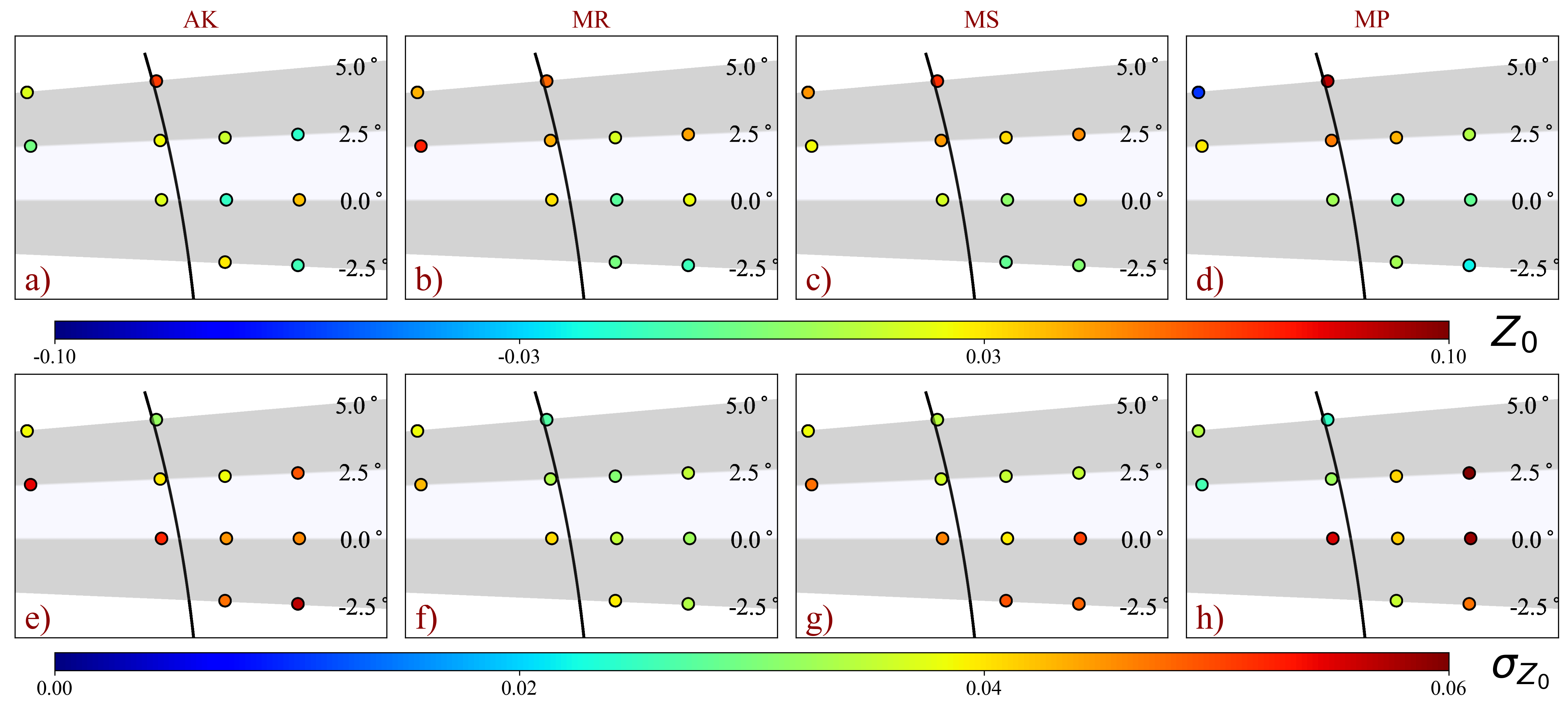}
    \caption{Similar to Fig.~\ref{fig:dk_anu_fan}, but for the maps of the position of the Galactic mid-plane.}
\label{fig:dk_z0_fan}
\end{figure*}

\begin{figure*}
    \centering
    \includegraphics[width=0.95\textwidth]{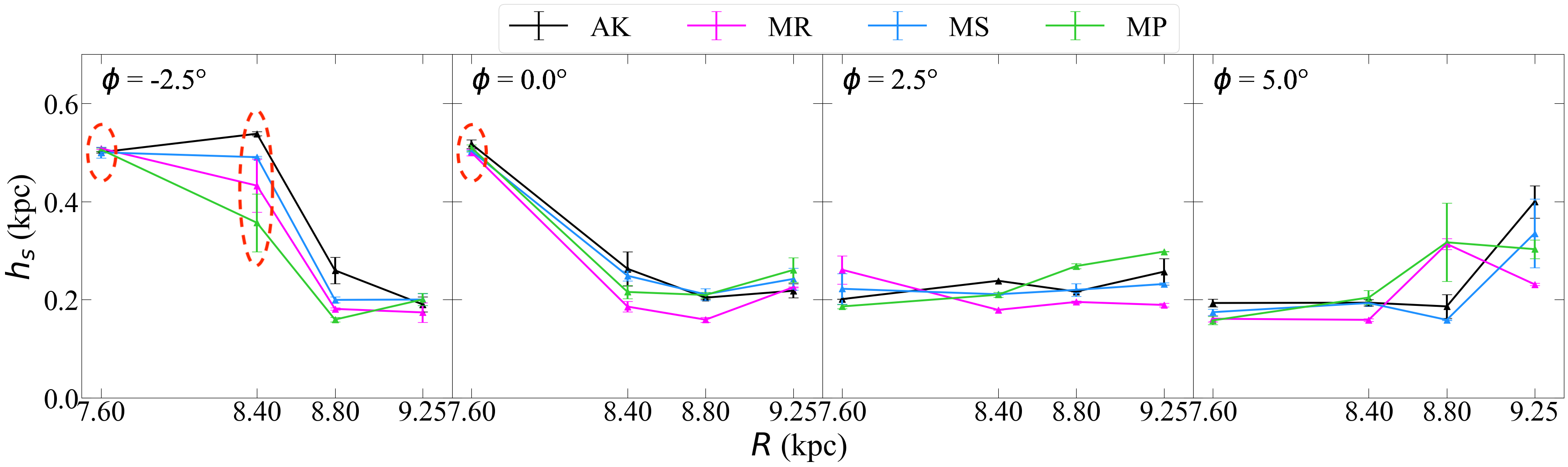} 
    \caption{Similar to Fig.~\ref{fig:dk_anu_R}, but for the variations of the scale heights on the south side of the Galactic disk.}
\label{fig:dk_hs_R}
\end{figure*}

\begin{figure}
    \centering
    \includegraphics[width=0.33\textwidth]{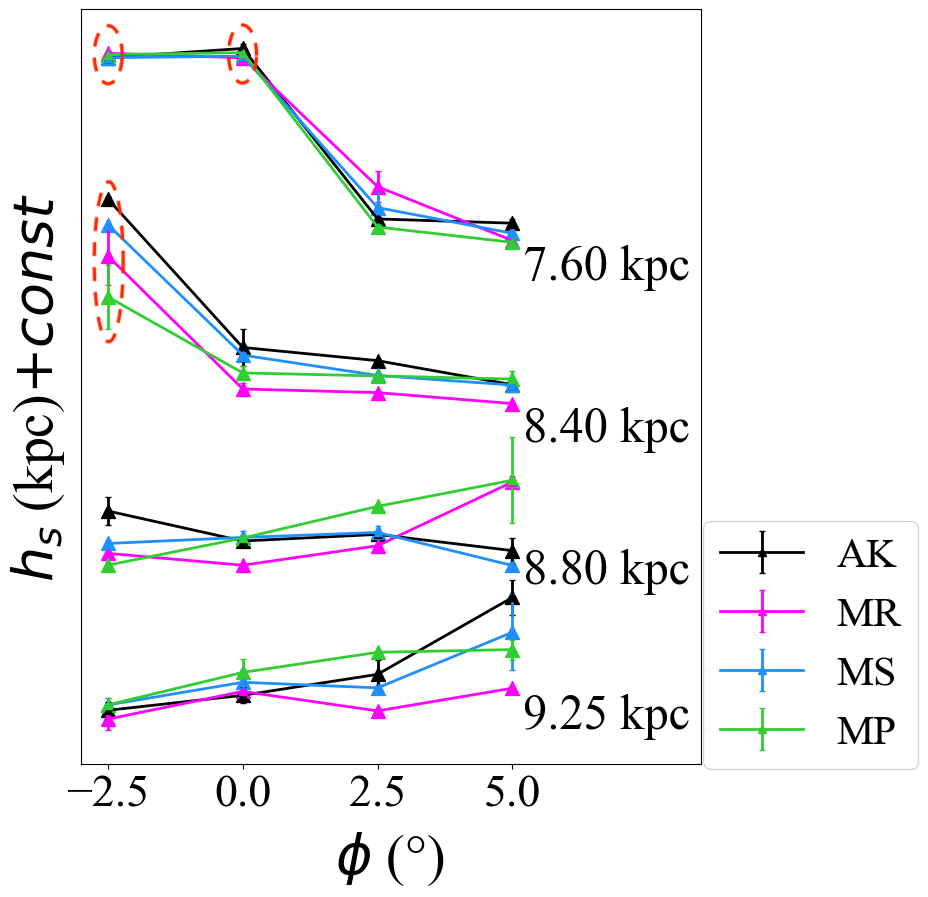} 
    \caption{Similar to Fig.~\ref{fig:dk_anu_phi}, but for the variations of the scale heights on the south side of the Galactic disk.}
\label{fig:dk_hs_phi}
\end{figure}

\begin{figure*}
    \centering
    \includegraphics[width=0.9\textwidth]{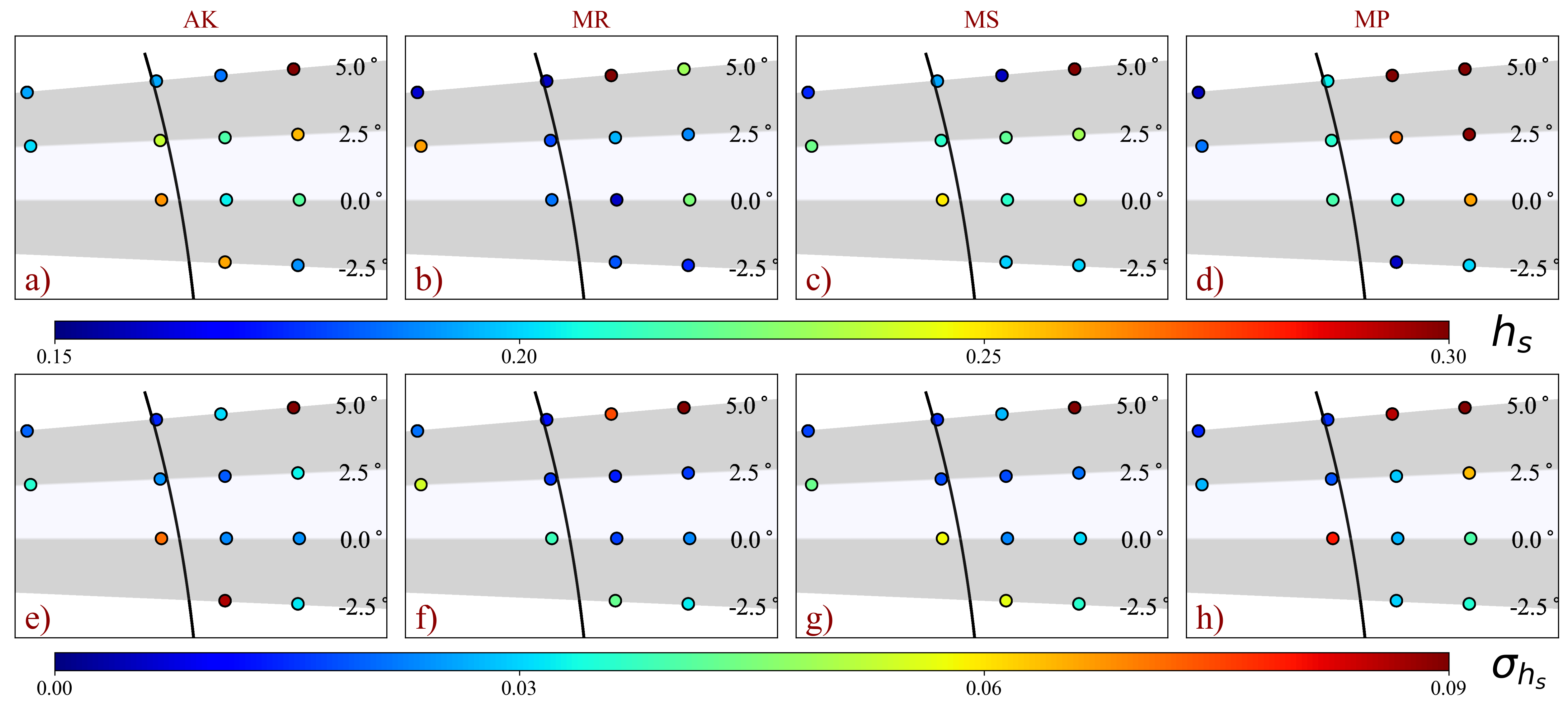} 
    \caption{Similar to Fig.~\ref{fig:dk_anu_fan}, but for the maps of the scale heights on the south side of the Galactic disk.}
\label{fig:dk_hs_fan}
\end{figure*}

\begin{figure*}
    \centering
    \includegraphics[width=0.95\textwidth]{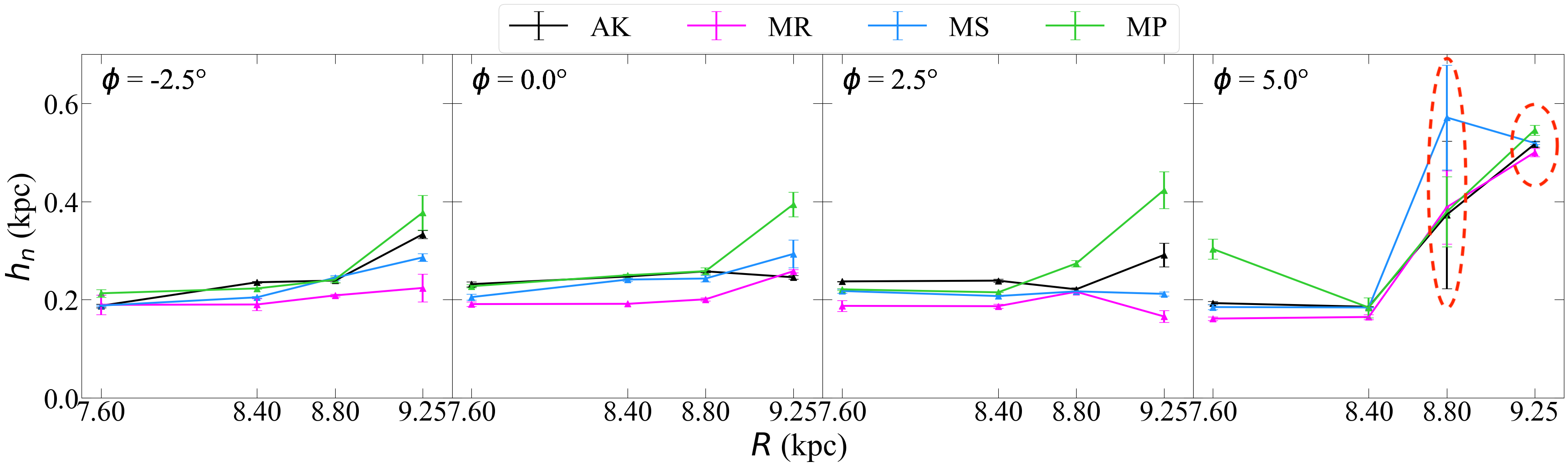}
    \caption{Similar to Fig.~\ref{fig:dk_anu_R}, but for the variations of the scale heights on the north side of the Galactic disk.}
\label{fig:dk_hn_R}
\end{figure*}

\begin{figure}
    \centering
    \includegraphics[width=0.33\textwidth]{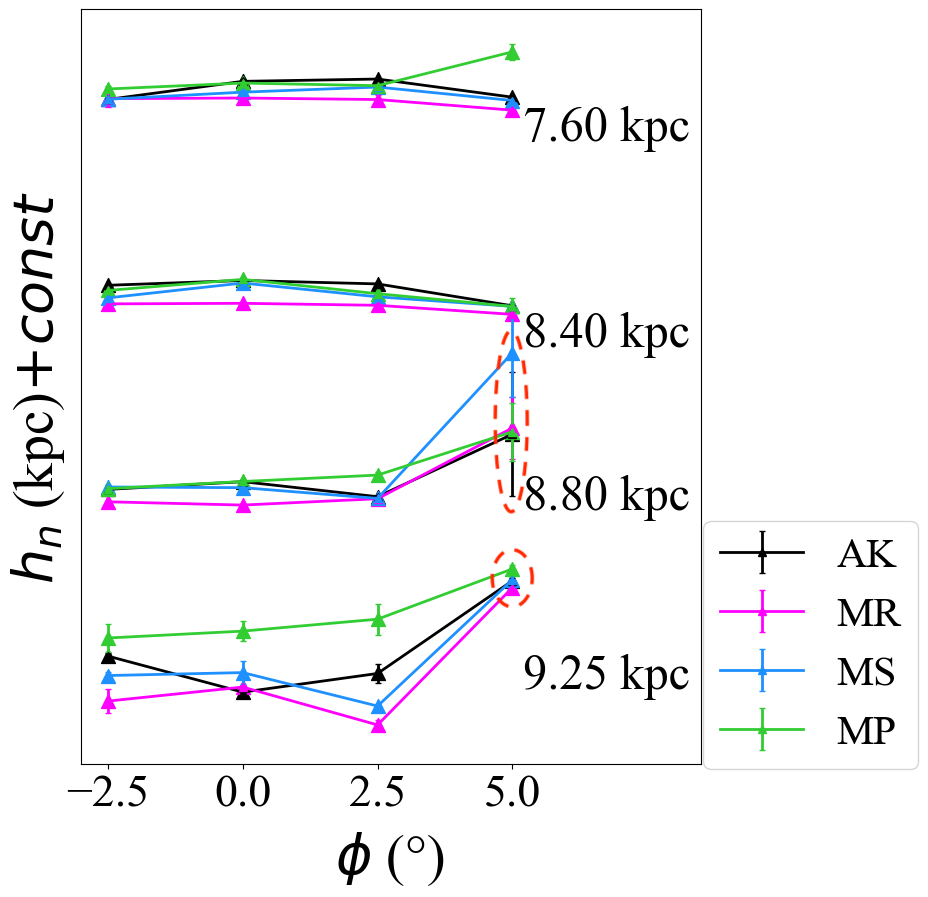}
    \caption{Similar to Fig.~\ref{fig:dk_anu_phi}, but for the variations of the scale heights on the north side of the Galactic disk.}
\label{fig:dk_hn_phi}
\end{figure}

\begin{figure*}
    \centering
    \includegraphics[width=0.9\textwidth]{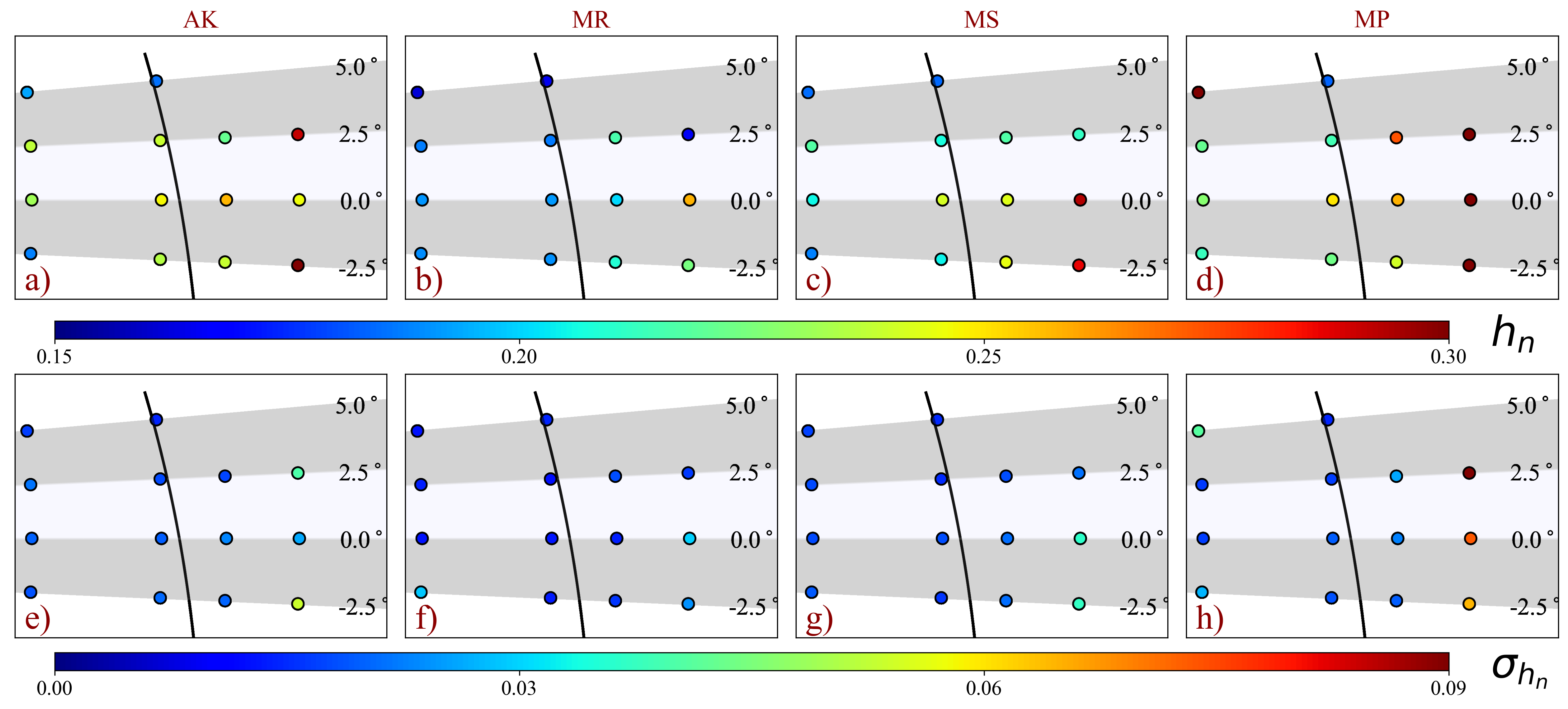} 
    \caption{Similar to Fig.~\ref{fig:dk_anu_fan}, but for the maps of the scale heights on the north side of the Galactic disk.}
\label{fig:dk_hn_fan}
\end{figure*}

\begin{figure*}
    \centering
    \includegraphics[width=0.7\textwidth]{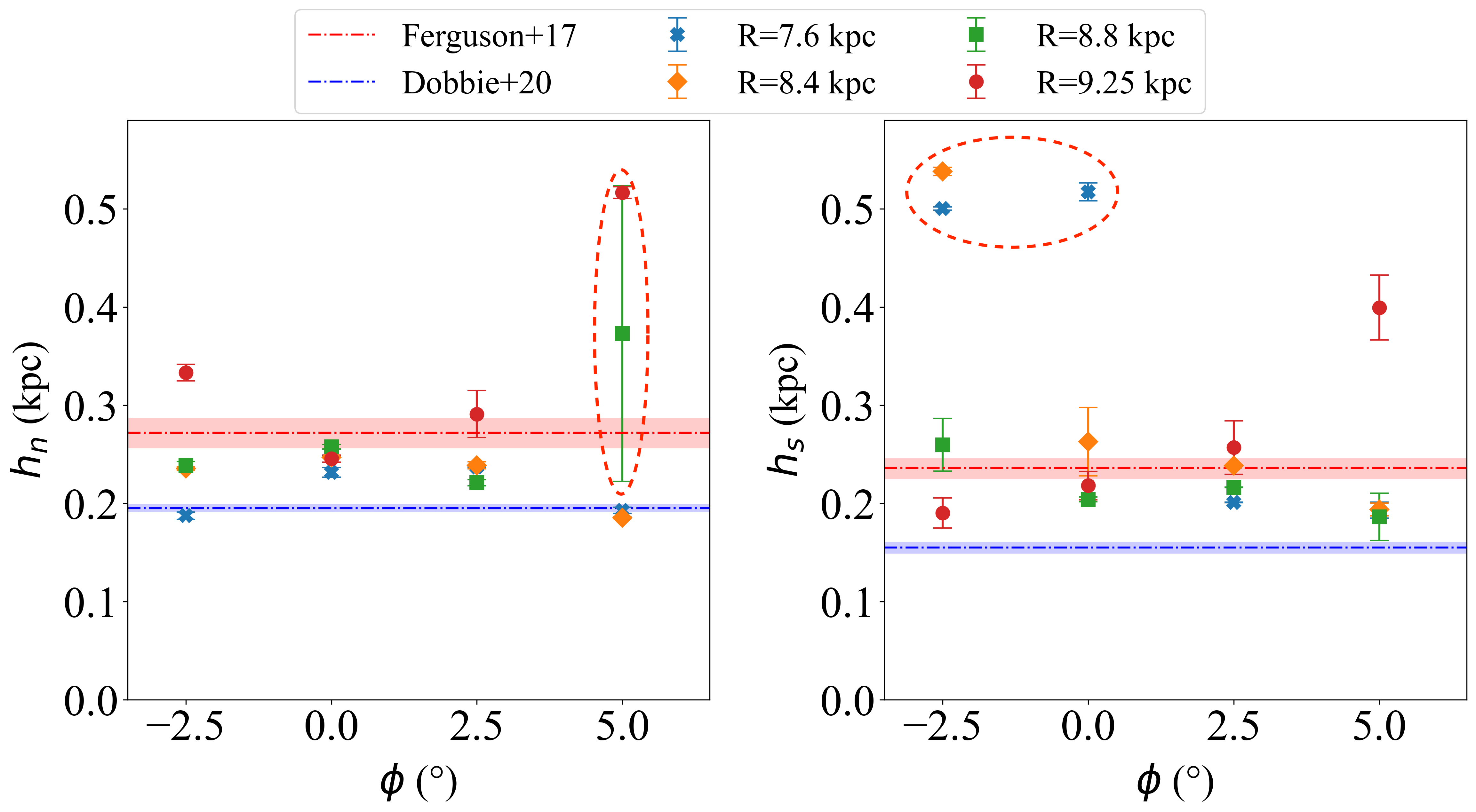} 
    \caption{The scale height of the northern and southern disks as a function of azimuthal angle. The values with error bars represent the scale heights (hs and hn) of the AK sample. The red and blue dashed lines depict the values from \citet{ferguson2017milky} and \citet{dobbie2020bayesian}, respectively.}
\label{fig:dk_hns_compare}
\end{figure*}

\subsection{Position of the Galactic Mid-plane} \label{subsec:mid}

In Figs.~\ref{fig:dk_z0_R} and~\ref{fig:dk_z0_phi}, we present the positional variations of the Galactic mid-plane ($Z_0$) with respect to Galactocentric distance ($R$) and azimuthal angle ($\phi$) for different metallicity K dwarf samples. In the solar neighbourhood, the $Z_0$ values remain close to zero ($\bar{Z_0} = 0.023^{+0.039}_{-0.042}\,\mathrm{kpc}$), which aligns with the findings reported by \citet{yu2021flare}. As either $R$ or $\phi$ increases, there is a slight upward trend observed in the $Z_0$ values. These trends are clearly visible when examining the spatial distributions of $Z_0$ for K dwarf subsamples with varying metallicities in the $R$-$\phi$ plane (see Fig.~\ref{fig:dk_z0_fan}). And there is a remarkable consistency in the mid-plane positions across the different metallicity samples. Finally, we find that despite minor variations in $Z_0$ near the Sun for different metallicities, $Z_0$ predominantly assumes positive values in most samples. 

\subsection{Scale Height} \label{subsec:sh}
The scale heights variations for the southern disk are illustrated in Figs.~\ref{fig:dk_hs_R}-\ref{fig:dk_hs_fan}, while for northern disk, they are in Figs.~\ref{fig:dk_hn_R}-\ref{fig:dk_hn_fan}. For bins with $\phi=-2.5^\circ$ and $0.0^\circ$, the scale heights of southern disk show a weakly decreasing trend as $R$ increases. Conversely, the bin with $\phi=5.0^\circ$ exhibits an increasing trend for $h_s$. And the bin at $\phi=2.5^\circ$ appears to lie in an intermediate transition region between these two scale height variations.  Meanwhile, the scale heights of the northern disk exhibit overall smooth variations, with $h_\mathrm{n}$ remaining nearly constant at about 0.2\,kpc, except for regions with relatively larger values of $R$ or $\phi$. We hypothesize that the pronounced variations observed in these regions may be attributed to the influence of the $north$ $near$ structure \citep{xu2015rings} or the $D8.5+0.2$ structure \citep{widrow2012galactoseismology,wang2018mapping}.

In this section, we endeavor to compare the differences in scale height between metal-rich (MR) and metal-poor (MP) samples. In the northern disk,  the scale height of the MP sample is slightly larger than that of the MR sample, i.e. the older disk has a relatively larger scale height claimed by \cite{bovy2012spatial}. However, the scale height in the southern disk $h_\mathrm{s}$ of the MP sample predominates in most slices (e.g., at $R=$9.25 kpc), instances where the $h_\mathrm{s}$ of the MR sample surpasses are still observed (e.g., at $R=$7.6 kpc). The fundamental reasons for this phenomenon remain unclear.

Figure~\ref{fig:dk_hns_compare} compares the 
scale heights of the northern and southern Galactic disks 
derived from our analysis of LAMOST K dwarfs with previous
studies. \citet{ferguson2017milky} and \citet{dobbie2020bayesian}
utilized the same SDSS K-M dwarf samples but employ different
methods and photometric distances to analyze the vertical 
asymmetrical feature in the $(l, b)$ directions. In terms 
of the overall distribution of scale heights, our results 
are more consistent with those of Ferguson et al. 
$(h_\mathrm{n,F}=0.272\pm0.016\,\mathrm{kpc}, 
h_\mathrm{s,F}=0.236\pm0.011\,\mathrm{kpc})$ 
and slightly larger than those of \citet{dobbie2020bayesian} 
$(h_\mathrm{n,D}=0.195\pm0.004\,\mathrm{kpc}, h_\mathrm{s,D}=0.155\pm0.006\,\mathrm{kpc})$. 
Specifically, the average value of the scale height for the northern disk is 
$\bar{h_\mathrm{n}} = 0.239\,\mathrm{kpc}$, while for the southern disk it is
$\bar{h_\mathrm{s}} = 0.216\,\mathrm{kpc}$ in our analysis. Consistent with 
previous observations, the scale height in the northern disk is larger than
that in the southern disk, by approximately 10.3\,\%. 

What is more, we find both of the scale heights are significantly offset from the averaged values.
In the left panel, the scale height in the northern disk varies around that from 
\citet{ferguson2017milky}. But the value with azimuthal angle $\phi\sim-2.5^\circ$ is significantly
larger. In the right panel for the scale height in the southern disk, we find clear increasing trend
for the scale height $h_\mathrm{s}$ with $R\sim9.25$ kpc, i.e. from
$0.19^{+0.02}_{-0.02}$ at $\phi=-2.5^\circ$ to $0.40^{+0.03}_{-0.03}$ at $\phi=5.0^\circ$,
and a decreasing trend with $R\sim8.8$ kpc from $0.26^{+0.03}_{-0.03}$ at $\phi=-2.5^\circ$
to $0.19^{+0.02}_{-0.03}$ at $\phi=5.0^\circ$.
This suggest that the disk is not axisymmetric.

\begin{figure*}
    \centering
    \includegraphics[width=0.8\textwidth]{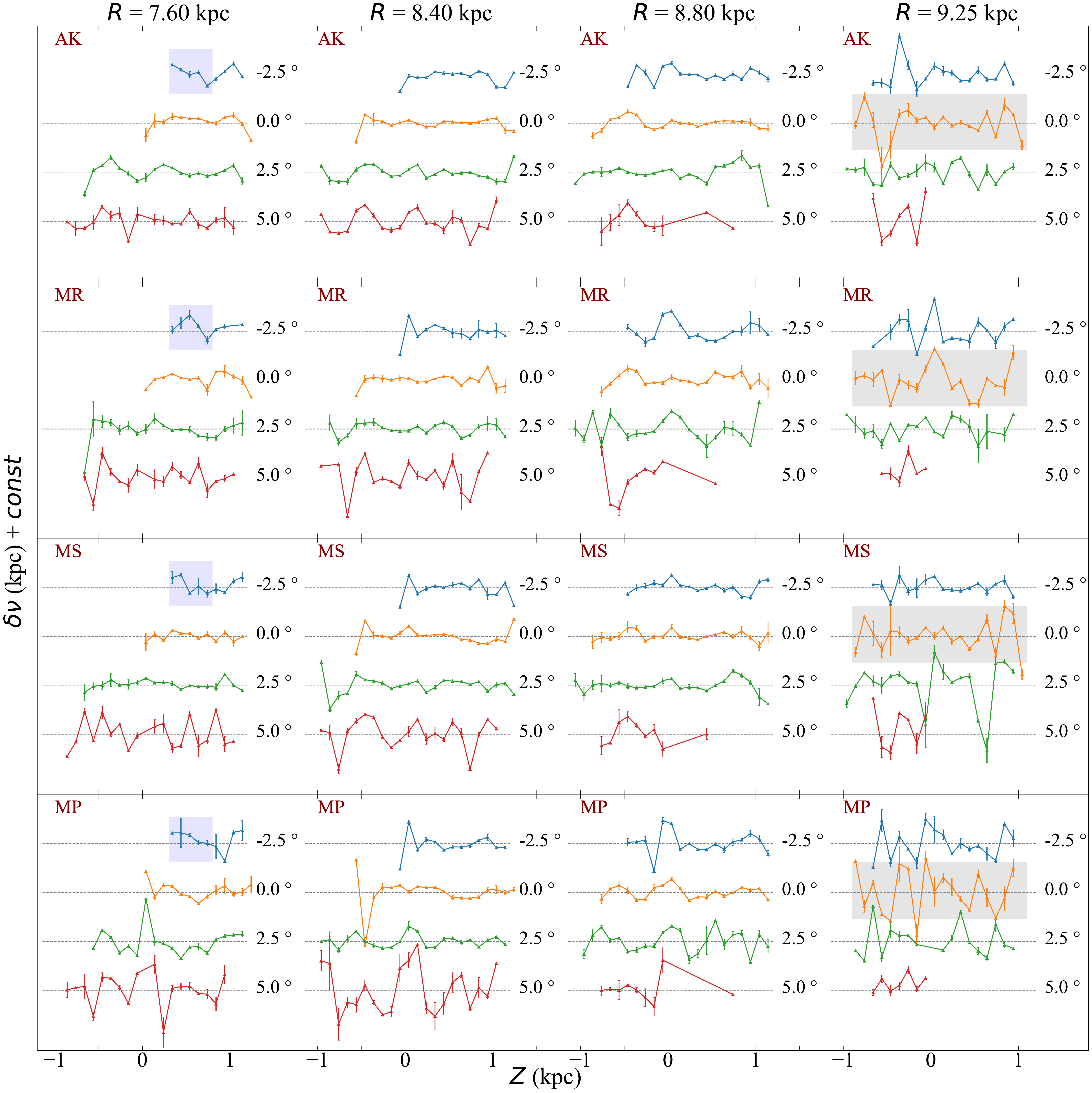} 
    \caption{Residuals $\delta \nu \equiv  \nu(Z) - \nu_{\text{model}}(Z)$ between the vertical stellar density profiles and the corresponding best-fit models are plotted as a function of $Z$ for different metallicity samples located at various ($R$, $\phi$) bins. The grey dashed line represents the zero-residual line, indicating where the residuals are zero. The shaded areas highlight subsamples with distinct density variation patterns compared to those with different metallicities.}
\label{fig:dk_z_phi}
\end{figure*}

\begin{figure*}
    \centering
    \includegraphics[width=0.8\textwidth]{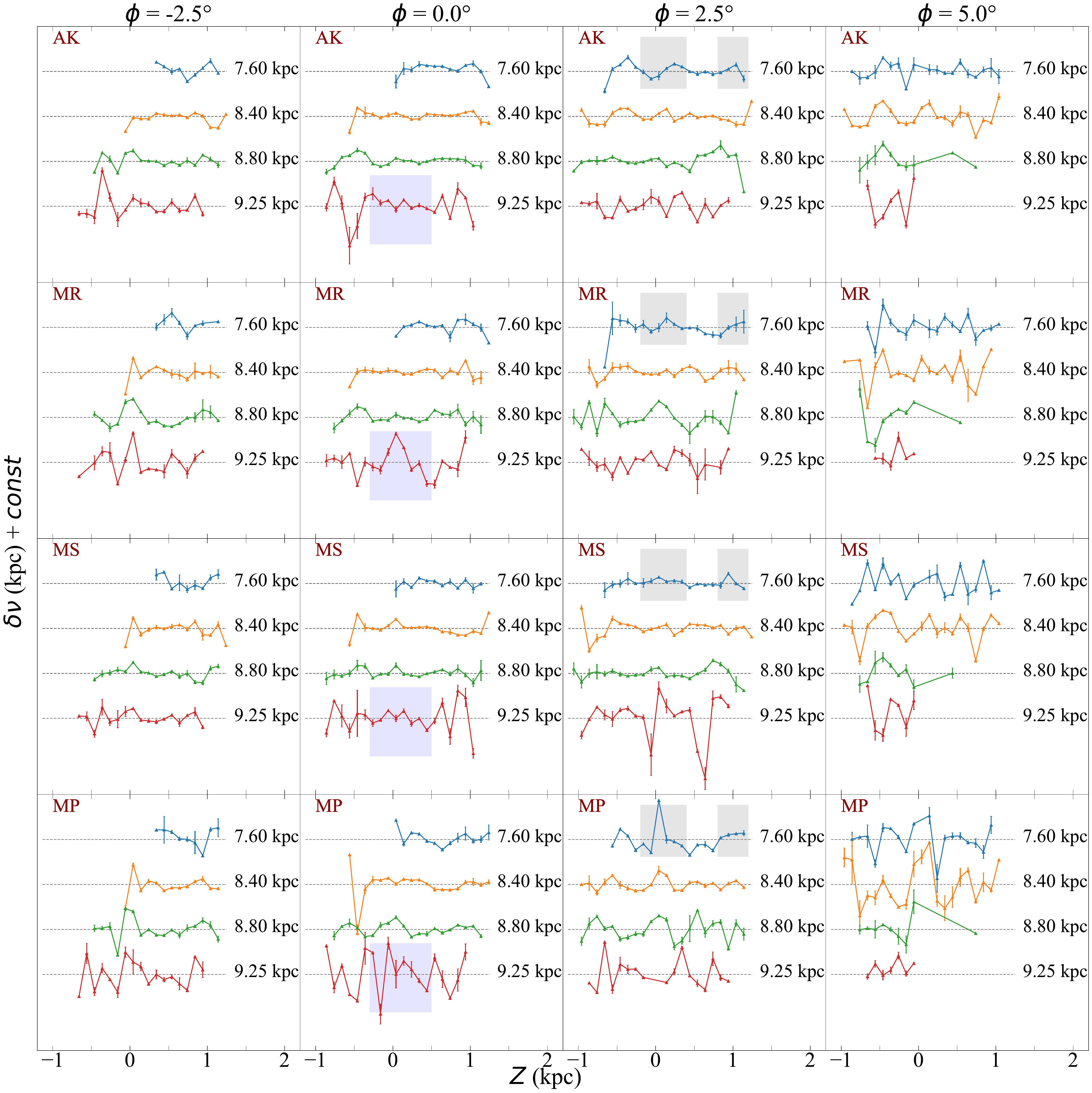} 
    \caption{Same contents as in Fig.~\ref{fig:dk_z_phi} but for the plots of subsamples located at identical $\phi$ values are grouped together.}
\label{fig:dk_z_R}
\end{figure*}

\section{Discussion} \label{sec:discussion}

\subsection{Residuals of the model fitting} \label{subsec:vd}
Figs.~\ref{fig:dk_z_phi} and \ref{fig:dk_z_R} show the residuals $\delta v \equiv v(Z) - v_{\text{model}}(Z)$ between the vertical stellar density profiles and their corresponding best-fit models as a function of $Z$ for different subsamples with varying metallicities and different ($R$, $\phi$) bins. It is observed that larger fluctuations are present in bins with larger values of $R$ (e.g., $R=9.25$ kpc) and larger $\phi$ (e.g., $\phi=5.0^\circ$).

The general trends in density residual variations along the vertical distances $Z$ 
remain consistent for various metallicity samples. Additionally,
we identify an overdensity at $Z=\pm 0.1$ kpc, which is consistent with the $O8.0-0.1$ 
structure reported in \citet{wang2018mapping}. 

However, at specific positions, the intensity and subtle trends in the $\delta v$-$Z$ variations may differ depending on the metallicity. For example, as shown in Fig. \ref{fig:dk_z_phi}, the subsamples at $(R, \phi) = (7.6\,\text{kpc}, -2.5^\circ)$ bin, marked by the blue shaded regions, show the gradual decrease in density for the AK, MS, and MP subsamples. In contrast, the MR sample exhibits an initial increase followed by a decrease. Furthermore, in terms of overall variation intensity, the MP subsample displays stronger fluctuation. 

Upon visual inspection, we observe a lack of data for LAMOST dataset at $Z<0$\,kpc in the bins of ($R,~\phi$) = (7.6 kpc, $-$2.5$^\circ$) and (7.6 kpc, 0.0$^\circ$). This absence of data may potentially result in overestimated scale heights for the southern disk in these bins. Similarly, for the (9.25 kpc, 5.0$^\circ$) bin, the fitted scale height for the northern disk may also be overestimated, as the scale height at this location is determined by prior values. Therefore, we have excluded these regions from our analysis of surface density and mid-plane considerations. Given that the Galactic mid-plane is nearly zero, our focus is specifically on removing the corresponding ($R, \phi$) bins to analyze the scale height accurately.

\begin{figure*}
    \centering
    \includegraphics[width=0.85\textwidth]{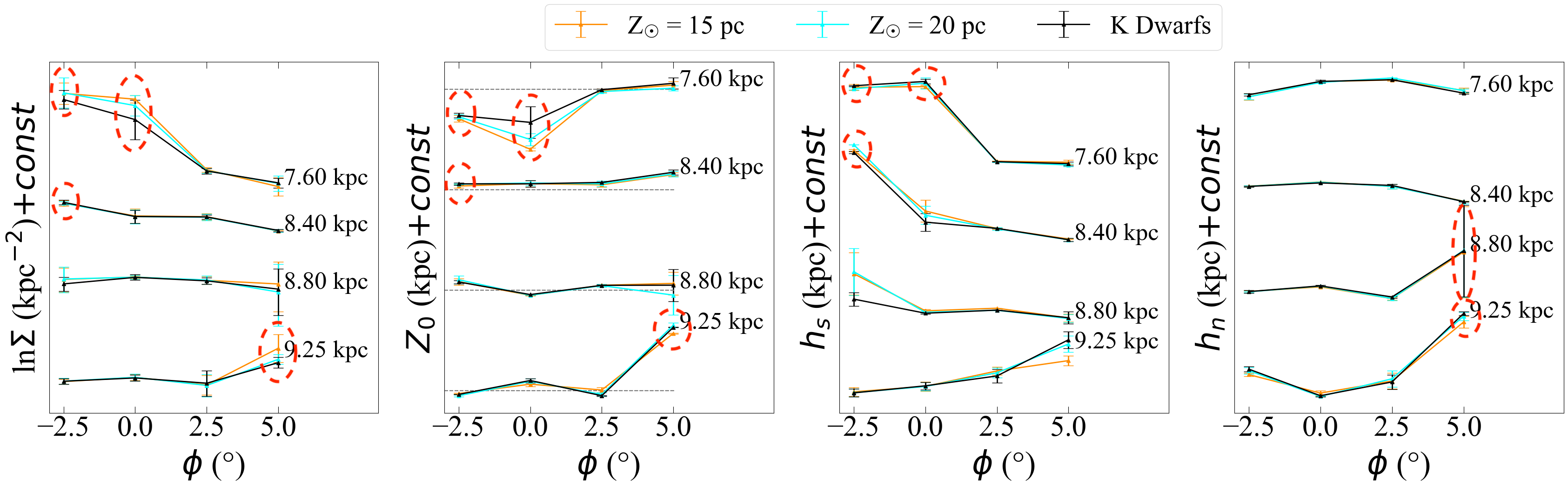} 
    \caption{Variation of the best-fit values of $\ln\Sigma$, $Z_0$, $h_\mathrm{s}$, and $h_\mathrm{n}$ for the AK sample at different $\phi$ values, while keeping the $R$ bins fixed. The vertical positions of the Sun $Z_{\odot}$ are assumed to vary, and the results obtained for different $Z_{\odot}$ values are distinguished by using various colors. The $Z_{\odot} =$ 15, 20, and $ 27$\,pc casea are represented by the orange, cyan and black colors, respectively. In the second panel from left to right, the dashed lines correspond to the location where $Z = 0$\,kpc.} 
\label{fig:dk_zsun_phi}
\end{figure*}

\begin{figure*}
    \centering
    \includegraphics[width=0.8\textwidth]{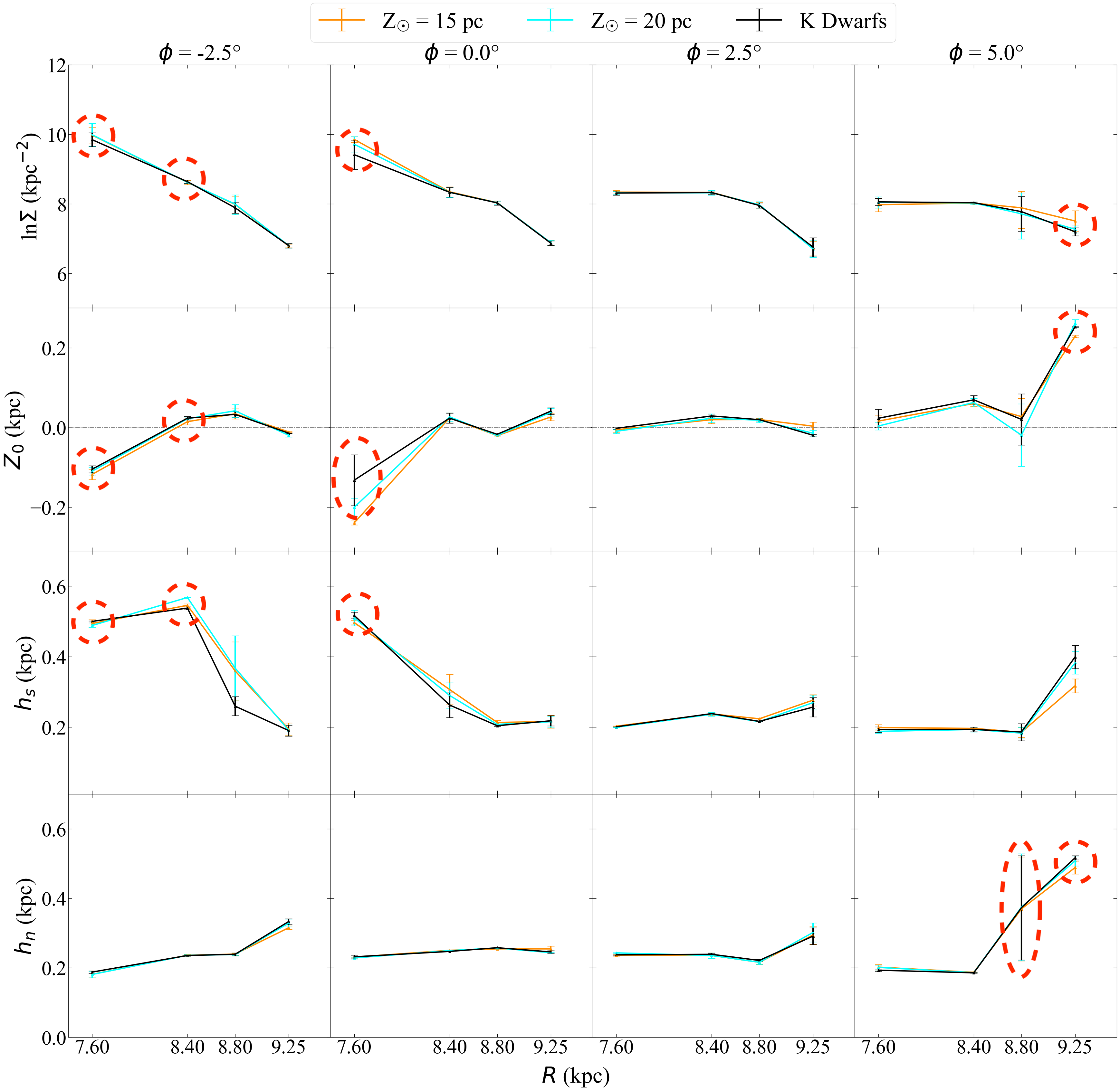} 
    \caption{Similar to Fig.~\ref{fig:dk_zsun_phi} but for the variation of the best-fit values at different $R$ values, while keeping the $\phi$ bins.} 
\label{fig:dk_zsun_R}
\end{figure*}

\subsection{Impact of Vertical Position of the Sun} \label{subsec:z_sun}

In this study, we have made an assumption that the vertical position of the Sun $Z_{\odot}$= 27\,pc \citep{reid2014trigonometric, chen2001stellar}. We here investigate the impact of $Z_{\odot}$ on the best-fit structure parameters. Previous estimates have placed the value of $Z_{\odot}$ within the range of 5 to 29\,pc \citep{2017MNRAS.465..472K,2019A&A...632L...1S,2019Sci...365..478S}. Therefore, in this section, we conduct tests using two different values: $Z_{\odot}$ = 15 and 20\,pc.

Figs.~\ref{fig:dk_zsun_phi} and \ref{fig:dk_zsun_R} show the influence of $Z_{\odot}$ on our results for the AK sample. Overall, the variations of $Z_{\odot}$ values have relatively minor impacts on the best-fit structure parameters. Notable discrepancies are observed for surface density in the (7.6 kpc, 2.5$^\circ$) bin,  Galactic mid-plane position in the (7.6 kpc, 0.0$^\circ$) and (8.8 kpc, 5.0$^\circ$) bins, and the scale height of southern disk in the (8.8 kpc, -2.5$^\circ$) bin. While the northern scale height remains consistent across all ($R,\phi$) bins. However, the maximum discrepancies obtained from different solar positions for $\delta_{max}(\mathrm{ln} \Sigma,Z_0,h_\mathrm{s},h_\mathrm{n})$ are only (0.105, 0.023, 0.108, 0.018), indicating the robustness of our results.

\section{Conclusion} \label{sec:conc}
In this study, we collect a sample of 174,443 K dwarfs observed by both LAMOST and
$Gaia$. Employing single exponential model including the asymmetry of structures on 
the north and south sides of the disk, we try to constrain the density peak
(ln$v_0$), midplane ($Z_\mathrm{0}$), and scale heights of the northern ($h_\mathrm{n}$) and southern ($h_\mathrm{s}$) 
of the disk in the solar neighborhood. With the best fitted results
from the $emcee$, we present the 3D spatial  distribution of the 
Milky Way disk and explore potential correlations between the asymmetric Galactic disk and stellar metallicity.

Firstly, from the distributions of the scale heights in both the north and south of the disk, we observe an asymmetric signal in the vertical direction. The scale height of the AK sample in the southern disk in bins with azimuthal angles of $-2.5^\circ$ and $0.0^\circ$ gradually decreases as $R$ increases. Conversely, bins with $\phi = 5.0^\circ$ show an increasing trend with $R$. Meanwhile, the scale height of the northern disk, $h_n$, increases with increasing $R$.

Secondly, significant signals of non-axisymmetry of the disk is represented by the analysis of the distributions versus azimuthal angle. Surface density and the scale height of the southern disk decrease with increasing $\phi$  at $R$ = 7.6 and 8.4 kpc. Meanwhile, the scale height increases with increasing $\phi$ at $R$ = 9.25 kpc, while the surface density remains almost constant at this radius. And we find that the variance of the density profile may be related to the known substructures such as the $north$ $near$ structure \citep{xu2015rings} or 
the $D8.5+0.2$ structure \citep{widrow2012galactoseismology,wang2018mapping}. 
Further investigation is required to elucidate the underlying causes of these two contrasting trends. 

Thirdly, different subsamples with different metallicity show asymmetric results in the vertical and azimuthal dimensions. And the scale height of the metal-poor sample is slightly higher than that of the metal-rich sample in the northern disk, whereas the relationship between the scale heights of the MR and MP samples in the southern disk is less clear. Additionally, the residuals of the vertical density model at various $(R, \phi)$ bins for the MP sample exhibit greater fluctuations.

Finally, we accomplish sensitivity tests on the influence of the Solar position by considering $Z_\odot$ = 15, 20 and 27 pc. The results do not show significant differences in the scale heights and surface density, and the Galactic midplane $Z_0$ remains close to zero. Combining all results, we find that the warp is not significant in the Solar neighborhood. Further studies on the asymmetry in both vertical and azimuthal directions in a larger volume will follow. The relation between density and dynamics will also be investigated.

\section*{Acknowledgement}
H.T. is supported by  the National Natural Science Foundation of China (NSFC) under grants Nos. 12103062, 12173046, U2031143.
J.L. thanks the National Science Foundation of China (NSFC)  with grant Nos.  12273027, the science research grants from the China Manned Space Project with NO. CMS-CSST-2021-B03, the Sichuan Youth Science and Technology Innovation Research Team (grant No. 21CXTD0038), and the Innovation Team Funds of China West Normal University (grant No. KCXTD2022-6).
B.Q.C. thanks the  National Natural Science Foundation of China (NSFC) with grant Nos. 12173034, 12322304.
C.L. thanks the National Natural Science Foundation of China (NSFC) with grant Nos.11835057 and 
the National Key R\&D Program of China No. 2019YFA0405501. 

Guoshoujing Telescope (the Large Sky Area Multi-Object Fiber Spectroscopic Telescope LAMOST) is a 
National Major Scientific Project built by the Chinese Academy of Sciences. Funding for the project has 
been provided by the National Development and Reform Commission. LAMOST is operated and 
managed by the National Astronomical Observatories, Chinese Academy of Sciences.

This work has made use of data from the European Space Agency (ESA) mission {\it Gaia} (\url{https://www.cosmos.esa.int/gaia}), processed by the {\it Gaia} Data Processing and Analysis Consortium (DPAC, \url{https://www.cosmos.esa.int/web/gaia/dpac/consortium}). Funding for the DPAC has been provided by national institutions, in particular the institutions participating in the {\it Gaia} Multilateral Agreement.

\section*{DATA AVAILABILITY}
The data presented in this article will be shared on reasonable request to the corresponding author.

\bibliography{aa_bib}{}

\begin{thebibliography}{}
\makeatletter
\relax
\def\mn@urlcharsother{\let\do\@makeother \do\$\do\&\do\#\do\^\do\_\do\%\do\~}
\def\mn@doi{\begingroup\mn@urlcharsother \@ifnextchar [ {\mn@doi@} {\mn@doi@[]}}
\def\mn@doi@[#1]#2{\def\@tempa{#1}\ifx\@tempa\@empty \href {http://dx.doi.org/#2} {doi:#2}\else \href {http://dx.doi.org/#2} {#1}\fi \endgroup}
\def\mn@eprint#1#2{\mn@eprint@#1:#2::\@nil}
\def\mn@eprint@arXiv#1{\href {http://arxiv.org/abs/#1} {{\tt arXiv:#1}}}
\def\mn@eprint@dblp#1{\href {http://dblp.uni-trier.de/rec/bibtex/#1.xml} {dblp:#1}}
\def\mn@eprint@#1:#2:#3:#4\@nil{\def\@tempa {#1}\def\@tempb {#2}\def\@tempc {#3}\ifx \@tempc \@empty \let \@tempc \@tempb \let \@tempb \@tempa \fi \ifx \@tempb \@empty \def\@tempb {arXiv}\fi \@ifundefined {mn@eprint@\@tempb}{\@tempb:\@tempc}{\expandafter \expandafter \csname mn@eprint@\@tempb\endcsname \expandafter{\@tempc}}}

\bibitem[\protect\citeauthoryear{Antoja et~al.,}{Antoja et~al.}{2018}]{antoja2018dynamically}
Antoja T.,  et~al., 2018, Nature, 561, 360

\bibitem[\protect\citeauthoryear{Bland-Hawthorn \& Gerhard}{Bland-Hawthorn \& Gerhard}{2016}]{bland2016galaxy}
Bland-Hawthorn J.,  Gerhard O.,  2016, Annual Review of Astronomy and Astrophysics, 54, 529

\bibitem[\protect\citeauthoryear{Bovy, Rix, Liu, Hogg, Beers  \& Lee}{Bovy et~al.}{2012}]{bovy2012spatial}
Bovy J.,  Rix H.-W.,  Liu C.,  Hogg D.~W.,  Beers T.~C.,   Lee Y.~S.,  2012, The Astrophysical Journal, 753, 148

\bibitem[\protect\citeauthoryear{Bovy, Rix, Schlafly, Nidever, Holtzman, Shetrone  \& Beers}{Bovy et~al.}{2016}]{bovy2016stellar}
Bovy J.,  Rix H.-W.,  Schlafly E.~F.,  Nidever D.~L.,  Holtzman J.~A.,  Shetrone M.,   Beers T.~C.,  2016, The Astrophysical Journal, 823, 30

\bibitem[\protect\citeauthoryear{Brown et~al.,}{Brown et~al.}{2021}]{brown2021gaia}
Brown A.~G.,  et~al., 2021, Astronomy \& Astrophysics, 649, A1

\bibitem[\protect\citeauthoryear{Carlin et~al.,}{Carlin et~al.}{2012}]{carlin2012algorithm}
Carlin J.~L.,  et~al., 2012, Research in Astronomy and Astrophysics, 12, 755

\bibitem[\protect\citeauthoryear{Casagrande \& VandenBerg}{Casagrande \& VandenBerg}{2018}]{casagrande2018use}
Casagrande L.,  VandenBerg D.~A.,  2018, Monthly Notices of the Royal Astronomical Society: Letters, 479, L102

\bibitem[\protect\citeauthoryear{Chen et~al.,}{Chen et~al.}{2001}]{chen2001stellar}
Chen B.,  et~al., 2001, The Astrophysical Journal, 553, 184

\bibitem[\protect\citeauthoryear{Chen, Wang, Deng, de Grijs, Liu  \& Tian}{Chen et~al.}{2019a}]{chen2019intuitive}
Chen X.,  Wang S.,  Deng L.,  de Grijs R.,  Liu C.,   Tian H.,  2019a, Nature Astronomy, 3, 320

\bibitem[\protect\citeauthoryear{Chen et~al.,}{Chen et~al.}{2019b}]{chen2019three}
Chen B.,  et~al., 2019b, Monthly Notices of the Royal Astronomical Society, 483, 4277

\bibitem[\protect\citeauthoryear{Chen et~al.,}{Chen et~al.}{2019c}]{chen2019galactic}
Chen B.,  et~al., 2019c, Monthly Notices of the Royal Astronomical Society, 487, 1400

\bibitem[\protect\citeauthoryear{Cui et~al.,}{Cui et~al.}{2012}]{cui2012large}
Cui X.-Q.,  et~al., 2012, Research in Astronomy and Astrophysics, 12, 1197

\bibitem[\protect\citeauthoryear{Dobbie \& Warren}{Dobbie \& Warren}{2020}]{dobbie2020bayesian}
Dobbie P.,  Warren S.~J.,  2020, arXiv preprint arXiv:2003.05757

\bibitem[\protect\citeauthoryear{Ferguson, Gardner  \& Yanny}{Ferguson et~al.}{2017}]{ferguson2017milky}
Ferguson D.,  Gardner S.,   Yanny B.,  2017, The Astrophysical Journal, 843, 141

\bibitem[\protect\citeauthoryear{Gilmore \& Reid}{Gilmore \& Reid}{1983}]{gilmore1983new}
Gilmore G.,  Reid N.,  1983, Monthly Notices of the Royal Astronomical Society, 202, 1025

\bibitem[\protect\citeauthoryear{Green, Schlafly, Zucker, Speagle  \& Finkbeiner}{Green et~al.}{2019}]{green20193d}
Green G.~M.,  Schlafly E.,  Zucker C.,  Speagle J.~S.,   Finkbeiner D.,  2019, The Astrophysical Journal, 887, 93

\bibitem[\protect\citeauthoryear{Hodgkin et~al.,}{Hodgkin et~al.}{2021}]{hodgkin2021gaia}
Hodgkin S.,  et~al., 2021, Astronomy \& Astrophysics, 652, A76

\bibitem[\protect\citeauthoryear{Juri{\'c} et~al.,}{Juri{\'c} et~al.}{2008}]{juric2008milky}
Juri{\'c} M.,  et~al., 2008, The Astrophysical Journal, 673, 864

\bibitem[\protect\citeauthoryear{{Karim} \& {Mamajek}}{{Karim} \& {Mamajek}}{2017}]{2017MNRAS.465..472K}
{Karim} T.,  {Mamajek} E.~E.,  2017, \mn@doi [\mnras] {10.1093/mnras/stw2772}, \href {https://ui.adsabs.harvard.edu/abs/2017MNRAS.465..472K} {465, 472}

\bibitem[\protect\citeauthoryear{Kent, Dame  \& Fazio}{Kent et~al.}{1991}]{kent1991galactic}
Kent S.,  Dame T.,   Fazio G.,  1991, The Astrophysical Journal, 378, 131

\bibitem[\protect\citeauthoryear{L{\'e}na, Lebrun  \& Mignard}{L{\'e}na et~al.}{2013}]{lena2013observational}
L{\'e}na P.,  Lebrun F.,   Mignard F.,  2013, Observational astrophysics.
Springer Science \& Business Media

\bibitem[\protect\citeauthoryear{{Li} \& {Shen}}{{Li} \& {Shen}}{2020}]{2020ApJ...890...85L}
{Li} Z.-Y.,  {Shen} J.,  2020, \mn@doi [\apj] {10.3847/1538-4357/ab6b21}, \href {https://ui.adsabs.harvard.edu/abs/2020ApJ...890...85L} {890, 85}

\bibitem[\protect\citeauthoryear{Lindegren et~al.,}{Lindegren et~al.}{2021}]{lindegren2021gaia}
Lindegren L.,  et~al., 2021, Astronomy \& Astrophysics, 649, A4

\bibitem[\protect\citeauthoryear{Liu \& van~de Ven}{Liu \& van~de Ven}{2012}]{liu2012chemo}
Liu C.,  van~de Ven G.,  2012, Monthly Notices of the Royal Astronomical Society, 425, 2144

\bibitem[\protect\citeauthoryear{Liu et~al.,}{Liu et~al.}{2017}]{liu2017mapping}
Liu C.,  et~al., 2017, Research in Astronomy and Astrophysics, 17, 096

\bibitem[\protect\citeauthoryear{L{\'o}pez-Corredoira \& Molg{\'o}}{L{\'o}pez-Corredoira \& Molg{\'o}}{2014}]{lopez2014flare}
L{\'o}pez-Corredoira M.,  Molg{\'o} J.,  2014, arXiv preprint arXiv:1405.7649

\bibitem[\protect\citeauthoryear{L{\'o}pez-Corredoira, Cabrera-Lavers, Garz{\'o}n  \& Hammersley}{L{\'o}pez-Corredoira et~al.}{2002}]{lopez2002old}
L{\'o}pez-Corredoira M.,  Cabrera-Lavers A.,  Garz{\'o}n F.,   Hammersley P.,  2002, Astronomy \& Astrophysics, 394, 883

\bibitem[\protect\citeauthoryear{Luo et~al.,}{Luo et~al.}{2015}]{luo2015first}
Luo A.-L.,  et~al., 2015, Research in Astronomy and Astrophysics, 15, 1095

\bibitem[\protect\citeauthoryear{{Momany}, {Zaggia}, {Gilmore}, {Piotto}, {Carraro}, {Bedin}  \& {de Angeli}}{{Momany} et~al.}{2006}]{2006A&A...451..515M}
{Momany} Y.,  {Zaggia} S.,  {Gilmore} G.,  {Piotto} G.,  {Carraro} G.,  {Bedin} L.~R.,   {de Angeli} F.,  2006, \mn@doi [\aap] {10.1051/0004-6361:20054081}, \href {https://ui.adsabs.harvard.edu/abs/2006A&A...451..515M} {451, 515}

\bibitem[\protect\citeauthoryear{Morganson et~al.,}{Morganson et~al.}{2016}]{morganson2016mapping}
Morganson E.,  et~al., 2016, The Astrophysical Journal, 825, 140

\bibitem[\protect\citeauthoryear{Newberg et~al.,}{Newberg et~al.}{2002}]{newberg2002ghost}
Newberg H.~J.,  et~al., 2002, The Astrophysical Journal, 569, 245

\bibitem[\protect\citeauthoryear{{Poggio} et~al.,}{{Poggio} et~al.}{2018}]{2018MNRAS.481L..21P}
{Poggio} E.,  et~al., 2018, \mn@doi [\mnras] {10.1093/mnrasl/sly148}, \href {https://ui.adsabs.harvard.edu/abs/2018MNRAS.481L..21P} {481, L21}

\bibitem[\protect\citeauthoryear{Price-Whelan, Johnston, Sheffield, Laporte  \& Sesar}{Price-Whelan et~al.}{2015}]{price2015reinterpretation}
Price-Whelan A.~M.,  Johnston K.~V.,  Sheffield A.~A.,  Laporte C.~F.,   Sesar B.,  2015, Monthly Notices of the Royal Astronomical Society, 452, 676

\bibitem[\protect\citeauthoryear{Reid et~al.,}{Reid et~al.}{2014}]{reid2014trigonometric}
Reid M.,  et~al., 2014, The Astrophysical Journal, 783, 130

\bibitem[\protect\citeauthoryear{{Siegert}}{{Siegert}}{2019}]{2019A&A...632L...1S}
{Siegert} T.,  2019, \mn@doi [\aap] {10.1051/0004-6361/201936659}, \href {https://ui.adsabs.harvard.edu/abs/2019A&A...632L...1S} {632, L1}

\bibitem[\protect\citeauthoryear{{Skowron} et~al.,}{{Skowron} et~al.}{2019}]{2019Sci...365..478S}
{Skowron} D.~M.,  et~al., 2019, \mn@doi [Science] {10.1126/science.aau3181}, \href {https://ui.adsabs.harvard.edu/abs/2019Sci...365..478S} {365, 478}

\bibitem[\protect\citeauthoryear{Skrutskie et~al.,}{Skrutskie et~al.}{2006}]{skrutskie2006two}
Skrutskie M.,  et~al., 2006, The Astronomical Journal, 131, 1163

\bibitem[\protect\citeauthoryear{Van Der~Kruit}{Van Der~Kruit}{1988}]{van1988three}
Van Der~Kruit P.,  1988, Astronomy and Astrophysics (ISSN 0004-6361), vol. 192, no. 1-2, March 1988, p. 117-127., 192, 117

\bibitem[\protect\citeauthoryear{Van~der Kruit \& Freeman}{Van~der Kruit \& Freeman}{2011}]{van2011galaxy}
Van~der Kruit P.,  Freeman K.,  2011, Annual Review of Astronomy and Astrophysics, 49, 301

\bibitem[\protect\citeauthoryear{Wan, Liu  \& Deng}{Wan et~al.}{2017}]{wan2017red}
Wan J.-C.,  Liu C.,   Deng L.-C.,  2017, Research in Astronomy and Astrophysics, 17, 079

\bibitem[\protect\citeauthoryear{Wang, Liu, Xu, Wan  \& Deng}{Wang et~al.}{2018}]{wang2018mapping}
Wang H.-F.,  Liu C.,  Xu Y.,  Wan J.-C.,   Deng L.,  2018, Monthly Notices of the Royal Astronomical Society, 478, 3367

\bibitem[\protect\citeauthoryear{Wang et~al.,}{Wang et~al.}{2020}]{wang2020mapping}
Wang H.-F.,  et~al., 2020, The Astrophysical Journal, 897, 119

\bibitem[\protect\citeauthoryear{Widrow, Gardner, Yanny, Dodelson  \& Chen}{Widrow et~al.}{2012}]{widrow2012galactoseismology}
Widrow L.~M.,  Gardner S.,  Yanny B.,  Dodelson S.,   Chen H.-Y.,  2012, The Astrophysical Journal Letters, 750, L41

\bibitem[\protect\citeauthoryear{Wu, Du, Luo, Zhao  \& Yuan}{Wu et~al.}{2014}]{wu2014automatic}
Wu Y.,  Du B.,  Luo A.,  Zhao Y.,   Yuan H.,  2014, Proceedings of the International Astronomical Union, 10, 340

\bibitem[\protect\citeauthoryear{{Xiang} et~al.,}{{Xiang} et~al.}{2018}]{Xiang2018msto}
{Xiang} M.,  et~al., 2018, \mn@doi [\apjs] {10.3847/1538-4365/aad237}, \href {https://ui.adsabs.harvard.edu/abs/2018ApJS..237...33X} {237, 33}

\bibitem[\protect\citeauthoryear{Xu, Newberg, Carlin, Liu, Deng, Li, Sch{\"o}nrich  \& Yanny}{Xu et~al.}{2015}]{xu2015rings}
Xu Y.,  Newberg H.~J.,  Carlin J.~L.,  Liu C.,  Deng L.,  Li J.,  Sch{\"o}nrich R.,   Yanny B.,  2015, The Astrophysical Journal, 801, 105

\bibitem[\protect\citeauthoryear{Xu et~al.,}{Xu et~al.}{2020}]{xu2020exploring}
Xu Y.,  et~al., 2020, The Astrophysical Journal, 905, 6

\bibitem[\protect\citeauthoryear{{Yu} et~al.,}{{Yu} et~al.}{2021a}]{Yu2021rc}
{Yu} Z.,  et~al., 2021a, \mn@doi [\apj] {10.3847/1538-4357/abf098}, \href {https://ui.adsabs.harvard.edu/abs/2021ApJ...912..106Y} {912, 106}

\bibitem[\protect\citeauthoryear{Yu et~al.,}{Yu et~al.}{2021b}]{yu2021flare}
Yu Y.,  et~al., 2021b, The Astrophysical Journal, 922, 80

\bibitem[\protect\citeauthoryear{Zhao, Zhao, Chu, Jing  \& Deng}{Zhao et~al.}{2012}]{zhao2012lamost}
Zhao G.,  Zhao Y.-H.,  Chu Y.-Q.,  Jing Y.-P.,   Deng L.-C.,  2012, Research in Astronomy and Astrophysics, 12, 723

\makeatother
\end{thebibliography}


\begin{thebibliography}{}
\makeatletter
\relax
\def\mn@urlcharsother{\let\do\@makeother \do\$\do\&\do\#\do\^\do\_\do\%\do\~}
\def\mn@doi{\begingroup\mn@urlcharsother \@ifnextchar [ {\mn@doi@}
  {\mn@doi@[]}}
\def\mn@doi@[#1]#2{\def\@tempa{#1}\ifx\@tempa\@empty \href
  {http://dx.doi.org/#2} {doi:#2}\else \href {http://dx.doi.org/#2} {#1}\fi
  \endgroup}
\def\mn@eprint#1#2{\mn@eprint@#1:#2::\@nil}
\def\mn@eprint@arXiv#1{\href {http://arxiv.org/abs/#1} {{\tt arXiv:#1}}}
\def\mn@eprint@dblp#1{\href {http://dblp.uni-trier.de/rec/bibtex/#1.xml}
  {dblp:#1}}
\def\mn@eprint@#1:#2:#3:#4\@nil{\def\@tempa {#1}\def\@tempb {#2}\def\@tempc
  {#3}\ifx \@tempc \@empty \let \@tempc \@tempb \let \@tempb \@tempa \fi \ifx
  \@tempb \@empty \def\@tempb {arXiv}\fi \@ifundefined
  {mn@eprint@\@tempb}{\@tempb:\@tempc}{\expandafter \expandafter \csname
  mn@eprint@\@tempb\endcsname \expandafter{\@tempc}}}

\bibitem[\protect\citeauthoryear{{Abazajian}, {Canac}, {Horiuchi}  \&
  {Kaplinghat}}{{Abazajian} et~al.}{2014}]{A+14}
{Abazajian} K.~N.,  {Canac} N.,  {Horiuchi} S.,   {Kaplinghat} M.,  2014,
  \mn@doi [\prd] {10.1103/PhysRevD.90.023526}, \href
  {https://ui.adsabs.harvard.edu/abs/2014PhRvD..90b3526A} {90, 023526}

\bibitem[\protect\citeauthoryear{{Ackermann} et~al.,}{{Ackermann}
  et~al.}{2017a}]{2017ApJ...836..208A}
{Ackermann} M.,  et~al., 2017a, \mn@doi [apj] {10.3847/1538-4357/aa5c3d}, \href
  {https://ui.adsabs.harvard.edu/\#abs/2017ApJ...836..208A} {836, 208}

\bibitem[\protect\citeauthoryear{{Ackermann} et~al.,}{{Ackermann}
  et~al.}{2017b}]{A+17}
{Ackermann} M.,  et~al., 2017b, \mn@doi [\apj] {10.3847/1538-4357/aa6cab},
  \href {https://ui.adsabs.harvard.edu/abs/2017ApJ...840...43A} {840, 43}

\bibitem[\protect\citeauthoryear{{Adamo}, {Kruijssen}, {Bastian}, {Silva-Villa}
   \& {Ryon}}{{Adamo} et~al.}{2015}]{2015MNRAS.452..246A}
{Adamo} A.,  {Kruijssen} J.~M.~D.,  {Bastian} N.,  {Silva-Villa} E.,   {Ryon}
  J.,  2015, \mn@doi [\mnras] {10.1093/mnras/stv1203}, \href
  {https://ui.adsabs.harvard.edu/\#abs/2015MNRAS.452..246A} {452, 246}

\bibitem[\protect\citeauthoryear{{Agarwal} \& {Milosavljevi{\'c}}}{{Agarwal} \&
  {Milosavljevi{\'c}}}{2011}]{2011ApJ...729...35A}
{Agarwal} M.,  {Milosavljevi{\'c}} M.,  2011, \mn@doi [\apj]
  {10.1088/0004-637X/729/1/35}, \href
  {https://ui.adsabs.harvard.edu/abs/2011ApJ...729...35A} {729, 35}

\bibitem[\protect\citeauthoryear{{Aharon} \& {Perets}}{{Aharon} \&
  {Perets}}{2015}]{2015ApJ...799..185A}
{Aharon} D.,  {Perets} H.~B.,  2015, \mn@doi [\apj]
  {10.1088/0004-637X/799/2/185}, \href
  {https://ui.adsabs.harvard.edu/abs/2015ApJ...799..185A} {799, 185}

\bibitem[\protect\citeauthoryear{{Ajello} et~al.,}{{Ajello}
  et~al.}{2016}]{A+16}
{Ajello} M.,  et~al., 2016, \mn@doi [ApJ] {10.3847/0004-637X/819/1/44}, \href
  {https://ui.adsabs.harvard.edu/\#abs/2016ApJ...819...44A} {819, 44}

\bibitem[\protect\citeauthoryear{{Albert} et~al.,}{{Albert}
  et~al.}{2017}]{2017ApJ...834..110A}
{Albert} A.,  et~al., 2017, \mn@doi [\apj] {10.3847/1538-4357/834/2/110}, \href
  {https://ui.adsabs.harvard.edu/abs/2017ApJ...834..110A} {834, 110}

\bibitem[\protect\citeauthoryear{{Alfaro-Cuello} et~al.,}{{Alfaro-Cuello}
  et~al.}{2019}]{2019ApJ...886...57A}
{Alfaro-Cuello} M.,  et~al., 2019, \mn@doi [\apj] {10.3847/1538-4357/ab1b2c},
  \href {https://ui.adsabs.harvard.edu/abs/2019ApJ...886...57A} {886, 57}

\bibitem[\protect\citeauthoryear{Ambartsumian, Goodman  \& Hut}{Ambartsumian
  et~al.}{1938}]{ambartsumian1938proc}
Ambartsumian V.,  Goodman J.,   Hut P.,  1938, Proc. IAU Symp. 113, Dynamics of
  Star Clusters

\bibitem[\protect\citeauthoryear{{Antonini}, {Capuzzo-Dolcetta},
  {Mastrobuono-Battisti}  \& {Merritt}}{{Antonini}
  et~al.}{2012}]{2012ApJ...750..111A}
{Antonini} F.,  {Capuzzo-Dolcetta} R.,  {Mastrobuono-Battisti} A.,   {Merritt}
  D.,  2012, \mn@doi [\apj] {10.1088/0004-637X/750/2/111}, \href
  {https://ui.adsabs.harvard.edu/abs/2012ApJ...750..111A} {750, 111}

\bibitem[\protect\citeauthoryear{{Antonini}, {Barausse}  \& {Silk}}{{Antonini}
  et~al.}{2015}]{2015ApJ...812...72A}
{Antonini} F.,  {Barausse} E.,   {Silk} J.,  2015, \mn@doi [\apj]
  {10.1088/0004-637X/812/1/72}, \href
  {https://ui.adsabs.harvard.edu/abs/2015ApJ...812...72A} {812, 72}

\bibitem[\protect\citeauthoryear{{Arca-Sedda} \&
  {Capuzzo-Dolcetta}}{{Arca-Sedda} \&
  {Capuzzo-Dolcetta}}{2014}]{2014ApJ...785...51A}
{Arca-Sedda} M.,  {Capuzzo-Dolcetta} R.,  2014, \mn@doi [\apj]
  {10.1088/0004-637X/785/1/51}, \href
  {https://ui.adsabs.harvard.edu/abs/2014ApJ...785...51A} {785, 51}

\bibitem[\protect\citeauthoryear{{Arca-Sedda} \& {Gualandris}}{{Arca-Sedda} \&
  {Gualandris}}{2018}]{2018MNRAS.477.4423A}
{Arca-Sedda} M.,  {Gualandris} A.,  2018, \mn@doi [\mnras]
  {10.1093/mnras/sty922}, \href
  {https://ui.adsabs.harvard.edu/abs/2018MNRAS.477.4423A} {477, 4423}

\bibitem[\protect\citeauthoryear{{Ashman} \& {Zepf}}{{Ashman} \&
  {Zepf}}{1992}]{1992ApJ...384...50A}
{Ashman} K.~M.,  {Zepf} S.~E.,  1992, \mn@doi [\apj] {10.1086/170850}, \href
  {https://ui.adsabs.harvard.edu/abs/1992ApJ...384...50A} {384, 50}

\bibitem[\protect\citeauthoryear{{Atwood} et~al.,}{{Atwood}
  et~al.}{2009}]{2009ApJ...697.1071A}
{Atwood} W.~B.,  et~al., 2009, \mn@doi [apj] {10.1088/0004-637X/697/2/1071},
  \href {https://ui.adsabs.harvard.edu/\#abs/2009ApJ...697.1071A} {697, 1071}

\bibitem[\protect\citeauthoryear{{Baumgardt}, {Hut}  \& {Heggie}}{{Baumgardt}
  et~al.}{2002}]{2002MNRAS.336.1069B}
{Baumgardt} H.,  {Hut} P.,   {Heggie} D.~C.,  2002, \mn@doi [\mnras]
  {10.1046/j.1365-8711.2002.05736.x}, \href
  {https://ui.adsabs.harvard.edu/abs/2002MNRAS.336.1069B} {336, 1069}

\bibitem[\protect\citeauthoryear{{Baumgardt}, {Sollima}, {Hilker}, {Bellini}
  \& {Vasiliev}}{{Baumgardt} et~al.}{2021}]{B+21}
{Baumgardt} H.,  {Sollima} A.,  {Hilker} M.,  {Bellini} A.,   {Vasiliev} E.,
  2021, {Fundamental parameters of Galactic globular clusters (as of May
  2021)}, \url {https://people.smp.uq.edu.au/HolgerBaumgardt/globular/}

\bibitem[\protect\citeauthoryear{{Behroozi}, {Wechsler}  \&
  {Conroy}}{{Behroozi} et~al.}{2013}]{2013ApJ...770...57B}
{Behroozi} P.~S.,  {Wechsler} R.~H.,   {Conroy} C.,  2013, \mn@doi [\apj]
  {10.1088/0004-637X/770/1/57}, \href
  {https://ui.adsabs.harvard.edu/\#abs/2013ApJ...770...57B} {770, 57}

\bibitem[\protect\citeauthoryear{{Bender} et~al.,}{{Bender}
  et~al.}{2005}]{2005ApJ...631..280B}
{Bender} R.,  et~al., 2005, \mn@doi [\apj] {10.1086/432434}, \href
  {https://ui.adsabs.harvard.edu/abs/2005ApJ...631..280B} {631, 280}

\bibitem[\protect\citeauthoryear{{Bhattacharya} \& {van den
  Heuvel}}{{Bhattacharya} \& {van den Heuvel}}{1991}]{1991PhR...203....1B}
{Bhattacharya} D.,  {van den Heuvel} E.~P.~J.,  1991, \mn@doi [\physrep]
  {10.1016/0370-1573(91)90064-S}, \href
  {https://ui.adsabs.harvard.edu/abs/1991PhR...203....1B} {203, 1}

\bibitem[\protect\citeauthoryear{{Bigiel} et~al.,}{{Bigiel}
  et~al.}{2011}]{2011ApJ...730L..13B}
{Bigiel} F.,  et~al., 2011, \mn@doi [\apj] {10.1088/2041-8205/730/2/L13}, \href
  {https://ui.adsabs.harvard.edu/\#abs/2011ApJ...730L..13B} {730, L13}

\bibitem[\protect\citeauthoryear{{B{\"o}ker}, {van der Marel}, {Mazzuca},
  {Rix}, {Rudnick}, {Ho}  \& {Shields}}{{B{\"o}ker}
  et~al.}{2001}]{2001AJ....121.1473B}
{B{\"o}ker} T.,  {van der Marel} R.~P.,  {Mazzuca} L.,  {Rix} H.-W.,  {Rudnick}
  G.,  {Ho} L.~C.,   {Shields} J.~C.,  2001, \mn@doi [\aj] {10.1086/319415},
  \href {https://ui.adsabs.harvard.edu/abs/2001AJ....121.1473B} {121, 1473}

\bibitem[\protect\citeauthoryear{{B{\"o}ker}, {Laine}, {van der Marel},
  {Sarzi}, {Rix}, {Ho}  \& {Shields}}{{B{\"o}ker}
  et~al.}{2002}]{2002AJ....123.1389B}
{B{\"o}ker} T.,  {Laine} S.,  {van der Marel} R.~P.,  {Sarzi} M.,  {Rix} H.-W.,
   {Ho} L.~C.,   {Shields} J.~C.,  2002, \mn@doi [\aj] {10.1086/339025}, \href
  {https://ui.adsabs.harvard.edu/abs/2002AJ....123.1389B} {123, 1389}

\bibitem[\protect\citeauthoryear{{Brandt} \& {Kocsis}}{{Brandt} \&
  {Kocsis}}{2015}]{2015ApJ...812...15B}
{Brandt} T.~D.,  {Kocsis} B.,  2015, \mn@doi [\apj]
  {10.1088/0004-637X/812/1/15}, \href
  {https://ui.adsabs.harvard.edu/abs/2015ApJ...812...15B} {812, 15}

\bibitem[\protect\citeauthoryear{{Bullock}, {Dekel}, {Kolatt}, {Kravtsov},
  {Klypin}, {Porciani}  \& {Primack}}{{Bullock}
  et~al.}{2001}]{2001ApJ...555..240B}
{Bullock} J.~S.,  {Dekel} A.,  {Kolatt} T.~S.,  {Kravtsov} A.~V.,  {Klypin}
  A.~A.,  {Porciani} C.,   {Primack} J.~R.,  2001, \mn@doi [\apj]
  {10.1086/321477}, \href
  {https://ui.adsabs.harvard.edu/\#abs/2001ApJ...555..240B} {555, 240}

\bibitem[\protect\citeauthoryear{{Caldwell} \& {Bothun}}{{Caldwell} \&
  {Bothun}}{1987}]{1987AJ.....94.1126C}
{Caldwell} N.,  {Bothun} G.~D.,  1987, \mn@doi [\aj] {10.1086/114550}, \href
  {https://ui.adsabs.harvard.edu/abs/1987AJ.....94.1126C} {94, 1126}

\bibitem[\protect\citeauthoryear{Calore, Cholis, McCabe  \& Weniger}{Calore
  et~al.}{2015a}]{PhysRevD.91.063003}
Calore F.,  Cholis I.,  McCabe C.,   Weniger C.,  2015a, \mn@doi [Phys. Rev. D]
  {10.1103/PhysRevD.91.063003}, 91, 063003

\bibitem[\protect\citeauthoryear{{Calore}, {Cholis}  \& {Weniger}}{{Calore}
  et~al.}{2015b}]{C+15}
{Calore} F.,  {Cholis} I.,   {Weniger} C.,  2015b, \mn@doi [\jcap]
  {10.1088/1475-7516/2015/03/038}, \href
  {https://ui.adsabs.harvard.edu/abs/2015JCAP...03..038C} {2015, 038}

\bibitem[\protect\citeauthoryear{{Capuzzo-Dolcetta}}{{Capuzzo-Dolcetta}}{1993}]{1993ApJ...415..616C}
{Capuzzo-Dolcetta} R.,  1993, \mn@doi [\apj] {10.1086/173189}, \href
  {https://ui.adsabs.harvard.edu/abs/1993ApJ...415..616C} {415, 616}

\bibitem[\protect\citeauthoryear{{Capuzzo-Dolcetta} \&
  {Mastrobuono-Battisti}}{{Capuzzo-Dolcetta} \&
  {Mastrobuono-Battisti}}{2009}]{2009A&A...507..183C}
{Capuzzo-Dolcetta} R.,  {Mastrobuono-Battisti} A.,  2009, \mn@doi [\aap]
  {10.1051/0004-6361/200912255}, \href
  {https://ui.adsabs.harvard.edu/abs/2009A&A...507..183C} {507, 183}

\bibitem[\protect\citeauthoryear{{Carson}, {Barth}, {Seth}, {den Brok},
  {Cappellari}, {Greene}, {Ho}  \& {Neumayer}}{{Carson}
  et~al.}{2015}]{2015AJ....149..170C}
{Carson} D.~J.,  {Barth} A.~J.,  {Seth} A.~C.,  {den Brok} M.,  {Cappellari}
  M.,  {Greene} J.~E.,  {Ho} L.~C.,   {Neumayer} N.,  2015, \mn@doi [\aj]
  {10.1088/0004-6256/149/5/170}, \href
  {https://ui.adsabs.harvard.edu/abs/2015AJ....149..170C} {149, 170}

\bibitem[\protect\citeauthoryear{{Chen} \& {Gnedin}}{{Chen} \&
  {Gnedin}}{2022}]{2022MNRAS.514.4736C}
{Chen} Y.,  {Gnedin} O.~Y.,  2022, \mn@doi [\mnras] {10.1093/mnras/stac1651},
  \href {https://ui.adsabs.harvard.edu/abs/2022MNRAS.514.4736C} {514, 4736}

\bibitem[\protect\citeauthoryear{{Chen} \& {Gnedin}}{{Chen} \&
  {Gnedin}}{2023}]{2023MNRAS.522.5638C}
{Chen} Y.,  {Gnedin} O.~Y.,  2023, \mn@doi [\mnras] {10.1093/mnras/stad1328},
  \href {https://ui.adsabs.harvard.edu/abs/2023MNRAS.522.5638C} {522, 5638}

\bibitem[\protect\citeauthoryear{{Chernoff} \& {Weinberg}}{{Chernoff} \&
  {Weinberg}}{1990}]{1990ApJ...351..121C}
{Chernoff} D.~F.,  {Weinberg} M.~D.,  1990, \mn@doi [apj] {10.1086/168451},
  \href {https://ui.adsabs.harvard.edu/\#abs/1990ApJ...351..121C} {351, 121}

\bibitem[\protect\citeauthoryear{{Choksi} \& {Gnedin}}{{Choksi} \&
  {Gnedin}}{2019}]{CG19}
{Choksi} N.,  {Gnedin} O.~Y.,  2019, \mn@doi [\mnras] {10.1093/mnras/stz811},
  \href {https://ui.adsabs.harvard.edu/abs/2019MNRAS.486..331C} {486, 331}

\bibitem[\protect\citeauthoryear{{Choksi}, {Gnedin}  \& {Li}}{{Choksi}
  et~al.}{2018}]{CGL18}
{Choksi} N.,  {Gnedin} O.~Y.,   {Li} H.,  2018, \mn@doi [\mnras]
  {10.1093/mnras/sty1952}, \href
  {https://ui.adsabs.harvard.edu/\#abs/2018MNRAS.480.2343C} {480, 2343}

\bibitem[\protect\citeauthoryear{{Cholis}, {Hooper}  \& {Linden}}{{Cholis}
  et~al.}{2015}]{2015JCAP...06..043C}
{Cholis} I.,  {Hooper} D.,   {Linden} T.,  2015, \mn@doi [\jcap]
  {10.1088/1475-7516/2015/06/043}, \href
  {https://ui.adsabs.harvard.edu/abs/2015JCAP...06..043C} {2015, 043}

\bibitem[\protect\citeauthoryear{{Cholis}, {Zhong}, {McDermott}  \&
  {Surdutovich}}{{Cholis} et~al.}{2022}]{2022PhRvD.105j3023C}
{Cholis} I.,  {Zhong} Y.-M.,  {McDermott} S.~D.,   {Surdutovich} J.~P.,  2022,
  \mn@doi [\prd] {10.1103/PhysRevD.105.103023}, \href
  {https://ui.adsabs.harvard.edu/abs/2022PhRvD.105j3023C} {105, 103023}

\bibitem[\protect\citeauthoryear{{Conroy} \& {Gunn}}{{Conroy} \&
  {Gunn}}{2010}]{2010ApJ...712..833C}
{Conroy} C.,  {Gunn} J.~E.,  2010, \mn@doi [\apj]
  {10.1088/0004-637X/712/2/833}, \href
  {https://ui.adsabs.harvard.edu/abs/2010ApJ...712..833C} {712, 833}

\bibitem[\protect\citeauthoryear{{Daylan}, {Finkbeiner}, {Hooper}, {Linden},
  {Portillo}, {Rodd}  \& {Slatyer}}{{Daylan} et~al.}{2016}]{D+16}
{Daylan} T.,  {Finkbeiner} D.~P.,  {Hooper} D.,  {Linden} T.,  {Portillo} S.
  K.~N.,  {Rodd} N.~L.,   {Slatyer} T.~R.,  2016, \mn@doi [Physics of the Dark
  Universe] {10.1016/j.dark.2015.12.005}, \href
  {https://ui.adsabs.harvard.edu/abs/2016PDU....12....1D} {12, 1}

\bibitem[\protect\citeauthoryear{{Deason} et~al.,}{{Deason}
  et~al.}{2021}]{2021MNRAS.501.5964D}
{Deason} A.~J.,  et~al., 2021, \mn@doi [\mnras] {10.1093/mnras/staa3984}, \href
  {https://ui.adsabs.harvard.edu/abs/2021MNRAS.501.5964D} {501, 5964}

\bibitem[\protect\citeauthoryear{{Dehnen} \& {Binney}}{{Dehnen} \&
  {Binney}}{1998}]{1998MNRAS.294..429D}
{Dehnen} W.,  {Binney} J.,  1998, \mn@doi [\mnras]
  {10.1046/j.1365-8711.1998.01282.x}, \href
  {https://ui.adsabs.harvard.edu/abs/1998MNRAS.294..429D} {294, 429}

\bibitem[\protect\citeauthoryear{{Di Mauro}, {Hou}, {Eckner}, {Zaharijas}  \&
  {Charles}}{{Di Mauro} et~al.}{2019}]{2019PhRvD..99l3027D}
{Di Mauro} M.,  {Hou} X.,  {Eckner} C.,  {Zaharijas} G.,   {Charles} E.,  2019,
  \mn@doi [\prd] {10.1103/PhysRevD.99.123027}, \href
  {https://ui.adsabs.harvard.edu/abs/2019PhRvD..99l3027D} {99, 123027}

\bibitem[\protect\citeauthoryear{{Do} et~al.,}{{Do}
  et~al.}{2019}]{2019Sci...365..664D}
{Do} T.,  et~al., 2019, \mn@doi [Science] {10.1126/science.aav8137}, \href
  {https://ui.adsabs.harvard.edu/abs/2019Sci...365..664D} {365, 664}

\bibitem[\protect\citeauthoryear{{Eckner} et~al.,}{{Eckner}
  et~al.}{2018}]{2018ApJ...862...79E}
{Eckner} C.,  et~al., 2018, \mn@doi [\apj] {10.3847/1538-4357/aac029}, \href
  {https://ui.adsabs.harvard.edu/abs/2018ApJ...862...79E} {862, 79}

\bibitem[\protect\citeauthoryear{{Eigenthaler} et~al.,}{{Eigenthaler}
  et~al.}{2018}]{2018ApJ...855..142E}
{Eigenthaler} P.,  et~al., 2018, \mn@doi [\apj] {10.3847/1538-4357/aaab60},
  \href {https://ui.adsabs.harvard.edu/abs/2018ApJ...855..142E} {855, 142}

\bibitem[\protect\citeauthoryear{{El-Badry}, {Quataert}, {Weisz}, {Choksi}  \&
  {Boylan-Kolchin}}{{El-Badry} et~al.}{2019}]{2019MNRAS.482.4528E}
{El-Badry} K.,  {Quataert} E.,  {Weisz} D.~R.,  {Choksi} N.,   {Boylan-Kolchin}
  M.,  2019, \mn@doi [\mnras] {10.1093/mnras/sty3007}, \href
  {https://ui.adsabs.harvard.edu/abs/2019MNRAS.482.4528E} {482, 4528}

\bibitem[\protect\citeauthoryear{{Fahrion}, {Leaman}, {Lyubenova}  \& {van de
  Ven}}{{Fahrion} et~al.}{2022}]{2022A&A...658A.172F}
{Fahrion} K.,  {Leaman} R.,  {Lyubenova} M.,   {van de Ven} G.,  2022, \mn@doi
  [\aap] {10.1051/0004-6361/202039778}, \href
  {https://ui.adsabs.harvard.edu/abs/2022A&A...658A.172F} {658, A172}

\bibitem[\protect\citeauthoryear{{Fakhouri} \& {Ma}}{{Fakhouri} \&
  {Ma}}{2008}]{2008MNRAS.386..577F}
{Fakhouri} O.,  {Ma} C.-P.,  2008, \mn@doi [\mnras]
  {10.1111/j.1365-2966.2008.13075.x}, \href
  {https://ui.adsabs.harvard.edu/abs/2008MNRAS.386..577F} {386, 577}

\bibitem[\protect\citeauthoryear{{Fakhouri}, {Ma}  \&
  {Boylan-Kolchin}}{{Fakhouri} et~al.}{2010}]{2010MNRAS.406.2267F}
{Fakhouri} O.,  {Ma} C.-P.,   {Boylan-Kolchin} M.,  2010, \mn@doi [\mnras]
  {10.1111/j.1365-2966.2010.16859.x}, \href
  {https://ui.adsabs.harvard.edu/abs/2010MNRAS.406.2267F} {406, 2267}

\bibitem[\protect\citeauthoryear{{Feldmann}}{{Feldmann}}{2013}]{2013MNRAS.433.1910F}
{Feldmann} R.,  2013, \mn@doi [\mnras] {10.1093/mnras/stt851}, \href
  {https://ui.adsabs.harvard.edu/\#abs/2013MNRAS.433.1910F} {433, 1910}

\bibitem[\protect\citeauthoryear{{Feng}, {Li}, {Su}, {Tam}  \& {Chen}}{{Feng}
  et~al.}{2019}]{2019RAA....19...46F}
{Feng} L.,  {Li} Z.-Y.,  {Su} M.,  {Tam} P.-H.~T.,   {Chen} Y.,  2019, \mn@doi
  [Research in Astronomy and Astrophysics] {10.1088/1674-4527/19/3/46}, \href
  {https://ui.adsabs.harvard.edu/abs/2019RAA....19...46F} {19, 046}

\bibitem[\protect\citeauthoryear{{Filippenko} \& {Ho}}{{Filippenko} \&
  {Ho}}{2003}]{2003ApJ...588L..13F}
{Filippenko} A.~V.,  {Ho} L.~C.,  2003, \mn@doi [\apjl] {10.1086/375361}, \href
  {https://ui.adsabs.harvard.edu/abs/2003ApJ...588L..13F} {588, L13}

\bibitem[\protect\citeauthoryear{{Forbes} et~al.,}{{Forbes}
  et~al.}{2018}]{2018RSPSA.47470616F}
{Forbes} D.~A.,  et~al., 2018, \mn@doi [Proceedings of the Royal Society of
  London Series A] {10.1098/rspa.2017.0616}, \href
  {https://ui.adsabs.harvard.edu/abs/2018RSPSA.47470616F} {474, 20170616}

\bibitem[\protect\citeauthoryear{{Ford} et~al.,}{{Ford}
  et~al.}{2013}]{2013ApJ...769...55F}
{Ford} G.~P.,  et~al., 2013, \mn@doi [\apj] {10.1088/0004-637X/769/1/55}, \href
  {https://ui.adsabs.harvard.edu/abs/2013ApJ...769...55F} {769, 55}

\bibitem[\protect\citeauthoryear{{Fragione}, {Antonini}  \&
  {Gnedin}}{{Fragione} et~al.}{2018}]{F+18MW}
{Fragione} G.,  {Antonini} F.,   {Gnedin} O.~Y.,  2018, \mn@doi [mnras]
  {10.1093/mnras/sty183}, \href
  {https://ui.adsabs.harvard.edu/\#abs/2018MNRAS.475.5313F} {475, 5313}

\bibitem[\protect\citeauthoryear{{Fragione}, {Antonini}  \&
  {Gnedin}}{{Fragione} et~al.}{2019}]{F+19M31}
{Fragione} G.,  {Antonini} F.,   {Gnedin} O.~Y.,  2019, \mn@doi [\apjl]
  {10.3847/2041-8213/aafc62}, \href
  {https://ui.adsabs.harvard.edu/abs/2019ApJ...871L...8F} {871, L8}

\bibitem[\protect\citeauthoryear{{Freire}, {Kramer}, {Lyne}, {Camilo},
  {Manchester}  \& {D'Amico}}{{Freire} et~al.}{2001}]{2001ApJ...557L.105F}
{Freire} P.~C.,  {Kramer} M.,  {Lyne} A.~G.,  {Camilo} F.,  {Manchester} R.~N.,
    {D'Amico} N.,  2001, \mn@doi [\apj] {10.1086/323248}, \href
  {https://ui.adsabs.harvard.edu/\#abs/2001ApJ...557L.105F} {557, L105}

\bibitem[\protect\citeauthoryear{{Galleti}, {Federici}, {Bellazzini}, {Fusi
  Pecci}, {Macrina}  \& {Buzzoni}}{{Galleti}
  et~al.}{2014}]{2014yCat.5143....0G}
{Galleti} S.,  {Federici} L.,  {Bellazzini} M.,  {Fusi Pecci} F.,  {Macrina}
  S.,   {Buzzoni} A.,  2014, VizieR Online Data Catalog, \href
  {https://ui.adsabs.harvard.edu/abs/2014yCat.5143....0G} {p. V/143}

\bibitem[\protect\citeauthoryear{{Genzel} et~al.,}{{Genzel}
  et~al.}{2015}]{2015ApJ...800...20G}
{Genzel} R.,  et~al., 2015, \mn@doi [apj] {10.1088/0004-637X/800/1/20}, \href
  {https://ui.adsabs.harvard.edu/\#abs/2015ApJ...800...20G} {800, 20}

\bibitem[\protect\citeauthoryear{{Georgiev} \& {B{\"o}ker}}{{Georgiev} \&
  {B{\"o}ker}}{2014}]{2014MNRAS.441.3570G}
{Georgiev} I.~Y.,  {B{\"o}ker} T.,  2014, \mn@doi [\mnras]
  {10.1093/mnras/stu797}, \href
  {https://ui.adsabs.harvard.edu/abs/2014MNRAS.441.3570G} {441, 3570}

\bibitem[\protect\citeauthoryear{{Georgiev}, {Puzia}, {Goudfrooij}  \&
  {Hilker}}{{Georgiev} et~al.}{2010}]{2010MNRAS.406.1967G}
{Georgiev} I.~Y.,  {Puzia} T.~H.,  {Goudfrooij} P.,   {Hilker} M.,  2010,
  \mn@doi [\mnras] {10.1111/j.1365-2966.2010.16802.x}, \href
  {https://ui.adsabs.harvard.edu/abs/2010MNRAS.406.1967G} {406, 1967}

\bibitem[\protect\citeauthoryear{{Georgiev}, {B{\"o}ker}, {Leigh},
  {L{\"u}tzgendorf}  \& {Neumayer}}{{Georgiev}
  et~al.}{2016}]{2016MNRAS.457.2122G}
{Georgiev} I.~Y.,  {B{\"o}ker} T.,  {Leigh} N.,  {L{\"u}tzgendorf} N.,
  {Neumayer} N.,  2016, \mn@doi [\mnras] {10.1093/mnras/stw093}, \href
  {https://ui.adsabs.harvard.edu/abs/2016MNRAS.457.2122G} {457, 2122}

\bibitem[\protect\citeauthoryear{{Gieles} \& {Baumgardt}}{{Gieles} \&
  {Baumgardt}}{2008}]{2008MNRAS.389L..28G}
{Gieles} M.,  {Baumgardt} H.,  2008, \mn@doi [\mnras]
  {10.1111/j.1745-3933.2008.00515.x}, \href
  {https://ui.adsabs.harvard.edu/abs/2008MNRAS.389L..28G} {389, L28}

\bibitem[\protect\citeauthoryear{{Gieles}, {Larsen}, {Scheepmaker}, {Bastian},
  {Haas}  \& {Lamers}}{{Gieles} et~al.}{2006}]{2006A&A...446L...9G}
{Gieles} M.,  {Larsen} S.~S.,  {Scheepmaker} R.~A.,  {Bastian} N.,  {Haas}
  M.~R.,   {Lamers} H.~J.~G.~L.~M.,  2006, \mn@doi [\aap]
  {10.1051/0004-6361:200500224}, \href
  {https://ui.adsabs.harvard.edu/\#abs/2006A&A...446L...9G} {446, L9}

\bibitem[\protect\citeauthoryear{{Gnedin}, {Ostriker}  \& {Tremaine}}{{Gnedin}
  et~al.}{2014}]{G+14}
{Gnedin} O.~Y.,  {Ostriker} J.~P.,   {Tremaine} S.,  2014, \mn@doi [\apj]
  {10.1088/0004-637X/785/1/71}, \href
  {https://ui.adsabs.harvard.edu/\#abs/2014ApJ...785...71G} {785, 71}

\bibitem[\protect\citeauthoryear{Gordon \& Mac\'{\i}as}{Gordon \&
  Mac\'{\i}as}{2013}]{G+13}
Gordon C.,  Mac\'{\i}as O.,  2013, \mn@doi [Phys. Rev. D]
  {10.1103/PhysRevD.88.083521}, 88, 083521

\bibitem[\protect\citeauthoryear{{Guillard}, {Emsellem}  \&
  {Renaud}}{{Guillard} et~al.}{2016}]{2016MNRAS.461.3620G}
{Guillard} N.,  {Emsellem} E.,   {Renaud} F.,  2016, \mn@doi [\mnras]
  {10.1093/mnras/stw1570}, \href
  {https://ui.adsabs.harvard.edu/abs/2016MNRAS.461.3620G} {461, 3620}

\bibitem[\protect\citeauthoryear{{Haggard}, {Heinke}, {Hooper}  \&
  {Linden}}{{Haggard} et~al.}{2017}]{2017JCAP...05..056H}
{Haggard} D.,  {Heinke} C.,  {Hooper} D.,   {Linden} T.,  2017, \mn@doi [\jcap]
  {10.1088/1475-7516/2017/05/056}, \href
  {https://ui.adsabs.harvard.edu/abs/2017JCAP...05..056H} {2017, 056}

\bibitem[\protect\citeauthoryear{{Harris}, {Harris}  \& {Hudson}}{{Harris}
  et~al.}{2015}]{2015ApJ...806...36H}
{Harris} W.~E.,  {Harris} G.~L.,   {Hudson} M.~J.,  2015, \mn@doi [\apj]
  {10.1088/0004-637X/806/1/36}, \href
  {https://ui.adsabs.harvard.edu/abs/2015ApJ...806...36H} {806, 36}

\bibitem[\protect\citeauthoryear{{Harris}, {Blakeslee}  \& {Harris}}{{Harris}
  et~al.}{2017}]{2017ApJ...836...67H}
{Harris} W.~E.,  {Blakeslee} J.~P.,   {Harris} G. L.~H.,  2017, \mn@doi [\apj]
  {10.3847/1538-4357/836/1/67}, \href
  {https://ui.adsabs.harvard.edu/abs/2017ApJ...836...67H} {836, 67}

\bibitem[\protect\citeauthoryear{{Hartmann}, {Debattista}, {Seth}, {Cappellari}
   \& {Quinn}}{{Hartmann} et~al.}{2011}]{2011MNRAS.418.2697H}
{Hartmann} M.,  {Debattista} V.~P.,  {Seth} A.,  {Cappellari} M.,   {Quinn}
  T.~R.,  2011, \mn@doi [\mnras] {10.1111/j.1365-2966.2011.19659.x}, \href
  {https://ui.adsabs.harvard.edu/abs/2011MNRAS.418.2697H} {418, 2697}

\bibitem[\protect\citeauthoryear{{Hinshaw} et~al.,}{{Hinshaw}
  et~al.}{2013}]{2013ApJS..208...19H}
{Hinshaw} G.,  et~al., 2013, \mn@doi [The Astrophysical Journal Supplement
  Series] {10.1088/0067-0049/208/2/19}, \href
  {https://ui.adsabs.harvard.edu/\#abs/2013ApJS..208...19H} {208, 19}

\bibitem[\protect\citeauthoryear{{Hooper} \& {Slatyer}}{{Hooper} \&
  {Slatyer}}{2013}]{H+13}
{Hooper} D.,  {Slatyer} T.~R.,  2013, \mn@doi [Physics of the Dark Universe]
  {10.1016/j.dark.2013.06.003}, \href
  {https://ui.adsabs.harvard.edu/abs/2013PDU.....2..118H} {2, 118}

\bibitem[\protect\citeauthoryear{{Hopkins} \& {Quataert}}{{Hopkins} \&
  {Quataert}}{2010}]{2010MNRAS.405L..41H}
{Hopkins} P.~F.,  {Quataert} E.,  2010, \mn@doi [\mnras]
  {10.1111/j.1745-3933.2010.00855.x}, \href
  {https://ui.adsabs.harvard.edu/abs/2010MNRAS.405L..41H} {405, L41}

\bibitem[\protect\citeauthoryear{{Hudson}, {Harris}  \& {Harris}}{{Hudson}
  et~al.}{2014}]{2014ApJ...787L...5H}
{Hudson} M.~J.,  {Harris} G.~L.,   {Harris} W.~E.,  2014, \mn@doi [\apjl]
  {10.1088/2041-8205/787/1/L5}, \href
  {https://ui.adsabs.harvard.edu/abs/2014ApJ...787L...5H} {787, L5}

\bibitem[\protect\citeauthoryear{{Hughes} et~al.,}{{Hughes}
  et~al.}{2005}]{2005AJ....130...73H}
{Hughes} M.~A.,  et~al., 2005, \mn@doi [\aj] {10.1086/430531}, \href
  {https://ui.adsabs.harvard.edu/abs/2005AJ....130...73H} {130, 73}

\bibitem[\protect\citeauthoryear{{Hurley}, {Pols}  \& {Tout}}{{Hurley}
  et~al.}{2000}]{2000MNRAS.315..543H}
{Hurley} J.~R.,  {Pols} O.~R.,   {Tout} C.~A.,  2000, \mn@doi [mnras]
  {10.1046/j.1365-8711.2000.03426.x}, \href
  {https://ui.adsabs.harvard.edu/\#abs/2000MNRAS.315..543H} {315, 543}

\bibitem[\protect\citeauthoryear{{Jiang}, {Jing}, {Faltenbacher}, {Lin}  \&
  {Li}}{{Jiang} et~al.}{2008}]{2008ApJ...675.1095J}
{Jiang} C.~Y.,  {Jing} Y.~P.,  {Faltenbacher} A.,  {Lin} W.~P.,   {Li} C.,
  2008, \mn@doi [\apj] {10.1086/526412}, \href
  {https://ui.adsabs.harvard.edu/\#abs/2008ApJ...675.1095J} {675, 1095}

\bibitem[\protect\citeauthoryear{{Johnson} et~al.,}{{Johnson}
  et~al.}{2017}]{2017ApJ...839...78J}
{Johnson} L.~C.,  et~al., 2017, \mn@doi [\apj] {10.3847/1538-4357/aa6a1f},
  \href {https://ui.adsabs.harvard.edu/\#abs/2017ApJ...839...78J} {839, 78}

\bibitem[\protect\citeauthoryear{{Kacharov}, {Neumayer}, {Seth}, {Cappellari},
  {McDermid}, {Walcher}  \& {B{\"o}ker}}{{Kacharov}
  et~al.}{2018}]{2018MNRAS.480.1973K}
{Kacharov} N.,  {Neumayer} N.,  {Seth} A.~C.,  {Cappellari} M.,  {McDermid} R.,
   {Walcher} C.~J.,   {B{\"o}ker} T.,  2018, \mn@doi [\mnras]
  {10.1093/mnras/sty1985}, \href
  {https://ui.adsabs.harvard.edu/abs/2018MNRAS.480.1973K} {480, 1973}

\bibitem[\protect\citeauthoryear{{Kafle}, {Sharma}, {Lewis}, {Robotham}  \&
  {Driver}}{{Kafle} et~al.}{2018}]{2018MNRAS.475.4043K}
{Kafle} P.~R.,  {Sharma} S.,  {Lewis} G.~F.,  {Robotham} A. S.~G.,   {Driver}
  S.~P.,  2018, \mn@doi [\mnras] {10.1093/mnras/sty082}, \href
  {https://ui.adsabs.harvard.edu/abs/2018MNRAS.475.4043K} {475, 4043}

\bibitem[\protect\citeauthoryear{{Kim}, {Sung}, {Park}  \& {Sung}}{{Kim}
  et~al.}{2004}]{2004ChJAA...4..299K}
{Kim} S.~C.,  {Sung} H.,  {Park} H.~S.,   {Sung} E.-C.,  2004, \mn@doi [\cjaa]
  {10.1088/1009-9271/4/4/299}, \href
  {https://ui.adsabs.harvard.edu/abs/2004ChJAA...4..299K} {4, 299}

\bibitem[\protect\citeauthoryear{{Kormendy}}{{Kormendy}}{1985}]{1985ApJ...292L...9K}
{Kormendy} J.,  1985, \mn@doi [\apjl] {10.1086/184463}, \href
  {https://ui.adsabs.harvard.edu/abs/1985ApJ...292L...9K} {292, L9}

\bibitem[\protect\citeauthoryear{{Kormendy}}{{Kormendy}}{2013}]{2013seg..book....1K}
{Kormendy} J.,  2013, in {Falc{\'o}n-Barroso} J.,  {Knapen} J.~H.,  eds, ,
  Secular Evolution of Galaxies.
p.~1

\bibitem[\protect\citeauthoryear{{Kroupa}}{{Kroupa}}{2001}]{2001MNRAS.322..231K}
{Kroupa} P.,  2001, \mn@doi [\mnras] {10.1046/j.1365-8711.2001.04022.x}, \href
  {https://ui.adsabs.harvard.edu/\#abs/2001MNRAS.322..231K} {322, 231}

\bibitem[\protect\citeauthoryear{{Kruijssen}, {Pelupessy}, {Lamers}, {Portegies
  Zwart}  \& {Icke}}{{Kruijssen} et~al.}{2011}]{2011MNRAS.414.1339K}
{Kruijssen} J.~M.~D.,  {Pelupessy} F.~I.,  {Lamers} H. J.~G.~L.~M.,  {Portegies
  Zwart} S.~F.,   {Icke} V.,  2011, \mn@doi [\mnras]
  {10.1111/j.1365-2966.2011.18467.x}, \href
  {https://ui.adsabs.harvard.edu/abs/2011MNRAS.414.1339K} {414, 1339}

\bibitem[\protect\citeauthoryear{Lacroix, Macias, Gordon, Panci, B\oe{}hm  \&
  Silk}{Lacroix et~al.}{2016}]{L+16}
Lacroix T.,  Macias O.,  Gordon C.,  Panci P.,  B\oe{}hm C.,   Silk J.,  2016,
  \mn@doi [Phys. Rev. D] {10.1103/PhysRevD.93.103004}, 93, 103004

\bibitem[\protect\citeauthoryear{{Larsen}}{{Larsen}}{2009}]{2009A&A...494..539L}
{Larsen} S.~S.,  2009, \mn@doi [aap] {10.1051/0004-6361:200811212}, \href
  {https://ui.adsabs.harvard.edu/\#abs/2009A&A...494..539L} {494, 539}

\bibitem[\protect\citeauthoryear{{Leveque}, {Giersz}, {Arca-Sedda}  \&
  {Askar}}{{Leveque} et~al.}{2022}]{2022MNRAS.514.5751L}
{Leveque} A.,  {Giersz} M.,  {Arca-Sedda} M.,   {Askar} A.,  2022, \mn@doi
  [\mnras] {10.1093/mnras/stac1694}, \href
  {https://ui.adsabs.harvard.edu/abs/2022MNRAS.514.5751L} {514, 5751}

\bibitem[\protect\citeauthoryear{{Li} \& {Gnedin}}{{Li} \&
  {Gnedin}}{2014}]{LG14}
{Li} H.,  {Gnedin} O.~Y.,  2014, \mn@doi [\apj] {10.1088/0004-637X/796/1/10},
  \href {https://ui.adsabs.harvard.edu/\#abs/2014ApJ...796...10L} {796, 10}

\bibitem[\protect\citeauthoryear{{Li}, {Gnedin}, {Gnedin}, {Meng}, {Semenov}
  \& {Kravtsov}}{{Li} et~al.}{2017}]{2017ApJ...834...69L}
{Li} H.,  {Gnedin} O.~Y.,  {Gnedin} N.~Y.,  {Meng} X.,  {Semenov} V.~A.,
  {Kravtsov} A.~V.,  2017, \mn@doi [\apj] {10.3847/1538-4357/834/1/69}, \href
  {https://ui.adsabs.harvard.edu/abs/2017ApJ...834...69L} {834, 69}

\bibitem[\protect\citeauthoryear{{Li}, {Gnedin}  \& {Gnedin}}{{Li}
  et~al.}{2018}]{2018ApJ...861..107L}
{Li} H.,  {Gnedin} O.~Y.,   {Gnedin} N.~Y.,  2018, \mn@doi [apj]
  {10.3847/1538-4357/aac9b8}, \href
  {https://ui.adsabs.harvard.edu/\#abs/2018ApJ...861..107L} {861, 107}

\bibitem[\protect\citeauthoryear{{Light}, {Danielson}  \&
  {Schwarzschild}}{{Light} et~al.}{1974}]{1974ApJ...194..257L}
{Light} E.~S.,  {Danielson} R.~E.,   {Schwarzschild} M.,  1974, \mn@doi [\apj]
  {10.1086/153241}, \href
  {https://ui.adsabs.harvard.edu/abs/1974ApJ...194..257L} {194, 257}

\bibitem[\protect\citeauthoryear{{Lilly}, {Carollo}, {Pipino}, {Renzini}  \&
  {Peng}}{{Lilly} et~al.}{2013}]{2013ApJ...772..119L}
{Lilly} S.~J.,  {Carollo} C.~M.,  {Pipino} A.,  {Renzini} A.,   {Peng} Y.,
  2013, \mn@doi [apj] {10.1088/0004-637X/772/2/119}, \href
  {https://ui.adsabs.harvard.edu/\#abs/2013ApJ...772..119L} {772, 119}

\bibitem[\protect\citeauthoryear{{Loose}, {Kruegel}  \& {Tutukov}}{{Loose}
  et~al.}{1982}]{1982A&A...105..342L}
{Loose} H.~H.,  {Kruegel} E.,   {Tutukov} A.,  1982, \aap, \href
  {https://ui.adsabs.harvard.edu/abs/1982A&A...105..342L} {105, 342}

\bibitem[\protect\citeauthoryear{{Lotz}, {Telford}, {Ferguson}, {Miller},
  {Stiavelli}  \& {Mack}}{{Lotz} et~al.}{2001}]{2001ApJ...552..572L}
{Lotz} J.~M.,  {Telford} R.,  {Ferguson} H.~C.,  {Miller} B.~W.,  {Stiavelli}
  M.,   {Mack} J.,  2001, \mn@doi [\apj] {10.1086/320545}, \href
  {https://ui.adsabs.harvard.edu/abs/2001ApJ...552..572L} {552, 572}

\bibitem[\protect\citeauthoryear{{Lyubenova} \& {Tsatsi}}{{Lyubenova} \&
  {Tsatsi}}{2019}]{2019A&A...629A..44L}
{Lyubenova} M.,  {Tsatsi} A.,  2019, \mn@doi [\aap]
  {10.1051/0004-6361/201833954}, \href
  {https://ui.adsabs.harvard.edu/abs/2019A&A...629A..44L} {629, A44}

\bibitem[\protect\citeauthoryear{{Macci{\`o}}, {Dutton}  \& {van den
  Bosch}}{{Macci{\`o}} et~al.}{2008}]{2008MNRAS.391.1940M}
{Macci{\`o}} A.~V.,  {Dutton} A.~A.,   {van den Bosch} F.~C.,  2008, \mn@doi
  [\mnras] {10.1111/j.1365-2966.2008.14029.x}, \href
  {https://ui.adsabs.harvard.edu/abs/2008MNRAS.391.1940M} {391, 1940}

\bibitem[\protect\citeauthoryear{{Magdis} et~al.,}{{Magdis}
  et~al.}{2012}]{2012ApJ...758L...9M}
{Magdis} G.~E.,  et~al., 2012, \mn@doi [\apj] {10.1088/2041-8205/758/1/L9},
  \href {https://ui.adsabs.harvard.edu/\#abs/2012ApJ...758L...9M} {758, L9}

\bibitem[\protect\citeauthoryear{{M{\'a}rquez}, {Lima Neto}, {Capelato},
  {Durret}  \& {Gerbal}}{{M{\'a}rquez} et~al.}{2000}]{2000A&A...353..873M}
{M{\'a}rquez} I.,  {Lima Neto} G.~B.,  {Capelato} H.,  {Durret} F.,   {Gerbal}
  D.,  2000, \mn@doi [\aap] {10.48550/arXiv.astro-ph/9911464}, \href
  {https://ui.adsabs.harvard.edu/abs/2000A&A...353..873M} {353, 873}

\bibitem[\protect\citeauthoryear{{Massari}, {Koppelman}  \& {Helmi}}{{Massari}
  et~al.}{2019}]{2019A&A...630L...4M}
{Massari} D.,  {Koppelman} H.~H.,   {Helmi} A.,  2019, \mn@doi [\aap]
  {10.1051/0004-6361/201936135}, \href
  {https://ui.adsabs.harvard.edu/abs/2019A&A...630L...4M} {630, L4}

\bibitem[\protect\citeauthoryear{{Matthews} \& {Gallagher}}{{Matthews} \&
  {Gallagher}}{1997}]{1997AJ....114.1899M}
{Matthews} L.~D.,  {Gallagher} John~S. I.,  1997, \mn@doi [\aj]
  {10.1086/118613}, \href
  {https://ui.adsabs.harvard.edu/abs/1997AJ....114.1899M} {114, 1899}

\bibitem[\protect\citeauthoryear{{McDaniel}, {Jeltema}  \&
  {Profumo}}{{McDaniel} et~al.}{2018}]{2018PhRvD..97j3021M}
{McDaniel} A.,  {Jeltema} T.,   {Profumo} S.,  2018, \mn@doi [prd]
  {10.1103/PhysRevD.97.103021}, \href
  {https://ui.adsabs.harvard.edu/\#abs/2018PhRvD..97j3021M} {97, 103021}

\bibitem[\protect\citeauthoryear{{McLaughlin}, {King}  \&
  {Nayakshin}}{{McLaughlin} et~al.}{2006}]{2006ApJ...650L..37M}
{McLaughlin} D.~E.,  {King} A.~R.,   {Nayakshin} S.,  2006, \mn@doi [\apjl]
  {10.1086/508627}, \href
  {https://ui.adsabs.harvard.edu/abs/2006ApJ...650L..37M} {650, L37}

\bibitem[\protect\citeauthoryear{{McMillan}}{{McMillan}}{2017}]{2017MNRAS.465...76M}
{McMillan} P.~J.,  2017, \mn@doi [\mnras] {10.1093/mnras/stw2759}, \href
  {https://ui.adsabs.harvard.edu/abs/2017MNRAS.465...76M} {465, 76}

\bibitem[\protect\citeauthoryear{{Miholics}, {Kruijssen}  \&
  {Sills}}{{Miholics} et~al.}{2017}]{2017MNRAS.470.1421M}
{Miholics} M.,  {Kruijssen} J.~M.~D.,   {Sills} A.,  2017, \mn@doi [\mnras]
  {10.1093/mnras/stx1312}, \href
  {https://ui.adsabs.harvard.edu/abs/2017MNRAS.470.1421M} {470, 1421}

\bibitem[\protect\citeauthoryear{{Mihos} \& {Hernquist}}{{Mihos} \&
  {Hernquist}}{1994}]{1994ApJ...437L..47M}
{Mihos} J.~C.,  {Hernquist} L.,  1994, \mn@doi [\apjl] {10.1086/187679}, \href
  {https://ui.adsabs.harvard.edu/abs/1994ApJ...437L..47M} {437, L47}

\bibitem[\protect\citeauthoryear{{Milosavljevi{\'c}}}{{Milosavljevi{\'c}}}{2004}]{2004ApJ...605L..13M}
{Milosavljevi{\'c}} M.,  2004, \mn@doi [\apjl] {10.1086/420696}, \href
  {https://ui.adsabs.harvard.edu/abs/2004ApJ...605L..13M} {605, L13}

\bibitem[\protect\citeauthoryear{{Mo}, {Mao}  \& {White}}{{Mo}
  et~al.}{1998}]{1998MNRAS.295..319M}
{Mo} H.~J.,  {Mao} S.,   {White} S. D.~M.,  1998, \mn@doi [\mnras]
  {10.1046/j.1365-8711.1998.01227.x}, \href
  {https://ui.adsabs.harvard.edu/\#abs/1998MNRAS.295..319M} {295, 319}

\bibitem[\protect\citeauthoryear{{Mo}, {van den Bosch}  \& {White}}{{Mo}
  et~al.}{2010}]{2010gfe..book.....M}
{Mo} H.,  {van den Bosch} F.~C.,   {White} S.,  2010, {Galaxy Formation and
  Evolution}.
Cambridge University Press

\bibitem[\protect\citeauthoryear{{Muratov} \& {Gnedin}}{{Muratov} \&
  {Gnedin}}{2010}]{2010ApJ...718.1266M}
{Muratov} A.~L.,  {Gnedin} O.~Y.,  2010, \mn@doi [\apj]
  {10.1088/0004-637X/718/2/1266}, \href
  {https://ui.adsabs.harvard.edu/\#abs/2010ApJ...718.1266M} {718, 1266}

\bibitem[\protect\citeauthoryear{{Navarro}, {Frenk}  \& {White}}{{Navarro}
  et~al.}{1997}]{NFW}
{Navarro} J.~F.,  {Frenk} C.~S.,   {White} S.~D.~M.,  1997, \mn@doi [Astrophys.
  J.] {10.1086/304888}, \href
  {http://adsabs.harvard.edu/abs/1997ApJ...490..493N} {490, 493}

\bibitem[\protect\citeauthoryear{{Nelson} et~al.,}{{Nelson}
  et~al.}{2015}]{2015A&C....13...12N}
{Nelson} D.,  et~al., 2015, \mn@doi [Astronomy and Computing]
  {10.1016/j.ascom.2015.09.003}, \href
  {https://ui.adsabs.harvard.edu/abs/2015A&C....13...12N} {13, 12}

\bibitem[\protect\citeauthoryear{{Neumayer}, {Seth}  \& {B{\"o}ker}}{{Neumayer}
  et~al.}{2020}]{2020A&ARv..28....4N}
{Neumayer} N.,  {Seth} A.,   {B{\"o}ker} T.,  2020, \mn@doi [\aapr]
  {10.1007/s00159-020-00125-0}, \href
  {https://ui.adsabs.harvard.edu/abs/2020A&ARv..28....4N} {28, 4}

\bibitem[\protect\citeauthoryear{{Nguyen} et~al.,}{{Nguyen}
  et~al.}{2019}]{2019ApJ...872..104N}
{Nguyen} D.~D.,  et~al., 2019, \mn@doi [\apj] {10.3847/1538-4357/aafe7a}, \href
  {https://ui.adsabs.harvard.edu/abs/2019ApJ...872..104N} {872, 104}

\bibitem[\protect\citeauthoryear{{Noyola}, {Gebhardt}  \& {Bergmann}}{{Noyola}
  et~al.}{2008}]{2008ApJ...676.1008N}
{Noyola} E.,  {Gebhardt} K.,   {Bergmann} M.,  2008, \mn@doi [\apj]
  {10.1086/529002}, \href
  {https://ui.adsabs.harvard.edu/abs/2008ApJ...676.1008N} {676, 1008}

\bibitem[\protect\citeauthoryear{{Oh} \& {Lin}}{{Oh} \&
  {Lin}}{2000}]{2000ApJ...543..620O}
{Oh} K.~S.,  {Lin} D.~N.~C.,  2000, \mn@doi [\apj] {10.1086/317118}, \href
  {https://ui.adsabs.harvard.edu/abs/2000ApJ...543..620O} {543, 620}

\bibitem[\protect\citeauthoryear{{Peebles}}{{Peebles}}{1980}]{1980lssu.book.....P}
{Peebles} P.~J.~E.,  1980, {The large-scale structure of the universe}.
Princeton University Press

\bibitem[\protect\citeauthoryear{{Peng}}{{Peng}}{2002}]{2002AJ....124..294P}
{Peng} C.~Y.,  2002, \mn@doi [\aj] {10.1086/340958}, \href
  {https://ui.adsabs.harvard.edu/abs/2002AJ....124..294P} {124, 294}

\bibitem[\protect\citeauthoryear{{Pfeffer}, {Kruijssen}, {Crain}  \&
  {Bastian}}{{Pfeffer} et~al.}{2018}]{2018MNRAS.475.4309P}
{Pfeffer} J.,  {Kruijssen} J.~M.~D.,  {Crain} R.~A.,   {Bastian} N.,  2018,
  \mn@doi [\mnras] {10.1093/mnras/stx3124}, \href
  {https://ui.adsabs.harvard.edu/abs/2018MNRAS.475.4309P} {475, 4309}

\bibitem[\protect\citeauthoryear{{Portegies Zwart}, {McMillan}  \&
  {Gieles}}{{Portegies Zwart} et~al.}{2010}]{2010ARA&A..48..431P}
{Portegies Zwart} S.~F.,  {McMillan} S. L.~W.,   {Gieles} M.,  2010, \mn@doi
  [Annual Review of Astronomy and Astrophysics]
  {10.1146/annurev-astro-081309-130834}, \href
  {https://ui.adsabs.harvard.edu/\#abs/2010ARA&A..48..431P} {48, 431}

\bibitem[\protect\citeauthoryear{{Prager}, {Ransom}, {Freire}, {Hessels},
  {Stairs}, {Arras}  \& {Cadelano}}{{Prager}
  et~al.}{2017}]{2017ApJ...845..148P}
{Prager} B.~J.,  {Ransom} S.~M.,  {Freire} P. C.~C.,  {Hessels} J. W.~T.,
  {Stairs} I.~H.,  {Arras} P.,   {Cadelano} M.,  2017, \mn@doi [\apj]
  {10.3847/1538-4357/aa7ed7}, \href
  {https://ui.adsabs.harvard.edu/\#abs/2017ApJ...845..148P} {845, 148}

\bibitem[\protect\citeauthoryear{{Prugniel} \& {Simien}}{{Prugniel} \&
  {Simien}}{1997}]{1997A&A...321..111P}
{Prugniel} P.,  {Simien} F.,  1997, \aap, \href
  {https://ui.adsabs.harvard.edu/abs/1997A&A...321..111P} {321, 111}

\bibitem[\protect\citeauthoryear{{Ransom}}{{Ransom}}{2008}]{2008IAUS..246..291R}
{Ransom} S.~M.,  2008, in {Vesperini} E.,  {Giersz} M.,   {Sills} A.,  eds, ~
  Vol. 246, Dynamical Evolution of Dense Stellar Systems. pp 291--300,
  \mn@doi{10.1017/S1743921308015810}

\bibitem[\protect\citeauthoryear{{Reina-Campos}, {Sills}  \&
  {Bichon}}{{Reina-Campos} et~al.}{2023}]{2023MNRAS.524..968R}
{Reina-Campos} M.,  {Sills} A.,   {Bichon} G.,  2023, \mn@doi [\mnras]
  {10.1093/mnras/stad1879}, \href
  {https://ui.adsabs.harvard.edu/abs/2023MNRAS.524..968R} {524, 968}

\bibitem[\protect\citeauthoryear{{S{\'a}nchez-Janssen}
  et~al.,}{{S{\'a}nchez-Janssen} et~al.}{2019}]{2019ApJ...878...18S}
{S{\'a}nchez-Janssen} R.,  et~al., 2019, \mn@doi [\apj]
  {10.3847/1538-4357/aaf4fd}, \href
  {https://ui.adsabs.harvard.edu/abs/2019ApJ...878...18S} {878, 18}

\bibitem[\protect\citeauthoryear{{Schechter}}{{Schechter}}{1976}]{1976ApJ...203..297S}
{Schechter} P.,  1976, \mn@doi [\apj] {10.1086/154079}, \href
  {https://ui.adsabs.harvard.edu/abs/1976ApJ...203..297S} {203, 297}

\bibitem[\protect\citeauthoryear{{Sch{\"o}del}, {Gallego-Cano}, {Dong},
  {Nogueras-Lara}, {Gallego-Calvente}, {Amaro-Seoane}  \&
  {Baumgardt}}{{Sch{\"o}del} et~al.}{2018}]{S+18}
{Sch{\"o}del} R.,  {Gallego-Cano} E.,  {Dong} H.,  {Nogueras-Lara} F.,
  {Gallego-Calvente} A.~T.,  {Amaro-Seoane} P.,   {Baumgardt} H.,  2018,
  \mn@doi [\aap] {10.1051/0004-6361/201730452}, \href
  {https://ui.adsabs.harvard.edu/abs/2018A&A...609A..27S} {609, A27}

\bibitem[\protect\citeauthoryear{{Schulz}, {Pflamm-Altenburg}  \&
  {Kroupa}}{{Schulz} et~al.}{2015}]{2015A&A...582A..93S}
{Schulz} C.,  {Pflamm-Altenburg} J.,   {Kroupa} P.,  2015, \mn@doi [\aap]
  {10.1051/0004-6361/201425296}, \href
  {https://ui.adsabs.harvard.edu/abs/2015A&A...582A..93S} {582, A93}

\bibitem[\protect\citeauthoryear{{S{\'e}rsic}}{{S{\'e}rsic}}{1963}]{1963BAAA....6...41S}
{S{\'e}rsic} J.~L.,  1963, Boletin de la Asociacion Argentina de Astronomia La
  Plata Argentina, \href
  {https://ui.adsabs.harvard.edu/abs/1963BAAA....6...41S} {6, 41}

\bibitem[\protect\citeauthoryear{{Shlosman}, {Begelman}  \& {Frank}}{{Shlosman}
  et~al.}{1990}]{1990Natur.345..679S}
{Shlosman} I.,  {Begelman} M.~C.,   {Frank} J.,  1990, \mn@doi [\nat]
  {10.1038/345679a0}, \href
  {https://ui.adsabs.harvard.edu/abs/1990Natur.345..679S} {345, 679}

\bibitem[\protect\citeauthoryear{{Spengler} et~al.,}{{Spengler}
  et~al.}{2018}]{2018ApJ...869...85S}
{Spengler} C.,  et~al., 2018, \mn@doi [\apj] {10.3847/1538-4357/aaf4bf}, \href
  {https://ui.adsabs.harvard.edu/abs/2018ApJ...869...85S} {869, 85}

\bibitem[\protect\citeauthoryear{{Spitler} \& {Forbes}}{{Spitler} \&
  {Forbes}}{2009}]{2009MNRAS.392L...1S}
{Spitler} L.~R.,  {Forbes} D.~A.,  2009, \mn@doi [\mnras]
  {10.1111/j.1745-3933.2008.00567.x}, \href
  {https://ui.adsabs.harvard.edu/abs/2009MNRAS.392L...1S} {392, L1}

\bibitem[\protect\citeauthoryear{{Tacconi} et~al.,}{{Tacconi}
  et~al.}{2013}]{2013ApJ...768...74T}
{Tacconi} L.~J.,  et~al., 2013, \mn@doi [\apj] {10.1088/0004-637X/768/1/74},
  \href {https://ui.adsabs.harvard.edu/\#abs/2013ApJ...768...74T} {768, 74}

\bibitem[\protect\citeauthoryear{{Tacconi} et~al.,}{{Tacconi}
  et~al.}{2018}]{2018ApJ...853..179T}
{Tacconi} L.~J.,  et~al., 2018, \mn@doi [apj] {10.3847/1538-4357/aaa4b4}, \href
  {https://ui.adsabs.harvard.edu/\#abs/2018ApJ...853..179T} {853, 179}

\bibitem[\protect\citeauthoryear{{Terzi{\'c}} \& {Graham}}{{Terzi{\'c}} \&
  {Graham}}{2005}]{2005MNRAS.362..197T}
{Terzi{\'c}} B.,  {Graham} A.~W.,  2005, \mn@doi [\mnras]
  {10.1111/j.1365-2966.2005.09269.x}, \href
  {https://ui.adsabs.harvard.edu/abs/2005MNRAS.362..197T} {362, 197}

\bibitem[\protect\citeauthoryear{{Tremaine}, {Ostriker}  \&
  {Spitzer}}{{Tremaine} et~al.}{1975}]{1975ApJ...196..407T}
{Tremaine} S.~D.,  {Ostriker} J.~P.,   {Spitzer} L. J.,  1975, \mn@doi [\apj]
  {10.1086/153422}, \href
  {https://ui.adsabs.harvard.edu/abs/1975ApJ...196..407T} {196, 407}

\bibitem[\protect\citeauthoryear{{Tsatsi}, {Mastrobuono-Battisti}, {van de
  Ven}, {Perets}, {Bianchini}  \& {Neumayer}}{{Tsatsi}
  et~al.}{2017}]{2017MNRAS.464.3720T}
{Tsatsi} A.,  {Mastrobuono-Battisti} A.,  {van de Ven} G.,  {Perets} H.~B.,
  {Bianchini} P.,   {Neumayer} N.,  2017, \mn@doi [\mnras]
  {10.1093/mnras/stw2593}, \href
  {https://ui.adsabs.harvard.edu/abs/2017MNRAS.464.3720T} {464, 3720}

\bibitem[\protect\citeauthoryear{{Vogelsberger} et~al.,}{{Vogelsberger}
  et~al.}{2014}]{2014Natur.509..177V}
{Vogelsberger} M.,  et~al., 2014, \mn@doi [\nat] {10.1038/nature13316}, \href
  {https://ui.adsabs.harvard.edu/abs/2014Natur.509..177V} {509, 177}

\bibitem[\protect\citeauthoryear{{Voss} \& {Gilfanov}}{{Voss} \&
  {Gilfanov}}{2007a}]{2007MNRAS.380.1685V}
{Voss} R.,  {Gilfanov} M.,  2007a, \mn@doi [\mnras]
  {10.1111/j.1365-2966.2007.12223.x}, \href
  {https://ui.adsabs.harvard.edu/abs/2007MNRAS.380.1685V} {380, 1685}

\bibitem[\protect\citeauthoryear{{Voss} \& {Gilfanov}}{{Voss} \&
  {Gilfanov}}{2007b}]{2007A&A...468...49V}
{Voss} R.,  {Gilfanov} M.,  2007b, \mn@doi [\aap] {10.1051/0004-6361:20066614},
  \href {https://ui.adsabs.harvard.edu/abs/2007A&A...468...49V} {468, 49}

\bibitem[\protect\citeauthoryear{{Walcher} et~al.,}{{Walcher}
  et~al.}{2005}]{2005ApJ...618..237W}
{Walcher} C.~J.,  et~al., 2005, \mn@doi [\apj] {10.1086/425977}, \href
  {https://ui.adsabs.harvard.edu/abs/2005ApJ...618..237W} {618, 237}

\bibitem[\protect\citeauthoryear{{Wang} \& {Lin}}{{Wang} \&
  {Lin}}{2023}]{2023ApJ...944..140W}
{Wang} L.,  {Lin} D.~N.~C.,  2023, \mn@doi [\apj] {10.3847/1538-4357/acac97},
  \href {https://ui.adsabs.harvard.edu/abs/2023ApJ...944..140W} {944, 140}

\bibitem[\protect\citeauthoryear{{Wilson}, {Harris}, {Longden}  \&
  {Scoville}}{{Wilson} et~al.}{2006}]{2006ApJ...641..763W}
{Wilson} C.~D.,  {Harris} W.~E.,  {Longden} R.,   {Scoville} N.~Z.,  2006,
  \mn@doi [\apj] {10.1086/500577}, \href
  {https://ui.adsabs.harvard.edu/abs/2006ApJ...641..763W} {641, 763}

\bibitem[\protect\citeauthoryear{{Ye} \& {Fragione}}{{Ye} \&
  {Fragione}}{2022}]{2022ApJ...940..162Y}
{Ye} C.~S.,  {Fragione} G.,  2022, \mn@doi [\apj] {10.3847/1538-4357/ac9cd0},
  \href {https://ui.adsabs.harvard.edu/abs/2022ApJ...940..162Y} {940, 162}

\bibitem[\protect\citeauthoryear{{Ye}, {Kremer}, {Chatterjee}, {Rodriguez}  \&
  {Rasio}}{{Ye} et~al.}{2019}]{2019ApJ...877..122Y}
{Ye} C.~S.,  {Kremer} K.,  {Chatterjee} S.,  {Rodriguez} C.~L.,   {Rasio}
  F.~A.,  2019, \mn@doi [\apj] {10.3847/1538-4357/ab1b21}, \href
  {https://ui.adsabs.harvard.edu/abs/2019ApJ...877..122Y} {877, 122}

\bibitem[\protect\citeauthoryear{{Zimmer}, {Macias}, {Ando}, {Crocker}  \&
  {Horiuchi}}{{Zimmer} et~al.}{2022}]{2022MNRAS.tmp.2330Z}
{Zimmer} F.,  {Macias} O.,  {Ando} S.,  {Crocker} R.~M.,   {Horiuchi} S.,
  2022, \mn@doi [\mnras] {10.1093/mnras/stac2464}, \href
  {https://ui.adsabs.harvard.edu/abs/2022MNRAS.tmp.2330Z} {}

\makeatother
\end{thebibliography}
\bibliographystyle{mnras}

\bsp
\clearpage
\setcounter{table}{0}
\renewcommand{\thetable}{A\arabic{table}}
\renewcommand{\arraystretch}{2}

\begin{sidewaystable}[ht]
  \centering
  \vspace{-60em}
  \caption{Best-ﬁt Results for the Metal-Poor(MP), Metal-Standard (MS), and Metal-Rich (MR) subsamples. }
    \begin{tabular}{@{}cc|rrrr|rrrr|rrrr}
    \hline
    \hline
          &       & \multicolumn{4}{c|}{Metal-Poor} & \multicolumn{4}{c|}{Metal-Standard} & \multicolumn{4}{c}{Metal-Rich} \bigstrut\\
\cline{3-14}    R     & Parameter & \multicolumn{4}{c|}{$\phi(^\circ)$} & \multicolumn{4}{c|}{$\phi(^\circ)$} & \multicolumn{4}{c}{$\phi(^\circ)$} \bigstrut[t]\\
    (kpc) &       & \multicolumn{1}{c}{(-4.0,-1.0)} & \multicolumn{1}{c}{(-1.0,1.0)} & \multicolumn{1}{c}{(1.0,4.0)} & \multicolumn{1}{c|}{(4.0,6.0)} & \multicolumn{1}{c}{(-4.0,-1.0)} & \multicolumn{1}{c}{(-1.0,1.0)} & \multicolumn{1}{c}{(1.0,4.0)} & \multicolumn{1}{c|}{(4.0,6.0)} & \multicolumn{1}{c}{(-4.0,-1.0)} & \multicolumn{1}{c}{(-1.0,1.0)} & \multicolumn{1}{c}{(1.0,4.0)} & \multicolumn{1}{c}{(4.0,6.0)} \bigstrut[b]\\
    \hline
    \multirow{4}[2]{*}{(7.0,8.2)} & ln$\Sigma$(kpc$^{\scriptscriptstyle -2}$) & 
8.55$^{+0.08}_{-0.08}$ & 9.47$^{+0.08}_{-0.09}$ & 7.86$^{+0.06}_{-0.06}$ & 6.98$^{+0.20}_{-0.21}$ & 9.43$^{+0.18}_{-0.19}$ & 9.21$^{+0.65}_{-0.65}$ & 8.10$^{+0.12}_{-0.13}$ & 7.78$^{+0.14}_{-0.13}$ & 8.90$^{+0.71}_{-0.70}$ & 9.72$^{+0.09}_{-0.10}$ & 7.85$^{+0.27}_{-0.27}$ & 7.54$^{+0.06}_{-0.05}$ \bigstrut[t]\\
          & Z$_0$(kpc) & -0.08$^{+0.01}_{-0.02}$ & -0.23$^{+0.02}_{-0.02}$ & 0.03$^{+0.01}_{-0.00}$ & -0.06$^{+0.00}_{-0.01}$ & -0.09$^{+0.01}_{-0.01}$ & -0.04$^{+0.13}_{-0.13}$ & 0.03$^{+0.01}_{-0.02}$ & 0.05$^{+0.01}_{-0.01}$ & -0.04$^{+0.02}_{-0.03}$ & -0.19$^{+0.00}_{-0.01}$ & 0.08$^{+0.01}_{-0.02}$ & 0.04$^{+0.02}_{-0.01}$ \\
          & h$_s$(kpc) & 0.51$^{+0.00}_{-0.01}$ & 0.51$^{+0.01}_{-0.01}$ & 0.19$^{+0.00}_{-0.01}$ & 0.16$^{+0.01}_{-0.01}$ & 0.50$^{+0.01}_{-0.01}$ & 0.50$^{+0.00}_{-0.00}$ & 0.22$^{+0.03}_{-0.03}$ & 0.17$^{+0.01}_{-0.00}$ & 0.51$^{+0.00}_{-0.00}$ & 0.50$^{+0.01}_{-0.01}$ & 0.26$^{+0.03}_{-0.03}$ & 0.16$^{+0.01}_{-0.01}$ \\
          & h$_n$(kpc) & 0.21$^{+0.01}_{-0.00}$ & 0.23$^{+0.00}_{-0.01}$ & 0.22$^{+0.00}_{-0.00}$ & 0.30$^{+0.02}_{-0.02}$ & 0.19$^{+0.00}_{-0.00}$ & 0.21$^{+0.00}_{-0.01}$ & 0.22$^{+0.00}_{-0.01}$ & 0.18$^{+0.01}_{-0.00}$ & 0.19$^{+0.02}_{-0.02}$ & 0.19$^{+0.01}_{-0.00}$ & 0.19$^{+0.01}_{-0.01}$ & 0.16$^{+0.01}_{-0.00}$ \bigstrut[b]\\
    \hline
    \multirow{4}[2]{*}{(8.2,8.6)} & ln$\Sigma$(kpc$^{\scriptscriptstyle -2}$) & 7.91$^{+0.11}_{-0.13}$ & 7.84$^{+0.04}_{-0.05}$ & 7.88$^{+0.03}_{-0.03}$ & 7.25$^{+0.17}_{-0.17}$ & 8.43$^{+0.21}_{-0.22}$ & 8.19$^{+0.06}_{-0.06}$ & 8.17$^{+0.08}_{-0.08}$ & 7.64$^{+0.12}_{-0.11}$ & 8.33$^{+0.14}_{-0.15}$ & 8.04$^{+0.06}_{-0.05}$ & 7.98$^{+0.06}_{-0.06}$ & 7.43$^{+0.04}_{-0.04}$\bigstrut[t]\\
          & Z$_0$(kpc) & 0.04$^{+0.00}_{-0.01}$ & 0.01$^{+0.00}_{-0.01}$ & 0.06$^{+0.00}_{-0.00}$ & 0.09$^{+0.03}_{-0.03}$ & 0.06$^{+0.02}_{-0.03}$ & 0.02$^{+0.00}_{-0.00}$ & 0.05$^{+0.00}_{-0.00}$ & 0.07$^{+0.00}_{-0.00}$ & 0.02$^{+0.03}_{-0.02}$ & 0.03$^{+0.01}_{-0.00}$ & 0.05$^{+0.00}_{-0.01}$ & 0.06$^{+0.01}_{-0.01}$ \\
          & h$_s$(kpc) & 0.36$^{+0.06}_{-0.06}$ & 0.22$^{+0.01}_{-0.02}$ & 0.21$^{+0.00}_{-0.00}$ & 0.20$^{+0.02}_{-0.01}$ & 0.49$^{+0.00}_{-0.00}$ & 0.25$^{+0.01}_{-0.01}$ & 0.21$^{+0.00}_{-0.00}$ & 0.19$^{+0.00}_{-0.00}$ & 0.43$^{+0.06}_{-0.05}$ & 0.19$^{+0.01}_{-0.01}$ & 0.18$^{+0.00}_{-0.00}$ & 0.16$^{+0.00}_{-0.00}$ \\
          & h$_n$(kpc) & 0.22$^{+0.00}_{-0.00}$ & 0.25$^{+0.00}_{-0.00}$ & 0.21$^{+0.00}_{-0.00}$ & 0.18$^{+0.02}_{-0.02}$ & 0.20$^{+0.01}_{-0.00}$ & 0.24$^{+0.00}_{-0.00}$ & 0.21$^{+0.00}_{-0.00}$ & 0.18$^{+0.00}_{-0.00}$ & 0.19$^{+0.01}_{-0.01}$ & 0.19$^{+0.00}_{-0.00}$ & 0.19$^{+0.00}_{-0.01}$ & 0.16$^{+0.01}_{-0.00}$ \bigstrut[b]\\
    \hline
    \multirow{4}[2]{*}{(8.6,9.0)} & ln$\Sigma$(kpc$^{\scriptscriptstyle -2}$) & 6.99$^{+0.13}_{-0.14}$ & 7.29$^{+0.03}_{-0.03}$ & 7.01$^{+0.04}_{-0.04}$ & 7.10$^{+0.90}_{-0.97}$ & 7.60$^{+0.08}_{-0.08}$ & 7.78$^{+0.17}_{-0.17}$ & 7.63$^{+0.15}_{-0.15}$ & 6.87$^{+0.50}_{-0.53}$ & 7.31$^{+0.01}_{-0.02}$ & 7.58$^{+0.06}_{-0.07}$ & 7.38$^{+0.03}_{-0.04}$ & 6.64$^{+0.32}_{-0.35}$   \bigstrut[t]\\
          & Z$_0$(kpc) & 0.01$^{+0.00}_{-0.00}$ & -0.00$^{+0.00}_{-0.01}$ & 0.04$^{+0.02}_{-0.01}$ & 0.14$^{+0.08}_{-0.08}$ & -0.01$^{+0.01}_{-0.00}$ & 0.00$^{+0.01}_{-0.00}$ & 0.04$^{+0.00}_{-0.01}$ & -0.14$^{+0.06}_{-0.07}$ & -0.00$^{+0.01}_{-0.01}$ & -0.01$^{+0.00}_{-0.00}$ & 0.02$^{+0.01}_{-0.00}$ & 0.12$^{+0.08}_{-0.08}$ \\
          & h$_s$(kpc) & 0.16$^{+0.00}_{-0.01}$ & 0.21$^{+0.01}_{-0.01}$ & 0.27$^{+0.00}_{-0.01}$ & 0.32$^{+0.08}_{-0.08}$ & 0.20$^{+0.01}_{-0.01}$ & 0.21$^{+0.01}_{-0.01}$ & 0.22$^{+0.01}_{-0.01}$ & 0.16$^{+0.00}_{-0.00}$ & 0.18$^{+0.00}_{-0.00}$ & 0.16$^{+0.00}_{-0.01}$ & 0.20$^{+0.00}_{-0.01}$ & 0.31$^{+0.01}_{-0.01}$ \\
          & h$_n$(kpc) & 0.24$^{+0.01}_{-0.00}$ & 0.26$^{+0.00}_{-0.01}$ & 0.27$^{+0.01}_{-0.00}$ & 0.38$^{+0.07}_{-0.07}$ & 0.24$^{+0.01}_{-0.00}$ & 0.24$^{+0.01}_{-0.00}$ & 0.22$^{+0.00}_{-0.00}$ & 0.57$^{+0.11}_{-0.11}$ & 0.21$^{+0.00}_{-0.00}$ & 0.20$^{+0.00}_{-0.00}$ & 0.22$^{+0.00}_{-0.01}$ & 0.39$^{+0.07}_{-0.08}$ \bigstrut[b]\\
    \hline
    \multirow{4}[2]{*}{(9.0,9.5)} & ln$\Sigma$(kpc$^{\scriptscriptstyle -2}$) & 
5.77$^{+0.08}_{-0.09}$ & 5.92$^{+0.10}_{-0.10}$ & 5.53$^{+0.07}_{-0.08}$ & 6.04$^{+0.12}_{-0.12}$ & 6.66$^{+0.23}_{-0.24}$ & 6.54$^{+0.28}_{-0.28}$ & 6.50$^{+0.08}_{-0.08}$ & 7.14$^{+0.45}_{-0.46}$ & 6.20$^{+0.16}_{-0.17}$ & 6.08$^{+0.03}_{-0.02}$ & 6.14$^{+0.17}_{-0.18}$ & 6.88$^{+0.06}_{-0.06}$  \bigstrut[t]\\
          & Z$_0$(kpc) & -0.03$^{+0.01}_{-0.00}$ & -0.01$^{+0.01}_{-0.00}$ & 0.01$^{+0.04}_{-0.04}$ & 0.19$^{+0.02}_{-0.03}$ & 0.00$^{+0.01}_{-0.00}$ & 0.03$^{+0.02}_{-0.01}$ & 0.05$^{+0.03}_{-0.02}$ & 0.22$^{+0.01}_{-0.01}$ & -0.01$^{+0.03}_{-0.04}$ & 0.03$^{+0.00}_{-0.01}$ & 0.05$^{+0.00}_{-0.01}$ & 0.17$^{+0.04}_{-0.04}$ \\
          & h$_s$(kpc) & 0.20$^{+0.01}_{-0.01}$ & 0.26$^{+0.03}_{-0.02}$ & 0.30$^{+0.00}_{-0.00}$ & 0.30$^{+0.02}_{-0.02}$ & 0.20$^{+0.01}_{-0.01}$ & 0.24$^{+0.02}_{-0.02}$ & 0.23$^{+0.00}_{-0.00}$ & 0.33$^{+0.08}_{-0.07}$ & 0.17$^{+0.02}_{-0.02}$ & 0.23$^{+0.00}_{-0.01}$ & 0.19$^{+0.00}_{-0.00}$ & 0.23$^{+0.00}_{-0.00}$ \\
          & h$_n$(kpc) & 0.38$^{+0.03}_{-0.04}$ & 0.39$^{+0.03}_{-0.02}$ & 0.42$^{+0.04}_{-0.03}$ & 0.55$^{+0.01}_{-0.02}$ & 0.29$^{+0.00}_{-0.01}$ & 0.29$^{+0.03}_{-0.03}$ & 0.21$^{+0.01}_{-0.00}$ & 0.52$^{+0.00}_{-0.00}$ & 0.22$^{+0.03}_{-0.02}$ & 0.26$^{+0.00}_{-0.01}$ & 0.17$^{+0.01}_{-0.02}$ & 0.50$^{+0.01}_{-0.01}$ \bigstrut[b]\\
    \hline
    \hline
    \end{tabular}%
  \label{tab:result_wl}%
\end{sidewaystable}%

\label{lastpage}
\end{document}